\documentclass[12pt]{article}
\usepackage{amsbsy,amsfonts,amsmath,amssymb,amsthm,euscript,enumerate}
\usepackage{graphicx,color,centernot,framed,extarrows}
\definecolor{shadecolor}{gray}{0.85}
\usepackage{url}
\usepackage{hyperref}
\hypersetup{colorlinks,citecolor=blue,linkcolor=blue,urlcolor=black}
\usepackage{bbm}
\newcommand{\ind}[1]{\mathbbm{1}{\raisebox{-2pt}{$\scriptstyle \{#1\}$}}}
\newcommand{\indic}[1]{\mathbbm{1}{\raisebox{-2pt}{$\scriptstyle #1$}}}

\allowdisplaybreaks[1]

\theoremstyle{plain}
\numberwithin{equation}{section}
\newtheorem{thm}{Theorem}[section]
\newtheorem{ass}[thm]{Assumption}
\newtheorem{cor}[thm]{Corollary}
\newtheorem{lmm}[thm]{Lemma}
\newtheorem{prp}[thm]{Proposition}

\newcommand{\B}{{\mathbb B}}

\newcommand{\betac}{\beta_\mathrm{c}}
\newcommand{\betamf}{\beta_\mathrm{MF}}

\newcommand{\cC}{\mathcal{C}}
\newcommand{\compl}{^\text{c}}

\newcommand{\cL}{\mathcal{L}}

\newcommand{\dc}{d_\mathrm{c}}
\newcommand{\DotU}{\overset{\sss\bullet}U{}}
\newcommand{\DotV}{\overset{\sss\bullet}V{}}
\newcommand{\DotX}{\overset{\sss\bullet}X{}}
\newcommand{\DDotU}{\overset{\sss\bullet\bullet}U{}}
\newcommand{\DDotV}{\overset{\sss\bullet\bullet}V{}}
\newcommand{\DDotX}{\overset{\sss\bullet\bullet}X{}}
\newcommand{\DDDotU}{\overset{\sss\bullet\bullet\bullet}U{}}
\newcommand{\DDDotV}{\overset{\sss\bullet\bullet\bullet}V{}}
\newcommand{\DDDotX}{\overset{\sss\bullet\bullet\bullet}X{}}
\newcommand{\dpst}{\displaystyle}
\newcommand{\Exp}[1]{\langle #1 \rangle}
\newcommand{\lbeq}[1]{\label{eq:#1}}

\newcommand{\mP}{{\mathbb P}}

\newcommand{\nn}{\nonumber}

\newcommand{\Proof}[1]{\paragraph{\it #1}}
\newcommand{\QED}{\hspace*{\fill}\rule{7pt}{7pt}\bigskip}

\newcommand{\refeq}[1]{(\ref{eq:#1})}
\newcommand{\sfL}{\textsf{L}}
\newcommand{\sss}{\scriptscriptstyle}

\newcommand{\veee}[1]{\|\hskip-1pt|#1\|\hskip-1pt|}

\newcommand{\vno}{\varnothing}
\newcommand{\vphi}{\varphi}
\newcommand{\vtri}{\vartriangle}
\newcommand{\Z}{\mathbb{Z}}
\newcommand{\Zd}{\Z^d}

\newcommand{\bk}{\boldsymbol{k}}
\newcommand{\bl}{\boldsymbol{\ell}}
\newcommand{\bm}{\boldsymbol{m}}
\newcommand{\bn}{\boldsymbol{n}}
\newcommand{\bN}{\boldsymbol{N}}
\newcommand{\bvphi}{\boldsymbol{\vphi}}

\makeatletter
\newcommand{\xRightarrow}[2][]{\ext@arrow 0359\Rightarrowfill@{#1}{#2}}
\newcommand{\Db}[2][]{\Leftarrow\hskip-10pt\ext@arrow 0359\Rightarrowfill@{#1}{#2}}
\makeatother

\newcommand{\cn}[2]{\underset{#1}{\overset{#2}{\longleftrightarrow}}}
\newcommand{\db}[2]{\underset{#1}{\overset{#2}\Longleftrightarrow}}

\title{Correct bounds on the Ising lace-expansion coefficients}
\author{
Akira Sakai\footnote{Faculty of Science, Hokkaido University,
Japan. \url{https://orcid.org/0000-0003-0943-7842}}
}
\date{February 22, 2022}

\begin{document}
\maketitle
\begin{abstract}
The lace expansion for the Ising two-point function was successfully derived in 
\cite[Proposition~1.1]{s07}.  It is an identity that involves an alternating 
series of lace-expansion coefficients.  In the same paper, we claimed that the 
expansion coefficients obey certain diagrammatic bounds which imply faster 
$x$-space decay (as the two-point function cubed) above the critical dimension 
$\dc$ ($=4$ for finite-variance models) if the spin-spin coupling is 
ferromagnetic, translation-invariant, summable and symmetric with respect to 
the underlying lattice symmetries.  However, we recently found a flaw in the 
proof of \cite[Lemma~4.2]{s07}, a key lemma to the aforementioned diagrammatic 
bounds.

In this paper, we no longer use the problematic \cite[Lemma~4.2]{s07}, and 
prove new diagrammatic bounds on the expansion coefficients that are slightly 
more complicated than those in \cite[Proposition~4.1]{s07} but nonetheless obey 
the same fast decay above the critical dimension $\dc$.~Consequently, the 
lace-expansion results for the Ising and $\vphi^4$ models in the literature are 
all saved.~The proof is based on the random-current representation and its 
source-switching technique of Griffiths, Hurst and Sherman, combined with a 
double expansion: a lace expansion for the lace-expansion coefficients.
\end{abstract}

\tableofcontents

\section{Background}
The (ferromagnetic) Ising model is a paradigmatic model in statistical physics 
that exhibits a phase transition and critical behavior.  One of the most 
powerful methods to investigate those phenomena is to use the random-current 
representation, which is a sophisticated version of the high-temperature 
expansion and provides a way to translate spin correlations into connectivity 
of the corresponding vertices via paths of bonds with positive current.  It was 
initiated by Griffiths, Hurst and Sherman \cite{ghs70} to prove the GHS 
inequality, and then made the most of it by Aizenman et al., in 1980s.  Since 
then, the random-current representation has given rise to many useful results 
for the Ising model, such as the uniqueness of the critical point and 
mean-field bounds on critical exponents \cite{abf87,ads15}, a sufficient 
condition, known as the bubble condition, for the mean-field behavior 
\cite{a82,af86,ag83} and a sufficient condition for the continuity of the 
spontaneous magnetization \cite{ads15}.  Those sufficient conditions hold in 
dimensions above 4 and 2, respectively, if the critical two-point function 
obeys an infrared bound on the underlying short-range random-walk Green 
function, which is true for reflection-positive models \cite{fss76}.  However, 
the reflection-positivity is too restrictive and may easily be violated by 
slight modification of the spin-spin coupling, such as introducing relatively 
large next-nearest-neighbor interaction.  Moreover, the reflection-positivity 
alone does not imply infrared asymptotics of the critical two-point function, 
i.e., the anomalous dimension $\eta=0$, even in high dimensions; only a 
one-sided inequality is proved.  To prove universal results, it is desirable to 
get rid of this strong symmetry condition.

The lace expansion is one of the few mathematically rigorous methods to 
prove mean-field critical behavior in high dimensions.  Since it does not 
require reflection-positivity, we can deal with a wider class of spin-spin 
couplings.  It is also applied to other models, such as percolation 
\cite{hs90p}, for which it is argued that reflection-positivity does not hold.  
The first lace expansion was invented by Brydges and Spencer \cite{bs85} 
for weakly self-avoiding walk.  Since then, it has been extended to strictly 
self-avoiding walk \cite{hs92}, lattice trees and lattice animals \cite{hs90l}, 
oriented percolation \cite{ny93}, the contact process \cite{s01}, the Ising 
model \cite{s07}, the $|\vphi|^4$ model \cite{bhh19,s15} and the 
random-connection model \cite{hhlm19}; see also \cite{s06} for the development 
of the subject until mid 2000s.  In general, the lace expansion gives rise to a 
recursion equation for the two-point function, which is almost identical to 
that for the Green function of the underlying random walk.  The difference 
between the two is the kernel: an alternating series of the lace-expansion 
coefficients for the former, and the 1-step distribution for the latter.  If 
the alternating series is absolutely convergent, then it can be treated as a 
1-step distribution (after normalization) and the critical two-point function 
exhibits the same infrared asymptotics as the Green function.  Therefore, 
absolute summability of the expansion coefficients (and existence of their 
lower-order moments) is a sufficient condition for the mean-field behavior. 

To prove this sufficient condition for all dimensions above the model-dependent 
upper critical dimension $\dc$, we need correlation inequalities, such as the 
famous BK inequality for percolation (see \cite{br06} for the ever simplest 
proof), with which the expansion coefficients can be bounded by optimal 
diagrams consisting of two-point functions.  For example, the 
$0^\text{th}$-order expansion coefficient for bond percolation is the 
probability that there are at least two bond-disjoint paths of occupied bonds 
from $o$ to $x$, and by the BK inequality, it is bounded by the two-point 
function squared: $\mP_p(o\Rightarrow x)\le\mP_p(o\to x)^2$.  The higher-order 
expansion coefficients for percolation are bounded similarly by diagrams that 
can be decomposed into triangles, which implies $\dc=6$ for percolation.

For the Ising model, there was no equivalent to the BK inequality to control 
the expansion coefficients that are defined by using the aforementioned 
random-current representation.  Inspired by the so-called Source-Switching 
Technique (SST) \cite{ghs70}, which is a way to exchange sources between two 
current configurations, we came up with \cite[Lemma~4.2]{s07} that was supposed 
to provide optimal diagrammatic bounds on the expansion coefficients.  However, 
as explained more in detail in Section~\ref{ss:problem}, we found a flaw in its 
proof, thanks to an inquiry by Duminil-Copin, and the diagrammatic bounds 
\cite[Proposition~4.1]{s07} became no longer reliable; directly affected 
are the proof of the bound on the $0^\text{th}$-order expansion coefficient 
in \cite[pp.306--307]{s07} and \cite[Lemma~4.4]{s07}; 
the rest of that paper is secure. 

In this paper, we prove new diagrammatic bounds on the Ising lace-expansion 
coefficients that are slightly more complicated than those in 
\cite[Proposition~4.1]{s07} but nonetheless obey the same $x$-space decay 
in high dimensions.  As an example, we demonstrate how to derive the wanted 
$x$-space decay from the new diagrammatic bounds for sufficiently spread-out 
(finite-variance) models in dimensions $d>4$; as a byproduct, we obtain better 
multiplicative constants in the $x$-space decay of the lace-expansion 
coefficients of order $j\ge2$ (see Corollary~\ref{cor:pige2bd} below).  The 
proof of those diagrammatic bounds is based on the standard SST and a double 
expansion, i.e., a lace expansion for the expansion coefficients along the 
``earliest'' path of odd current joining the two sources.  See 
Section~\ref{ss:thm1} for more details.

The rest of this paper is organized as follows.  In Section~\ref{ss:ising}, 
we define the model and introduce some notation.  In Section~\ref{ss:rc}, 
we explain the random-current representation.   In Section~\ref{ss:sst}, 
the aforementioned SST and its implications are summarized; one of them 
(Lemma~\ref{lmm:lmm2}) is nontrivial and its proof shares the key idea (i.e., 
the use of the earliest path of odd current) used in the proof of the 
aforementioned new diagrammatic bounds on the expansion coefficients.  
In Section~\ref{ss:le}, we briefly review the lace expansion and its results 
obtained by assuming bounds on the expansion coefficients.  In 
Section~\ref{ss:problem}, we explain why the proof of \cite[Lemma~4.2]{s07}, 
on which the previous diagrammatic bounds in \cite[Proposition~4.1]{s07} rely, 
does not work.  Then, in Section~\ref{s:main}, we present the new diagrammatic 
bounds on four main building blocks of the expansion coefficients.  Those 
bounds are proven in Sections~\ref{ss:thm1}--\ref{ss:thm5}, respectively.  
Finally, in Section~\ref{s:appl2spread-out}, we demonstrate how to use 
the new diagrammatic bounds for the spread-out model in $d>4$.

\section{Definition}
\subsection{The Ising model}\label{ss:ising}
For simplicity, we consider the $d$-dimensional integer lattice $\Zd$ as space 
(this assumption is not essential as long as the concerned graph is transitive, 
as in \cite{s07}).  Let $J:\Zd\to[0,\infty)$ be symmetric in such a way that 
$J(x)=J(y)$ as long as $|x|=|y|$.  Let $\{J_{x,y}\}$ be a collection of 
spin-spin couplings that satisfy $J_{x,y}=J(y-x)$.  
We say that a subset $\Lambda\subset\Zd$ is a connected domain 
if any pair of distinct vertices $x,y\in\Lambda$ are connected by a path of 
bonds in $\B_\Lambda=\{\{x,y\}\subset\Lambda:J_{x,y}>0\}$, i.e., there 
is a sequence $\{v_j\}_{j=0}^n\subset\Lambda$ such that $v_0=x$, $v_n=y$ and 
$\{v_{j-1},v_j\}\in\B_\Lambda$ for all $j=1,\dots,n$.  
We assume $J(o)=0$, i.e., there are no self-bonds.
Given a finite $\Lambda\subset\Zd$, we define the Ising Hamiltonian as
\begin{align}\lbeq{hamiltonian}
H_\Lambda(\bvphi)=-\sum_{\{x,y\}\subset\Lambda}J_{x,y}\vphi_x\vphi_y,
\end{align}
where $\bvphi=\{\vphi_x\}_{x\in\Lambda}\in\{\pm1\}^\Lambda$ is a spin 
configuration.  
Then, we define the finite-volume two-point function and its infinite-volume 
limit at the inverse temperature $\beta\in[0,\infty)$ as
\begin{align}\lbeq{2pt-def}
\Exp{\vphi_x\vphi_y}_{\beta,\Lambda}=\frac{\dpst\sum_{\bvphi\in\{\pm1\}^\Lambda}
 \vphi_x\vphi_ye^{-\beta H_\Lambda(\bvphi)}}{\dpst\sum_{\bvphi\in\{\pm1
 \}^\Lambda}e^{-\beta H_\Lambda(\bvphi)}},&&
G_\beta(x)=\lim_{\Lambda\uparrow\Zd}\Exp{\vphi_o\vphi_x}_{\beta,\Lambda},
\end{align}
where the limit is nonnegative and unique due to monotonicity in terms of 
volume-increasing limits (see Lemma~\ref{lmm:G} below).  
The summable model (i.e., $\sum_xJ(x)<\infty$) is known to exhibit 
a phase transition at the critical point defined by
\begin{align}
\betac=\sup\bigg\{\beta\ge0:\sum_{x\in\Zd}G(x)<\infty\bigg\}.
\end{align}

\subsection{The random-current representation}\label{ss:rc}
\begin{figure}[t]
\begin{center}
\includegraphics[scale=0.44]{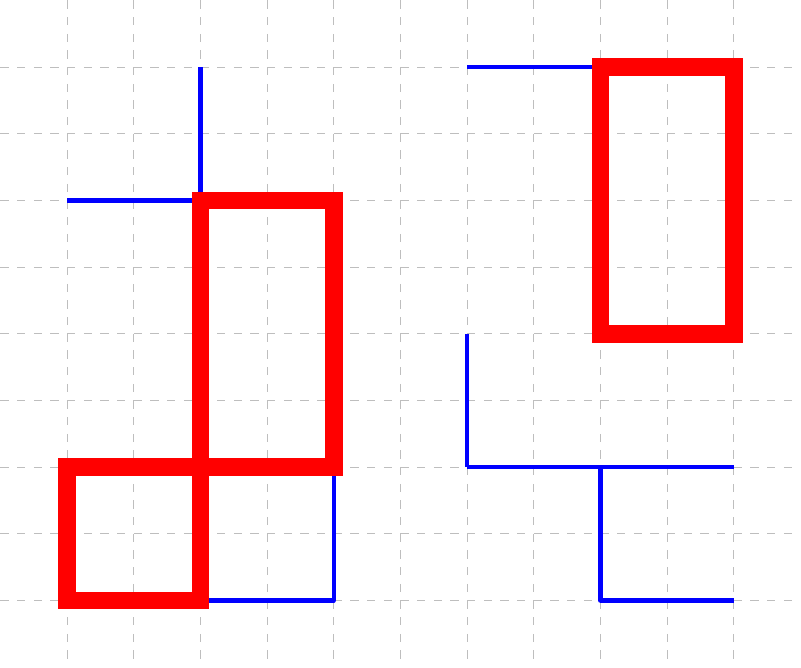}\hskip4pc
\includegraphics[scale=0.44]{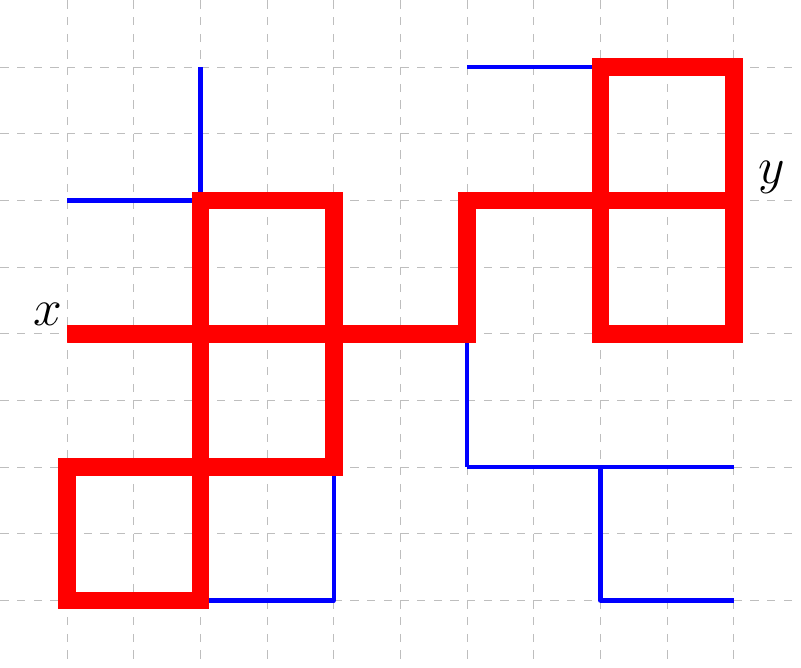}
\end{center}
\caption{Current configurations satisfying the source constraint in \refeq{RCD} 
(left) and that in \refeq{RCN} (right).  Bonds with odd current are bold (in 
red), while those with positive-even current are thin-solid (in blue).}
\label{fig:RC}
\end{figure}

Let $\Z_+=\{0,1,2,\dots\}$.  Given a finite $\Lambda\subset\Zd$ and 
a current configuration $\bn=\{n_b\}_{b\in\B_\Lambda}\in\Z_+^{\B_\Lambda}$ and 
a bond set $B\subset\B_\Lambda$, we define the source set and the weight on 
$B$ as
\begin{align}
\partial\bn=\bigg\{x:\sum_{b\ni x}n_b\text{ is odd}\bigg\},&&
w_B(\bn)=\prod_{\{x,y\}\in B}
 \frac{(\beta J_{x,y})^{n_{x,y}}}{n_{x,y}!}.
\end{align}
Then, by simple arithmatic, we obtain the rewrite
\begin{gather}
\sum_{\bvphi\in\{\pm1\}^\Lambda}e^{-\beta H_\Lambda(\bvphi)}
=\sum_{\bvphi\in\{\pm1\}^\Lambda}\prod_{\{x,y\}\in\B_\Lambda}e^{J_{x,y}\vphi_x
 \vphi_y}
=\sum_{\bvphi\in\{\pm1\}^\Lambda}\prod_{\{x,y\}\in\B_\Lambda}\sum_{n_{x,y}\in
 \Z_+}\frac{(\beta J_{x,y}\vphi_x\vphi_y)^{n_{x,y}}}{n_{x,y}!}\nn\\[5pt]
=\sum_{\bn\in\Z_+^{\B_\Lambda}}w_{\B_\Lambda}(\bn)\prod_{x\in\Lambda}
 \underbrace{\sum_{\vphi_x=\pm1}\vphi_x^{\sum_{b\ni x}n_b}}_{2\times\indic{\sss
 \{x\notin\partial\bn\}}}
=2^{|\Lambda|}\sum_{\substack{\bn\in\Z_+^{\B_\Lambda}:\\ \partial\bn=\vno}}
 w_{\B_\Lambda}(\bn),\lbeq{RCD}
\end{gather}
where $\ind{\cdots}$ is the indicator function.  By this representation, we can 
interpret the partition function (= the denominator in the definition of the 
two-point function) as the sum of the weight over the current configurations 
in which bonds with odd current form loops (see the left of 
Figure~\ref{fig:RC}).  
Similarly, we have
\begin{align}\lbeq{RCN}
\sum_{\bvphi\in\{\pm1\}^\Lambda}\vphi_x\vphi_ye^{-\beta H_\Lambda(\bvphi)}
 =2^{|\Lambda|}\sum_{\substack{\bn\in\Z_+^{\B_\Lambda}:\\ \partial\bn=x\vtri
 y}}w_{\B_\Lambda}
 (\bn),
\end{align}
where $x\vtri y$ is an abbreviation for the heavier notation of symmetric 
difference $\{x\}\triangle\{y\}$.  By this representation, we can interpret 
the numerator in the definition of the two-point function as the sum of  the 
weight over the current configurations in which there is a path of 
bonds with odd current between $x$ and $y$ in the sea of loops with odd 
current (see the right of Figure~\ref{fig:RC}).  As a result, we obtain the 
random-current representation for the two-point function: for any 
$B\subset\B_\Lambda$,
\begin{align}\lbeq{RCrepr}
\Exp{\vphi_x\vphi_y}_B=\sum_{\substack{\bn\in\Z_+^B:\\ \partial\bn
 =x\vtri y}}\frac{w_B(\bn)}{Z_B},&&
\Exp{\vphi_x\vphi_y}_{\B_\Lambda}=\Exp{\vphi_x\vphi_y}_{\beta,\Lambda},
\end{align}
where
\begin{align}
Z_B=\sum_{\substack{\bn\in\Z_+^B:\\ \partial\bn=\vno}}w_B(\bn).
\end{align}
From now on, we omit the $\beta$-dependence if unnecessary, such as 
$G(x)=G_\beta(x)$.  Similarly, we have the random-current representation for 
the four-point function:
\begin{align}\lbeq{4pt}
\Exp{\vphi_x\vphi_y\vphi_u\vphi_v}_B=\sum_{\substack{\bn\in\Z_+^B:\\
 \partial\bn=x\vtri y\vtri u\vtri v}}\frac{w_B(\bn)}{Z_B}.
\end{align}

\subsection{The source-switching technique (SST)}\label{ss:sst}
One of the advantages of the random-current representation (as compared to 
other similar representations, such as the high-temperature expansion) is the 
following source-switching technique (SST) of Griffiths, Hurst and Sherman 
\cite{ghs70}.  It provides a way to exchange sources between two current 
configurations.  

\begin{shaded}
\begin{lmm}[SST, e.g., Lemma~2.3 in \cite{s07}]\label{lmm:SST}
For any finite $B\subset B'\subset\B_{\Zd}$, $x,y\in V(B)$ (= the set of end vertices 
of bonds in $B$) and $\bN\in\Z_+^{B'}$,
\begin{align}\lbeq{originalSST}
\sum_{\substack{\bn\in\Z_+^B:\\ \partial\bn=x\vtri y}}\prod_{b\in B}\binom{N_b}
 {n_b}=\ind{x\cn{\bN}{}y\text{ in }B}\sum_{\substack{\bn\in\Z_+^B:\\ \partial\bn
 =\vno}}\prod_{b\in B}\binom{N_b}{n_b},
\end{align}
where $x\cn{\bN}{}y$ in $B$ means either $x=y$ or there is a path from $x$ 
to $y$ consisting of bonds $b\in B$ with positive current $N_b>0$.
\end{lmm}
\end{shaded}

Using the random-current representation and the SST, we can easily show the 
following consequences of Griffiths' inequalities \cite{g67a} for Ising ferromagnets.

\begin{shaded}
\begin{lmm}\label{lmm:G}
For every $x,y\in\Zd$ and $\beta\ge0$, the two-point function 
$\Exp{\vphi_x\vphi_y}_B$, provided $x,y\in V(B)$, is nonnegative and 
nondecreasing in terms of the bond set $B$.  As a result, there is a unique 
translation-invariant infinite-volume limit 
$G(y-x)=\lim_{\Lambda\uparrow\Zd}\Exp{\vphi_x\vphi_y}_{\B_\Lambda}$.
\end{lmm}
\end{shaded}

\Proof{Proof.}
Since $J_b\ge0$ for any $b\in\B_{\Zd}$, the weight $w_B(\bn)$ in \refeq{RCrepr} 
is nonnegative for any $B\subset\B_{\Zd}$, and so are $\Exp{\vphi_x\vphi_y}_B$ 
and its infinite-volume limit as $B\uparrow\B_{\Zd}$ (if it exists).  To prove monotonicity, 
we consider $B\subset B'\subset\B_{\Zd}$.  By the random-current representation 
\refeq{RCrepr}, the difference of the two-point functions on $B'$ and $B$ is 
written as
\begin{align}
\Exp{\vphi_x\vphi_y}_{B'}-\Exp{\vphi_x\vphi_y}_B
&=\sum_{\substack{\bm\in\Z_+^{B'}:\\ \partial\bm=x\vtri y}}\frac{w_{B'}(\bm)}
 {Z_{B'}}-\sum_{\substack{\bn\in\Z_+^B:\\ \partial\bn=x\vtri y}}\frac{w_B(\bn)}
 {Z_B}\nn\\
&=\sum_{\substack{\bm\in\Z_+^{B'}:\\ \partial\bm=x\vtri y}}\sum_{\substack{\bn
 \in\Z_+^B:\\ \partial\bn=\vno}}\frac{w_{B'}(\bm)}{Z_{B'}}\frac{w_B(\bn)}{Z_B}
 -\sum_{\substack{\bm\in\Z_+^{B'}:\\ \partial\bm=\vno}}\sum_{\substack{\bn\in
 \Z_+^B:\\ \partial\bn=x\vtri y}}\frac{w_{B'}(\bm)}{Z_{B'}}\frac{w_B(\bn)}{Z_B}
 \nn\\
&=\sum_{\substack{\bN\in\Z_+^{B'}:\\ \partial\bN=x\vtri y}}\frac{w_{B'}(\bN)}
 {Z_{B'}Z_B}\Bigg(\sum_{\substack{\bn\in\Z_+^B:\\ \partial\bn=\vno}}\prod_{b
 \in B}\binom{N_b}{n_b}-\sum_{\substack{\bn\in\Z_+^B:\\ \partial\bn=x\vtri y}}
 \prod_{b\in B}\binom{N_b}{n_b}\Bigg).
\end{align}
However, by \refeq{originalSST}, the last line equals 
\begin{align}
&\sum_{\substack{\bN\in\Z_+^{B'}:\\ \partial\bN=x\vtri y}}\frac{w_{B'}(\bN)}
 {Z_{B'}Z_B}\sum_{\substack{\bn\in\Z_+^B:\\ \partial\bn=\vno}}\prod_{b\in B}
 \binom{N_b}{n_b}\Big(1-\ind{x\cn{\bN}{}y\text{ in }B}\Big)\nn\\
&=\sum_{\substack{\bm\in\Z_+^{B'}:\\ \partial\bm=x\vtri y}}\sum_{\substack{\bn
 \in\Z_+^B:\\ \partial\bn=\vno}}\frac{w_{B'}(\bm)}{Z_{B'}}\frac{w_B(\bn)}{Z_B}
 \Big(1-\ind{x\cn{\bm+\bn}{}y\text{ in }B}\Big),
\end{align}
which is nonnegative.  This proves monotonicity of $\Exp{\vphi_x\vphi_y}_B$ in 
$B$ and the uniqueness of the infinite-volume limit.  To prove 
translation-invariance of the limit, we take two hypercubes 
$\Lambda(o)\subset\Lambda'(o)\subset\Zd$, both centered at $o$, 
such that $\B_{\Lambda(x)}\subset B\subset\B_{\Lambda'(x)}$, where 
$\Lambda(x)$ is the translation of $\Lambda(o)$ by $x$.  Then, by the 
monotonicity, we have 
\begin{align}
\Exp{\vphi_o\vphi_{y-x}}_{\B_{\Lambda(o)}}=\Exp{\vphi_x\vphi_y}_{\B_{\Lambda
 (x)}}\le\Exp{\vphi_x\vphi_y}_B\le\Exp{\vphi_x\vphi_y}_{\B_{\Lambda'(x)}}=
 \Exp{\vphi_o\vphi_{y-x}}_{\B_{\Lambda'(o)}}.
\end{align}
Since both ends go to the same limit $G(y-x)$, this proves 
translation-invariance of the limit.
\QED

Other relevant results already proven in the previous work by using the 
random-current representation and the SST are summarized as follows:

\begin{shaded}
\begin{lmm}\label{lmm:lmm1}
\begin{enumerate}[(i)]
\item
For any $x\ne o$,
\begin{align}\lbeq{tildeGdef}
G(x)\le\tilde G(x)=(\tau*G)(x),
\end{align}
where $*$ represents a convolution in $\Zd$:~
$(\tau*G)(x)=\sum_{y\in\Zd}\tau(y)\,G(x-y)$.
\item
For any finite  $B\subset\B_{\Zd}$ with $o,x,y\in V(B)$,
\begin{align}\lbeq{lmm1}
\sum_{\substack{\bn\in\Z_+^B:\\ \partial\bn=o\vtri x}}\frac{w_B(\bn)}{Z_B}\,
 \ind{o\cn{\bn}{}y}&\le G(y)\,G(x-y).
\end{align}
In particular, when $x=o$,
\begin{align}\lbeq{lmm0}
\sum_{\substack{\bn\in\Z_+^B:\\ \partial\bn=\vno}}\frac{w_B(\bn)}{Z_B}\,\ind{o
 \cn{\bn}{}x}&\le G(x)^2.
\end{align}
\item
For any finite $B,B'\subset\B_{\Zd}$ with $o,x,y\in V(B)$,
\begin{align}\lbeq{bubblechain}
\sum_{\substack{\bm\in\Z_+^{B'}:\\ \partial\bm=\vno}}\sum_{\substack{\bn\in
 \Z_+^B:\\ \partial\bn=o\vtri x}}\frac{w_{B'}(\bm)}{Z_{B'}}\frac{w_B(\bn)}{Z_B}
 \,\ind{o\cn{\bm+\bn}{}y\text{ in }B}\le\sum_vG(v)\,G(x-v)
 \sum_{j=0}^\infty(\tilde G^2)^{*j}(y-v),
\end{align}
where $(\tilde G^2)^{*j}$ is the $j$-fold convolution of 
$\tilde G^2$ (recall the definition of $\tilde G$ in \refeq{tildeGdef}).
\end{enumerate}
\end{lmm}
\end{shaded}

\Proof{Sketch proof.}
The result (i) is obtained by taking the infinite-volume limit of a 
finite-volume version in \cite[(2.34)--(2.35)]{cs15}.  The result (ii) is 
obtained by multiplying the left-hand side of \refeq{lmm1} by 
$1=\sum_{\partial\bm=\vno}w_B(\bm)/Z_B$, then using 
$\ind{o\cn{\bn}{}y}\le\ind{o\cn{\bm+\bn}{}y}$, and finally applying the SST.  
The result (iii) is a simple extension of \cite[(4.51)--(4.62)]{s07}.
\QED

\begin{figure}[t]
\[ T(o,x,y)=~~\raisebox{-2.3pc}{\includegraphics[scale=0.5]{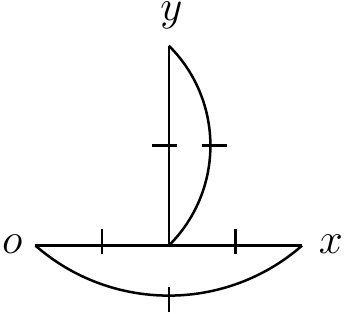}}~~
 +~~\raisebox{-2pc}{\includegraphics[scale=0.5]{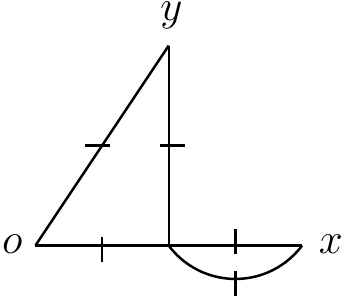}}~~
 +~~\raisebox{-2pc}{\includegraphics[scale=0.5]{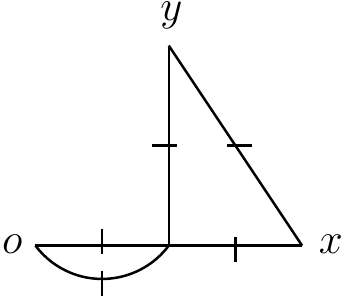}} \]
\caption{A schematic representation of $T(o,x,y)$.  The slashed line segments 
represent $G$.}
\label{fig:Tdef}
\end{figure}

Let (cf., Figure~\ref{fig:Tdef})
\begin{align}\lbeq{Tdef}
T(o,x,y)&=\sum_zG(z)\,G(x-z)\,G(z-y)\nn\\
&\qquad\times\Big(G(x)\,G(z-y)+G(y)\,G(z-x)+G(z)\,
 G(y-x)\Big).
\end{align}
We will later use the following lemma to bound the $1^\text{st}$- and 
higher-order expansion coefficients (cf., 
\refeq{ddotUdef}--\refeq{ddotVdef} and \refeq{dddotUdef}--\refeq{dddotVdef}).

\begin{shaded}
\begin{lmm}\label{lmm:lmm2}
For any finite $B\subset\B_\Lambda$ with $o,x,y\in V(B)$,
\begin{align}\lbeq{lmm2}
\sum_{\substack{\bn\in\Z_+^B:\\ \partial\bn=\vno}}\frac{w_B(\bn)}{Z_B}\ind{o
 \cn{\bn}{}x\}\cap\{o\cn{\bn}{}y}\le T(o,x,y).
\end{align}
\end{lmm}
\end{shaded}
The proof of \refeq{lmm2} turns out to require the same idea as in the 
proof of new diagrammatic bounds on the expansion coefficients, as well as 
Lebowitz' inequality \cite{l74}, by which a four-point function is bounded by 
the three terms in the second line of \refeq{Tdef}.  We prove 
Lemma~\ref{lmm:lmm2} in Section~\ref{ss:thm1}.

\subsection{The lace expansion}\label{ss:le}
By heavy use of the random-current representation and the SST, we derived 
in \cite[Section~2.2]{s07} the lace expansion (= the recursion equation 
\refeq{laceexp} below) for the two-point function.  To explain it, we first define
\begin{align}
\tau(x)=\tanh\big(\beta J(x)\big),
\end{align}
which is zero for $x=o$, as $J(o)=0$ (see above \refeq{hamiltonian}).
Then, for any $\beta\ge0$, there exist lace-expansion coefficients 
$\{\pi_{\B_\Lambda}^{\sss(i)}\}_{i\in\Z_+}$, which are nonnegative 
functions on $\Lambda$ for ferromagnetic models, such that the following 
identity holds for every $j\in\Z_+$ \cite[Proposition~1.1]{s07}:
\begin{align}\lbeq{laceexp}
\Exp{\vphi_o\vphi_x}_{\B_\Lambda}=\Pi_{\B_\Lambda}^{\sss(j)}(x)
 +\sum_{v\in\Lambda}(\Pi_{\B_\Lambda}^{\sss(j)}*\tau)(o,v)\,
 \Exp{\vphi_v\vphi_x}_{\B_\Lambda}+(-1)^{j+1}R_{\B_\Lambda}^{\sss(j+1)}(x),
\end{align}
where the remainder $R_{\B_\Lambda}^{\sss(j+1)}$ is nonnegative 
for ferromagnetic models, and
\begin{align}
\Pi_{\B_\Lambda}^{\sss(j)}(x)=\sum_{i=0}^j(-1)^i\pi_{\B_\Lambda}^{\sss(i)}(x).
\end{align}
In fact, the identity \refeq{laceexp} holds independently of the signs of the 
spin-spin couplings.  However, as defined in the beginning of this section, 
if we restrict our attention to ferromagnetic models, then 
$\pi_{\B_\Lambda}^{\sss(j)}(x)$ and $R_{\B_\Lambda}^{\sss(j+1)}(x)$ 
are proven at the end of \cite[Section~2.2.3]{s07} to obey the bounds
\begin{align}
\pi_{\B_\Lambda}^{\sss(j)}(x)\ge\delta_{j,0}\delta_{o,x},&&
R_{\B_\Lambda}^{\sss(j+1)}(x)\le\sum_{v\in\Lambda}(\pi_{\B_\Lambda}^{\sss(j)}
 *\tau)(o,v)\,\Exp{\vphi_v\vphi_x}_{\B_\Lambda}.
\end{align}

We note that the lace expansion \refeq{laceexp} looks similar to the recursion 
equation for the random-walk Green function $S_p$ generated by the 1-step 
distribution $D$ with fugacity $p$:
\begin{align}
S_p(x)=\delta_{o,x}+pD*S_p(x).
\end{align}
If $D(x)$ decays faster than $|x|^{-d-\alpha}$ for some $\alpha>2$ (hence 
$\sigma^2=\sum_x|x|^2D(x)<\infty$), then the critical Green function $S_1(x)$ 
exists in dimensions $d>2$ and exhibits the asymptotic behavior 
$\sim\frac{a_d}{\sigma^2}|x|^{2-d}$ as $|x|\uparrow\infty$, where 
$a_d=\frac{d}2\Gamma(\frac{d-2}2)\pi^{-d/2}$, which is obtained by the 
time-integral of the $d$-dimensional heat kernel defined by the diagonal 
covariance matrix with all entries $1/d$ (see, e.g., \cite{cs15}).

Suppose that $J$ is the spread-out interaction with parameter 
$L\in[1,\infty)$ of the form
\begin{align}\lbeq{spread-out}
J(x)=L^{-d}h(x/L)\qquad[x\ne o],
\end{align}
where $h:[-1,1]^d\to[0,\infty)$ is a bounded probability distribution, 
which is piecewise continuous and symmetric with respect to rotations 
by $\pi/2$ and reflections against cordinate hyperplanes.  
It has been shown (the latest reference is \cite[Theorem~1.3]{cs19}, 
where we regard $\alpha=\infty$) that, if $d>2$, then
\begin{align}\lbeq{theta-def}
\theta=\sup_{x\ne o}\frac{S_1(x)}{(|x|\vee L)^{2-d}}=O(L^{-2}),
\end{align}
where $S_1$ is generated by the 1-step distribution $D=\tau/\|\tau\|_1$.  
It has also been shown (the latest reference is \cite[Section~3.3]{cs15}, 
where we regard $\alpha=\infty$) that, if $d>4$, $\theta\ll1$ and 
\begin{align}\lbeq{aboveassumption}
\sup_{j\in\Z_+}|\Pi_{\B_\Lambda}^{\sss(j)}(x)-\delta_{o,x}|\le O(L^{-d})\,
 \delta_{o,x}+\frac{O(\theta^3)}{(|x|\vee L)^{3(d-2)}},&&
R_{\B_\Lambda}^{\sss(j)}(x)\xrightarrow[j\uparrow\infty]{}0
\end{align}
hold uniformly in $x\in\Lambda\subset\Zd$ and $\beta\le\betac$, then 
the aforementioned similarity to random walk is justified and that 
$G_{\betac}(x)\sim Aa_d|x|^{2-d}$ as $|x|\uparrow\infty$, 
where, by denoting the limit $\lim_{\beta\uparrow\betac}\lim_{\Lambda\uparrow
\Zd}\lim_{j\uparrow\infty}\Pi_{\beta,\B_\Lambda}^{\sss(j)}$ by $\Pi_{\betac}$, 
\begin{align}\lbeq{A-def}
A=\frac1{\|\tau_{\betac}\|_1}\Bigg(\sum_{x\in\Zd}|x|^2\bigg(D(x)
 +\frac{\Pi_{\betac}(x)}{\sum_y\Pi_{\betac}(y)}\bigg)\Bigg)^{-1}.
\end{align}

For the nearest-neighbor model, where $J(x)=\ind{|x|=1}$, the same asymptotic 
behavior $G_{\betac}(x)\sim Aa_d|x|^{2-d}$ as $|x|\uparrow\infty$ has been 
shown for $d\gg4$ \cite[Proposition~3.3(i)]{s07} provided that
\begin{align}
&\sum_x|\Pi_{\B_\Lambda}^{\sss(j)}(x)|=1+O(\theta),&
&\sum_x|x|^2|\Pi_{\B_\Lambda}^{\sss(j)}(x)|=O(\theta),\lbeq{nncond1}\\
&|\Pi_{\B_\Lambda}^{\sss(j)}(x)|=\frac{O(1)}{(|x|\vee1)^{d+2}},&
&R_{\B_\Lambda}^{\sss(j)}(x)\xrightarrow[j\uparrow\infty]{}0\lbeq{nncond2}
\end{align}
hold uniformly in $j\in\Z_+$, $x\in\Lambda\subset\Zd$ and $\beta\le\betac$, 
where 
\begin{align}\lbeq{nntheta}
\theta=\max\Big\{\|D*S_1^{*2}\|_\infty,~\sup_xx_1^2S_1(x)\Big\}=O\big((d
 -4)^{-1}\big).
\end{align} 
The implicit constants in $O(\theta)$ in \refeq{nncond1} and in $O((d-4)^{-1})$ 
in \refeq{nntheta} are independent of $d$, but that in $O(1)$ in \refeq{nncond2} 
may be large depending on $d$.  Using the obtained asymptotics of 
$G_{\betac}(x)$, we can improve the power exponent $d+2$ in \refeq{nncond2} to 
the same in \refeq{aboveassumption}, i.e., $3(d-2)=d+2+4(d-4)$, and thanks to 
this excess power, the correction term to the leading asymptotics of 
$G_{\betac}(x)$ can be evaluated (see, e.g., \cite[Theorem~1.4]{h08}).  Due to 
the extra assumption \refeq{nncond1}, which is to ensure convergence of the 
lace expansion, the nearest-neighbor model is more cumbersome than the 
spread-out model\footnote{The proof for the nearest-neighbor model is slightly 
different from the spread-out model.  It goes roughly as follows.  First, by 
assuming \refeq{nncond1} and convergence of the remainder term in 
\refeq{nncond2}, we can show that, for $d\gg4$, 
\begin{align}
\max\Big\{\|\tau_\beta\|_1-1,~\|D*G_\beta^{*2}\|_\infty,~\sup_xx_1^2
 G_\beta(x)\Big\}=O\big((d-4)^{-1}\big),
\end{align}
uniformly in $\beta<\betac$.  Applying this to the desired diagrammatic bounds 
on the expansion coefficients, which we prove in this paper, we can show 
that \refeq{nncond1} and convergence of the remainder term in \refeq{nncond2} 
indeed hold for all $d\gg4$.  By the standard continuity argument, the above 
bounds can be extended all the way to $\beta=\betac$.  Since 
$\sum_x|x|^r|\Pi_{\B_\Lambda}(x)|$ for $r=0,2$ 
is bounded, we can show that $\bar G^{\sss(r)}$ is finite as long as $r<d-2$ 
and that $\bar W^{\sss(r)}$ is finite as long as $r<d-4$ 
\cite[Proposition~3.3(ii)]{s07}, where
\begin{align}
\bar G^{\sss(r)}=\sup_x|x|^rG_{\betac}(x),&&
\bar W^{\sss(r)}=\sup_x\sum_y|y|^rG_{\betac}(y)\,G_{\betac}(x-y).
\end{align}
Then, by using $\bar G^{\sss(2)}<\infty$ and $\bar W^{\sss(r)}<\infty$, 
we can show that $\sum_x|x|^{r+2}|\Pi_{\B_\Lambda}(x)|$ is bounded 
\cite[Proposition~3.3(iii)]{s07}.  Repeat this procedure until $r$ reaches 
$(d+2)/3~(<d-2)$, so that $G_{\betac}(x)=O(|x|^{-(d+2)/3})$.  Applying 
this to the desired diagrammatic bounds on the expansion coefficients, 
we can show the pointwise bound in \refeq{nncond2}.  For those who 
want to know more, please refer to \cite[Section~3.2]{s07}.}.  
But, nonetheless, it is doable.  So far, so good...

\subsection{Problematic bounds on the expansion coefficients}\label{ss:problem}
To verify the above assumptions in \refeq{aboveassumption} and 
\refeq{nncond1}--\refeq{nncond2}, we want to bound the 
lace-expansion coefficients by diagrams consisting of two-point functions $G$.  
For example, we claimed in \cite{s07} that
\begin{align}\lbeq{wrongbd0}
\pi_{\B_\Lambda}^{\sss(0)}(x)=\sum_{\substack{\bn\in\Z_+^{\B_\Lambda}:\\
 \partial\bn=o\vtri x}}\frac{w_{\B_\Lambda}(\bn)}{Z_{\B_\Lambda}}\ind{o\db{\bn}
 {}x}\stackrel{(*)}\le\Exp{\vphi_o\vphi_x}_{\B_\Lambda}^3\le G(x)^3,
\end{align}
where $o\db{\bn}{}x$ means either $o=x$ or there are at least two bond-disjoint 
paths from $o$ to $x$ consisting of bonds $b$ with positive current 
$n_b>0$.  The last inequality is due to the monotonicity in volume mentioned 
below \refeq{2pt-def}.  The inequality $(*)$ is the issue to be 
discussed in this paper.

We also claimed in \cite{s07} that the higher-order expansion coefficients 
$\pi_{\B_\Lambda}^{\sss(j)}(x)$, $j\ge1$, were bounded similarly, but by more 
involved diagrams.  There are two key lemmas to show those diagrammatic 
bounds: \cite[Lemmas~4.3 \& 4.4]{s07}.  The former is affirmatively obtained by 
repeated use of the SST.  On the other hand, the latter (= 
\cite[Lemma~4.4]{s07}) is based on the same idea used in showing the inequality 
$(*)$ in \refeq{wrongbd0}.  As a result, we cannot justify the credibility of the 
diagrammatic bounds in \cite[Proposition~4.1]{s07}.

The common culprit is \cite[Lemma~4.2]{s07}, which was supposed to be an 
extension of the SST and to derive a similar inequality to the BK inequality 
for percolation.  As seen in \refeq{originalSST}, the identity due to the SST 
holds if and only if $o$ is connected to $y$ via a path of bonds in $B$ with 
positive current in the superposition 
$\bm+\bn=\{m_b+n_b\}_{b\in\B_\Lambda}$.  Any such path can be used to 
define a bijection\footnote{In fact, what we do is to consider the multigraph 
$G_{\bm+\bn}=(\Lambda,\B_\Lambda^{\bm+\bn})$, where each bond 
$b\in\B_\Lambda$ is duplicated $m_b+n_b$ times.  Choose any path 
$(o=v_0,v_1,\dots,v_n=y)$ of bonds $b_j=\{v_{j-1},v_j\}$ with 
$m_{b_j}+n_{b_j}>0$, let $e_j=\{v_{j-1},v_j\}$ be one of those 
$m_{b_j}+n_{b_j}$ edges, and set $\omega=\{e_j\}_{j=1}^n$.  Then, the 
aforementioned bijection is defined by taking the symmetric difference between 
$\omega$ and $E\subset\B_\Lambda^{\bm+\bn}$ with the set of odd-degree 
vertices $\partial E=o\vtri x$; the image satisfies the required condition 
$\partial(E\,\triangle\,\omega)=y\vtri x$.} between two sets of pairs 
$(\bm,\bn)$ with $\bm+\bn$ fixed: one with $\partial\bm=\vno$, 
$\partial\bn=o\vtri x$ and the other with $\partial\bm=o\vtri y$, 
$\partial\bn=y\vtri x$.  To generalize this idea to deal with bond-disjoint 
connections, such as $o\db{\bn}{}x$, and exchange sources among more than two 
current configurations simultaneously, in the proof of \cite[Lemma~4.2]{s07} we 
used the ``earliest'' path from $o$ to $x$ and another disjoint one to define 
a bijection between two sets of triples $(\bn,\bm_1,\bm_2)$ with $\bn+\bm_1$ 
and $\bn+\bm_2$ fixed: one with the source constraint $\partial\bn=o\vtri x$, 
$\partial\bm_1=\partial\bm_2=\vno$ and the other with 
$\partial\bn=\partial\bm_1=\partial\bm_2=o\vtri x$.  The ordering used in 
defining the earliest path was non-local, so as to ensure existence of another 
unaffected path after removal of the earliest.  It turns out that this 
non-local rule disrupts construction of a bijection; for some cases, the image 
is empty because the first two earliest paths used to define the bijection are 
no longer the first two earliest in the image.  Therefore, we have decided to 
abandon \cite[Lemma~4.2]{s07} and to seek alternative diagrammatic bounds on 
the expansion coefficients.

\section{Main results}\label{s:main}
\subsection{Results for the $0^\text{th}$-order expansion coefficient}
\begin{figure}[t]
\begin{align*}
U^m(y,z;y',z')=\sum_{j=0}^m~\raisebox{-1.7pc}{\includegraphics[scale=0.3]
 {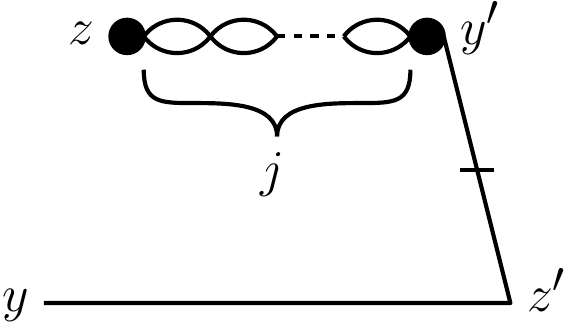}}&&&&
V^m(y,z;x)=\sum_{j=1}^m~\raisebox{-1.8pc}{\includegraphics[scale=0.3]
 {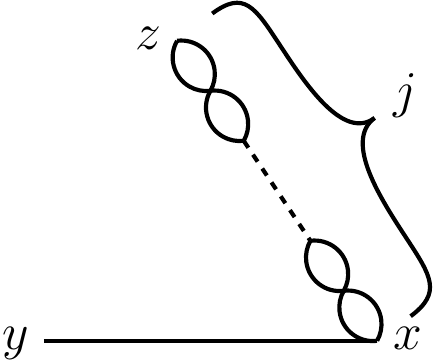}}
\end{align*}
\caption{Schematic representations of $U^m(y,z;y',z')$ and $V^m(y,z;x)$ for 
$m\ge1$.  The slashed line segment represents $G$ (as in 
Figure~\ref{fig:Tdef}), while the other unslashed ones represent 
$\tilde G=\tau*G$.  In addition, the small filled discs represent 
$\delta+\tau^2$.}
\label{fig:UVdef}
\end{figure}

First we recall \refeq{tildeGdef}:
\begin{align}
\tilde G(x)=(\tau*G)(x).
\end{align}
To explain the new diagrammatic bounds on the expansion coefficients, we define
(see Figure~\ref{fig:UVdef})
\begin{align}
U^m(y,z;y',z')&=\tilde G(z'-y)\,G(z'-y')\nn\\
&\quad\times
 \begin{cases}
 G(y'-z)^2&(m=0),\\
 \Big((\delta+\tau^2)*\sum_{j=0}^m(\tilde G^2)^{*j}*(\delta+\tau^2)\Big)(y'-z)
  &(m\ge1),
 \end{cases}\lbeq{Um-def}\\
V^m(y,z;x)&=\tilde G(x-y)\sum_{j=1}^m(\tilde G^2)^{*j}(x-z),
\end{align}
where $(\tilde G^2)^{*0}(x)=\delta_{o,x}$ by convention.  
Also, we define 
\begin{align}
X^m_{o,x}&=\sum_{i=0}^\infty\sum_{\substack{u_0,\dots,u_i,\\ v_0,\dots,v_i:\\
 u_0=v_0=o}}\prod_{j=1}^iU^m(u_{j-1},v_{j-1};u_j,v_j)\,V^m(u_i,v_i;x),
\end{align}
where the empty product $\prod_{j=1}^0$ for $i=0$ is regarded as 1.  
To make it short, we abbreviate this to
\begin{align}\lbeq{Xdef}
X^m_{o,x}=\sum_{i=0}^\infty\big((U^m)^{\star i}\star V^m\big)_{o,x}.
\end{align}
In the following diagrammatic bounds (in Theorems~\ref{thm:pi0bd}, 
\ref{thm:Theta'bd}, \ref{thm:pi0'bd} and \ref{thm:Theta''bd}), we only 
use the two extremes: $m=1$ or $\infty$ (cf., e.g., \refeq{Theta'bd}); 
$U^0$ is used only for quantitative estimates (cf., e.g., \refeq{UmVmbds}).

\begin{shaded}
\begin{thm}\label{thm:pi0bd}
Let $x\ne o$.  For the ferromagnetic models defined above \refeq{hamiltonian}, 
\begin{align}\lbeq{pi0diagbd}
\pi_{\B_\Lambda}^{\sss(0)}(x)&\le2X^1_{o,x}\nn\\
&=2\bigg(\raisebox{-6pt}{\includegraphics[scale=0.27]{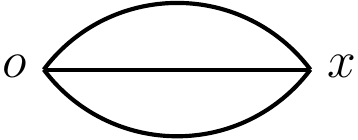}}
 +\sum_{j=0}^1~\raisebox{-16pt}{\includegraphics[scale=0.27]{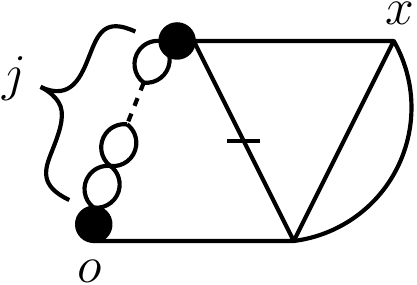}}~
 +\sum_{i,j=0}^1\raisebox{-16pt}{\includegraphics[scale=0.27]{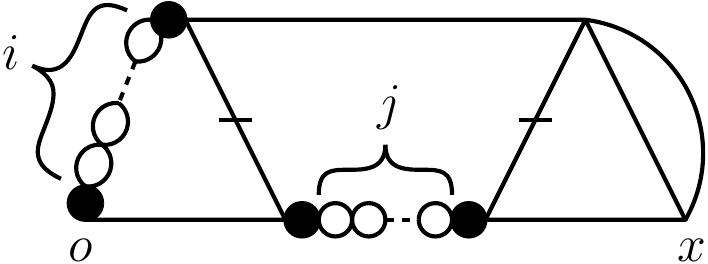}}~
 +\cdots\bigg).
\end{align}
\end{thm}
\end{shaded}

We will prove this theorem in Section~\ref{ss:thm1}.

The bound in \refeq{pi0diagbd} may look awful, but in fact, it is not so.  
Presumably a nonzero bubble $\tilde G^2$
is small in sufficiently high dimensions or with sufficiently large spread-out 
parameter in $d>4$, the dominant term is 
$2\times\raisebox{-6pt}{\includegraphics[scale=0.27]{pi0diag1}}
=2V^1(o,o;x)=2\tilde G(x)^3$ 
and the other terms are geometrically small.  We will demonstrate this 
for the spread-out model in $d>4$ in Section~\ref{ss:pi0}.

\subsection{Results for the $1^\text{st}$-order expansion coefficient}
\begin{figure}[t]
\begin{align*}
\DotU^m_a(y,z;y',z')&=\sum_{j=0}^m\Bigg(~\raisebox{-2.6pc}
 {\includegraphics[scale=0.3]{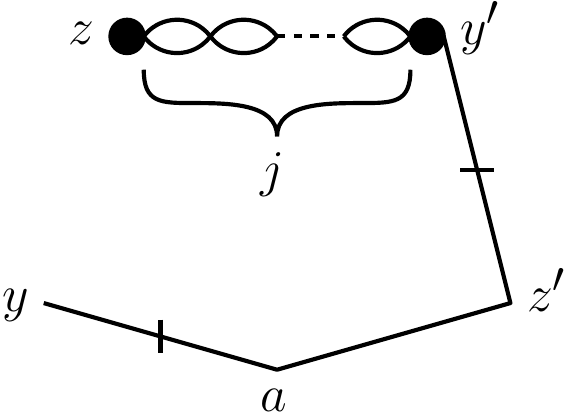}}~+~\raisebox{-1.7pc}
 {\includegraphics[scale=0.3]{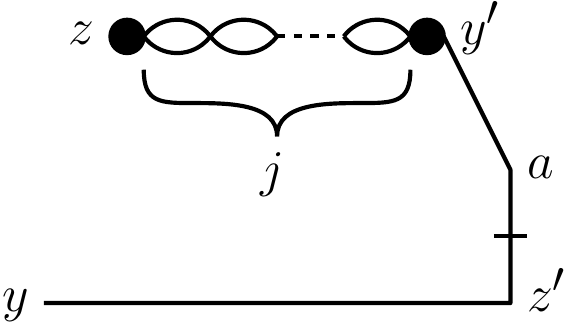}}~\Bigg)\\[5pt]
\DotV^m_a(y,z;x)&=\sum_{j=1}^m~\raisebox{-2.6pc}{\includegraphics
 [scale=0.3]{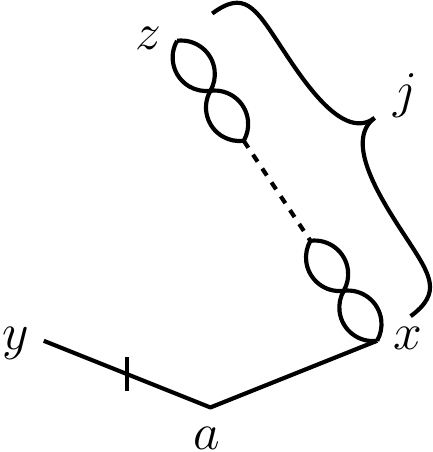}}
\end{align*}
\caption{Schematic representations of $\DotU^m_a(y,z;y',z')$ and 
$\DotV^m_a(y,z;x)$ for $m\ge1$.}
\label{fig:tildeUVdef}
\end{figure}

Next we present a diagrammatic bound on the following $\Theta'_{o,x;A}$, which 
shows up in the bounds \cite[Lemma~4.3]{s07} on the higher-order expansion 
coefficients $\pi_{\B_\Lambda}^{\sss(j)}(x)$ for $j\ge1$ (see \refeq{pi1prebd} 
below).  Given a vertex set $A\subset\Lambda$, we denote by $\B_{A\compl}$ 
the set of bonds whose end vertices are both in $A\compl$, and define
\begin{align}\lbeq{Theta'def}
\Theta'_{o,x;A}=\sum_{\substack{\bm\in\Z_+^{\B_{A\compl}}:\\ \partial\bm=\vno}}
 \frac{w_{\B_{A\compl}}(\bm)}{Z_{\B_{A\compl}}}\sum_{\substack{\bn\in
 \Z_+^{\B_\Lambda}:\\ \partial\bn=o\vtri x}}\frac{w_{\B_\Lambda}(\bn)}
 {Z_{\B_\Lambda}}\ind{o\db{\bm+\bn}{A}x},
\end{align}
where $o\db{\bm+\bn}{A}x$ means that $o\db{\bm+\bn}{}x$ and that all paths 
from $o$ to $x$ with positive current in $\bm+\bn$ must go through the set $A$ 
(i.e., every path from $o$ to $x$ consisting of bonds with positive current in 
$\bm+\bn$ has at least one bond with an endpoint in $A$).  Then we define (see 
Figure~\ref{fig:tildeUVdef})
\begin{align}
\DotU^m_a(y,z;y',z')&=\Big(G(a-y)\,\tilde G(z'-a)\,G(z'-y')+\tilde G(z'
 -y)\,\tilde G(a-y')\,G(z'-a)\Big)\nn\\
&\quad\times
 \begin{cases}
 G(y'-z)^2&(m=0),\\
 \Big((\delta+\tau^2)*\sum_{j=0}^m(\tilde G^2)^{*j}*(\delta+\tau^2)\Big)(y'-z)
  &(m\ge1),
 \end{cases}\lbeq{dotUdef}\\
\DotV^m_a(y,z;x)&=G(a-y)\,\tilde G(x-a)\sum_{j=1}^m(\tilde G^2)^{*j}(x
 -z),\lbeq{dotVdef}
\end{align}
and let
\begin{align}\lbeq{dotXdef}
\DotX^m_{o,x;a}=\sum_{i=0}^\infty\Big((U^m)^{\star i}\star\DotV^m_a\Big)_{o,x}
 +\sum_{i,j=0}^\infty\Big((U^m)^{\star i}\star\DotU^m_a\star(U^m)^{\star j}\star
 V^m\Big)_{o,x}.
\end{align}

\begin{shaded}
\begin{thm}[cf., Lemma~4.4 of \cite{s07}]\label{thm:Theta'bd}
Let $x\ne o$.  For the ferromagnetic models defined above \refeq{hamiltonian},
\begin{align}\lbeq{Theta'bd}
\Theta'_{o,x;A}\le2\sum_{a\in A}\Big(X^\infty_{o,x}\,\delta_{x,a}
 +\DotX^\infty_{o,x;a}\Big).
\end{align}
\end{thm}
\end{shaded}

We will prove this theorem in Section~\ref{ss:thm2}.  As discussed at the end 
of the previous subsection, if a nonzero bubble $\tilde G^2$ is small, then the 
dominant terms in $X^\infty_{o,x}$ and $\DotX^\infty_{o,x;a}$ in 
\refeq{Theta'bd} are $V^1(o,o;x)$ and $\DotV^1_a(o,o;x)$, respectively, 
and the others are geometrically small.  We will demonstrate this statement for 
the spread-out model in $d>4$ in Section~\ref{ss:pi1}.

\begin{figure}[t]
\begin{align*}
\DDotU^m_a(y,z;y',z')=\sum_{j_1,j_2,j_3=0}^{m-1}\raisebox{-2.2pc}
 {\includegraphics[scale=0.32]{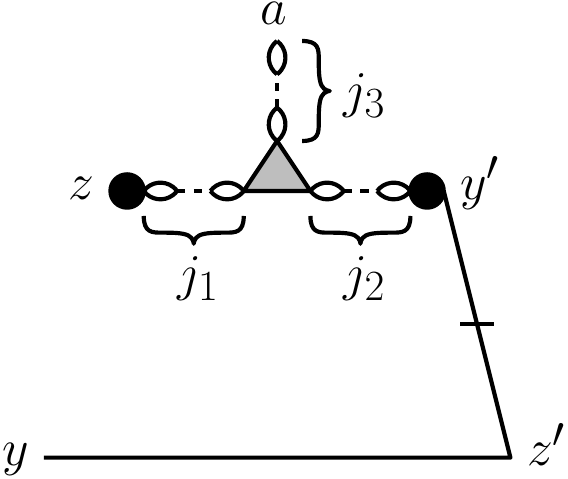}}\hskip3pc
\DDotV^m_a(y,z;x)=\sum_{j_1,j_2,j_3=0}^{m-1}\raisebox{-2.2pc}
 {\includegraphics[scale=0.32]{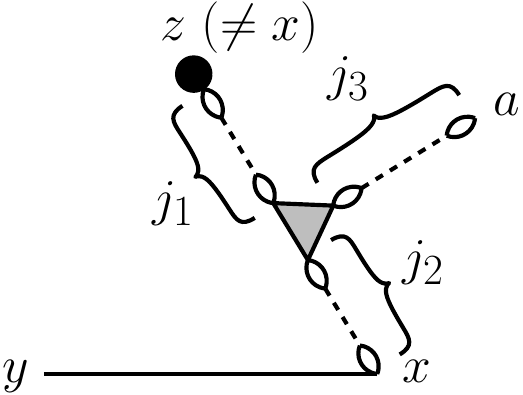}}
\end{align*}
\caption{Schematic representations of $\DDotU^m_a(y,z;y',z')$ and 
$\DDotV^m_a(y,z;x)$ for $m\ge1$.  The shaded triangles represent 
$T(v_1,v_2,v_3)$ in \refeq{ddotUdef}--\refeq{ddotVdef}.}
\label{fig:ddotUV}
\end{figure}

Next we present a diagrammatic bound on 
$\tilde\pi_{\B_\Lambda;y}^{\sss(0)}(x)$, which is a variant of 
$\pi_{\B_\Lambda}^{\sss(0)}(x)$ and is defined as 
\begin{align}\lbeq{tildepi0def}
\tilde\pi_{\B_\Lambda;y}^{\sss(0)}(x)=\sum_{\substack{\bn\in
 \Z_+^{\B_\Lambda}:\\ \partial\bn=o\vtri x}}\frac{w_{\B_\Lambda}(\bn)}
 {Z_{\B_\Lambda}}\ind{o\db{\bn}{}x\}\cap\{o\cn{\bn}{}y}.
\end{align}
This shows up in the bounds on the higher-order expansion coefficients 
$\pi_{\B_\Lambda}^{\sss(j)}(x)$ for $j\ge1$.  For example, by using 
$\tilde\cC_{\bn}^b(o)=\{z\in\Lambda:o\cn{\bn}{}z$ in $\B_\Lambda\setminus b\}$ 
(here and in the rest of the paper, we abbreviate $\B_\Lambda\setminus\{b\}$ to 
$\B_\Lambda\setminus b$ for any $b\in\B_\Lambda$) and substituting 
\cite[(4.33)]{s07} to the first line in \cite[(4.37)]{s07}, we can get
\begin{align}
\pi_{\B_\Lambda}^{\sss(1)}(x)\le\sum_{u,v}\sum_{\substack{\bn\in
 \Z_+^{\B_\Lambda}:\\ \partial\bn=o\vtri u}}\frac{w_{\B_\Lambda}(\bn)}
 {Z_{\B_\Lambda}}\ind{o\db{\bn}{}u}\tau(v-u)\sum_y\big(\delta_{v,y}+\tilde
 G(y-v)\big)\Theta'_{y,x;\tilde\cC_{\bn}^{\{u,v\}}(o)}.
\end{align}
Now, by Theorem~\ref{thm:Theta'bd} and $\Theta'_{x,x;A}=\ind{x\in A}$, we obtain
\begin{align}\lbeq{pi1prebd}
\pi_{\B_\Lambda}^{\sss(1)}(x)\le\sum_{a,u,y}\underbrace{\sum_{\substack{\bn
 \in\Z_+^{\B_\Lambda}:\\ \partial\bn=o\vtri u}}\frac{w_{\B_\Lambda}(\bn)}{Z_{
 \B_\Lambda}}\ind{o\db{\bn}{}u\}\cap\{o\cn{\bn}{}a}}_{\tilde\pi_{\B_\Lambda;
 a}^{\sss(0)}(u)}\big(\tau*(\delta+\tilde G)\big)(y-u)\nn\\
\times2\Big(\delta_{y,x}\,\delta_{x,a}+X^\infty_{y,x}\,\delta_{x,a}
 +\DotX^\infty_{y,x;a}\Big).
\end{align}
In the previous work, we claimed that $\tilde\pi_{\B_\Lambda;y}^{\sss(0)}(x)$ 
obeys the diagrammatic bound \cite[(4.16)]{s07}, but its 
proof given around \cite[(4.23)--(4.26)]{s07} is based on the problematic 
\cite[Lemma~4.2]{s07}.  We no longer use it and prove the following 
theorem instead, in which we use for $m\ge1$ (see Figure~\ref{fig:ddotUV})
\begin{align}
\DDotU^m_a(y,z;y',z')=\tilde G(z'-y)\,G(z'-y')&\sum_{\substack{u_1,u_2,u_3,\\
 v_1,v_2,v_3}}(\delta+\tau^2)(z-u_1)\,(\delta+\tau^2)(y'-u_2)\,\delta_{u_3,a}
 \nn\\
&\times\prod_{i=1}^3\sum_{j_i=0}^{m-1}(\tilde G^2)^{*j_i}(u_i-v_i)\,T(v_1,v_2,
 v_3),\lbeq{ddotUdef}\\
\DDotV^m_a(y,z;x)=\ind{z\ne x}\,\tilde G(x-y)&\sum_{\substack{u_1,u_2,u_3,\\
 v_1,v_2,v_3}}(\delta+\tau^2)(z-u_1)\,\delta_{u_2,x}\,\delta_{u_3,a}\nn\\
&\times\prod_{i=1}^3\sum_{j_i=0}^{m-1}(\tilde G^2)^{*j_i}(u_i-v_i)\,T(v_1,v_2,
 v_3),\lbeq{ddotVdef}
\end{align}
and
\begin{align}\lbeq{ddotXdef}
\DDotX^m_{o,x;y}=\sum_{i=0}^\infty\Big((U^m)^{\star i}\star\tfrac12\DDotV^m_y
 \Big)_{o,x}+\sum_{i,j=0}^\infty\Big((U^m)^{\star i}\star\DDotU^m_y\star
 (U^m)^{\star j}\star V^m\Big)_{o,x}.
\end{align}

\begin{shaded}
\begin{thm}\label{thm:pi0'bd}
Let $x\ne o$.  For the ferromagnetic models defined above \refeq{hamiltonian},
\begin{align}\lbeq{tildepibd}
\tilde\pi_{\B_\Lambda;y}^{\sss(0)}(x)\le2\Big(X^1_{o,x}\,\delta_{x,y}
 +\DotX^1_{o,x;y}+\DDotX^1_{o,x;y}\Big).
\end{align}
\end{thm}
\end{shaded}

We will prove this theorem in Section~\ref{ss:thm3}.  In addition to what we 
have stated below Theorem~\ref{thm:Theta'bd}, we will also demonstrate in 
Section~\ref{ss:pi1} that the dominant term in $\DDotX^1_{o,u;a}$ is 
$\DDotV^1_a(o,o;u)$ for the spread-out model in $d>4$.

Applying \refeq{tildepibd} and $\tilde\pi_{\B_\Lambda;y}^{\sss(0)}(o)\le G(y)^2$ 
(cf., \refeq{lmm0}) to \refeq{pi1prebd} yields the following diagrammatic bound 
on $\pi_{\B_\Lambda}^{\sss(1)}(x)$.

\begin{shaded}
\begin{cor}\label{cor:pi1bd}
For the ferromagnetic models defined above \refeq{hamiltonian},
\begin{align}\lbeq{cor:pi1bd}
\pi_{\B_\Lambda}^{\sss(1)}(x)\le4\sum_{a,u,y}&\Big(G(y)^2\,\delta_{o,u}
 +X^1_{o,u}\,\delta_{u,a}+\DotX^1_{o,u;a}+\DDotX^1_{o,u;a}\Big)\nn\\
&\times\big(\tau*(\delta+\tilde G)\big)(y-u)\,\Big(\delta_{y,x}\,\delta_{x,a}
 +X^\infty_{y,x}\,\delta_{x,a}+\DotX^\infty_{y,x;a}\Big).
\end{align}
\end{cor}
\end{shaded}

\subsection{Results for the higher-order expansion coefficients}
Finally we present a diagrammatic bound on $\Theta''_{o,x,y;A}$, which is a 
variant of $\Theta'_{o,x;A}$ and is defined as
\begin{align}\lbeq{Theta''def}
\Theta''_{o,x,y;A}=\sum_{\substack{\bm\in\Z_+^{\B_{A\compl}}:\\ \partial\bm
 =\vno}}\frac{w_{\B_{A\compl}}(\bm)}{Z_{\B_{A\compl}}}\sum_{\substack{\bn\in
 \Z_+^{\B_\Lambda}:\\ \partial\bn=o\vtri x}}\frac{w_{\B_\Lambda}(\bn)}
 {Z_{\B_\Lambda}}\ind{o\db{\bm+\bn}{A}x\}\cap\{o\cn{\bm+\bn}{}y}.
\end{align}
This shows up in the bounds on the higher-order expansion coefficients 
$\pi_{\B_\Lambda}^{\sss(j)}$ for $j\ge2$.  For example, 
$\pi_{\B_\Lambda}^{\sss(2)}(x)$ is bounded in a similar way to \refeq{pi1prebd} 
as (cf., \cite[(4.38)]{s07})
\begin{align}\lbeq{pi2prebd}
\pi_{\B_\Lambda}^{\sss(2)}(x)&\le2\sum_u\sum_{\substack{\bn\in\Z_+^{
 \B_\Lambda}:\\ \partial\bn=o\vtri u}}\frac{w_{\B_\Lambda}(\bn)}{Z_{\B_\Lambda}}
 \,\ind{o\db{\bn}{}u}\nn\\
&\qquad\times\sum_{\substack{v,z,y,\\ u',a'}}\tau(v-u)\,G(z-v)\bigg(\tilde
 G(y-z)\,\delta_{z,a'}\,\Theta'_{y,u';\tilde\cC_{\bn}^{\{u,v\}}(o)}+\delta_{y,z}
 \,\Theta''_{y,u',a';\tilde\cC_{\bn}^{\{u,v\}}(o)}\bigg)\nn\\
&\hskip3.3pc\times\sum_{y'}\big(\tau*(\delta+\tilde G)\big)(y'-u')\,\Big(
 X^\infty_{y',x}\,\delta_{x,a'}+\DotX^\infty_{y',x;a'}\Big).
\end{align}

To show a diagrammatic bound on $\Theta''_{o,x,y;A}$, we introduce the 
following building blocks of the diagrams: for $m\ge1$,
\begin{align}\lbeq{dddotUdef}
\DDDotU^m_{a,v}(y,z;y',z')&=\Big(G(a-y)\,\tilde G(z'-a)\,G(z'-y')+\tilde
 G(z'-y)\,\tilde G(a-y')\,G(z'-a)\Big)\nn\\
&\hskip3pc\times\sum_{\substack{u_1,u_2,u_3,\\ v_1,v_2,v_3}}(\delta+\tau^2)(z-u_1)
 \,(\delta+\tau^2)(y'-u_2)\,\delta_{u_3,v}\nn\\
&\hskip5pc\times\prod_{i=1}^3\sum_{j_i=0}^{m-1}(\tilde G^2)^{*j_i}(u_i-v_i)\,T(v_1,
 v_2,v_3),
\end{align}
\begin{align}\lbeq{dddotVdef}
\DDDotV^m_{a,v}(y,z;x)=\ind{z\ne x}\,G(a-y)\,\tilde G(x-a)\sum_{
 \substack{u_1,u_2,u_3,\\ v_1,v_2,v_3}}(\delta+\tau^2)(z-u_1)\,\delta_{u_2,x}
 \,\delta_{u_3,v}\nn\\
\times\prod_{i=1}^3\sum_{j_i=0}^{m-1}(\tilde G^2)^{*j_i}(u_i-v_i)\,T(v_1,v_2,
 v_3).
\end{align}
Depending on which of the six terms in $\DDotX^m_{o,x;y}$ in \refeq{ddotXdef} 
the extra vertex $a$ is added, we define
\begin{align}\lbeq{DDDotXdef}
\DDDotX^m_{o,x;a,y}&=\sum_{i=0}^\infty\Big((U^m)^{\star i}\star\tfrac12
 \DDDotV^m_{a,y}\Big)_{o,x}+\sum_{i,j=0}^\infty\Big((U^m)^{\star i}\star
 \DDDotU^m_{a,y}\star(U^m)^{\star j}\star V^m\Big)_{o,x}\nn\\
&~+\sum_{i,j=0}^\infty\Big((U^m)^{\star i}\star\DotU^m_a\star(U^m)^{\star
 j}\star\tfrac12\DDotV^m_y\Big)_{o,x}+\sum_{i,j=0}^\infty\Big((U^m)^{\star i}
 \star\DDotU^m_y\star(U^m)^{\star j}\star\DotV^m_a\Big)_{o,x}\nn\\
&~+\sum_{i,j,k=0}^\infty\Big((U^m)^{\star i}\star\DotU^m_a\star(U^m)^{\star
 j}\star\DDotU^m_y\star(U^m)^{\star k}\star V^m\Big)_{o,x}\nn\\
&~+\sum_{i,j,k=0}^\infty\Big((U^m)^{\star i}\star\DDotU^m_y\star(U^m)^{\star
 j}\star\DotU^m_a\star(U^m)^{\star k}\star V^m\Big)_{o,x}.
\end{align}
\begin{shaded}
\begin{thm}[cf., Lemma~4.4 of \cite{s07}]\label{thm:Theta''bd}
Let $x\ne o$.  For the ferromagnetic models defined above \refeq{hamiltonian},
\begin{align}\lbeq{Theta''bd}
\Theta''_{o,x,y;A}\le2\sum_{a\in A}\bigg(\DDotX^\infty_{o,x;y}\delta_{a,x}
 +\DDDotX^\infty_{o,x;a,y}+\sum_{y'}\Big(\DDotX^\infty_{o,x;a}\delta_{y',x}
 +\DDDotX^\infty_{o,x;y',a}\Big)\sum_{i=0}^\infty(\tilde G^2)^{*i}(y-y')\bigg).
\end{align}
\end{thm}
\end{shaded}

We will prove this theorem in Section~\ref{ss:thm5} and also demonstrate in 
Section~\ref{ss:higher} that the main contribution to $\DDDotX^m_{o,x;a,y}$ 
comes from $\DDDotV^m_{a,y}(o,o;x)$ for the spread-out model in $d>4$.

Applying \refeq{tildepibd} and \refeq{Theta''bd} to \refeq{pi2prebd} and using 
$\Theta''_{x,x,y;A}\le\ind{x\in A}\sum_{j=0}^\infty(\tilde G^2)^{*j}(y)$ would 
yield a diagrammatic bound on $\pi_{\B_\Lambda}^{\sss(2)}(x)$.  We refrain 
from stating it explicitly.  The higher-order expansion coefficients can be 
bounded in a similar way.

\section{Proofs of the main results}\label{s:proofs}
In Section~\ref{ss:thm1}, we first prove Theorem~\ref{thm:pi0bd} in detail, 
as it provides a common foundation for the other three theorems.  
We prove those three theorems in Sections~\ref{ss:thm2}--\ref{ss:thm5}, 
respectively, only focusing on differences from Theorem~\ref{thm:pi0bd}.  
In the course of the proof of Theorem~\ref{thm:pi0bd} (at the end of Step 1 
in Section~\ref{ss:thm1}), we also prove Lemma~\ref{lmm:lmm2}.

\subsection{Proof of Theorem~\ref{thm:pi0bd}}\label{ss:thm1}
The proof is progressed along the following five steps.
\begin{enumerate}
\item
Rewrite $\pi_{\B_\Lambda}^{\sss(0)}(x)$ for $x\ne o$ by identifying the 
``earliest'' path from $o$ to $x$ with odd current.
\item
A double expansion: a sort of lace expansion along the earliest path chosen 
above.  Let $N\ge1$ be the number of lace edges in the expansion. 
\item
Proof for the $N=1$ case.
\item
Proof for the $N\ge2$ case, part 1: bounds on the contributions from lace edges.
\item
Proof for the $N\ge2$ case, part 2: bound on the contribution from the earliest path.
\end{enumerate}

\begin{figure}[t]
\begin{center}
\includegraphics[scale=0.9]{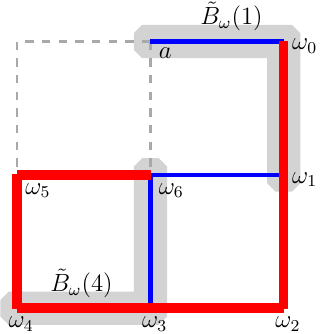}
\end{center}
\caption{Suppose that $\Lambda$ consists of nine vertices as depicted, 
and that bonds incident on each vertex $x$ are ordered in a counter-clockwise 
way as $\{x,x+e_1\}\preceq\{x,x+e_2\}\preceq\{x,x-e_1\}\preceq\{x,x-e_2\}$, 
where $e_1=(1,0)$ and $e_2=(0,1)$.  Given 
$\omega=(\omega_0,\omega_1,\dots,\omega_6)$ as depicted, we have 
$\tilde B_\omega(0)=\vno$, 
$\tilde B_\omega(1)=\{\{\omega_0,\omega_1\},\{\omega_0,a\}\}$ (in grey), 
$\tilde B_\omega(2)=\{\{\omega_1,\omega_2\},\{\omega_1,\omega_6\}\}$, 
$\tilde B_\omega(3)=\{\{\omega_2,\omega_3\}\}$, 
$\tilde B_\omega(4)=\{\{\omega_3,\omega_4\},\{\omega_3,\omega_6\}\}$ (in grey), 
$\tilde B_\omega(5)=\{\{\omega_4,\omega_5\}\}$, 
$\tilde B_\omega(6)=\{\{\omega_5,\omega_6\}\}$, so that $B_\omega$ is the set 
of bold bonds (in red) and $\tilde B_\omega\setminus B_\omega$ is the set of 
thin-solid bonds (in blue).}
\label{fig:Vdef1}
\end{figure}

\paragraph{Step 1.}
First we note that, due to the source constraint in the definition of 
$\pi_{\B_\Lambda}^{\sss(0)}$, there must be a path from $o$ to $x$ (assumed 
not to be $o$) of bonds with odd current.  To identify a unique one among those 
paths, we introduce a fixed (e.g., lexicographic) ordering in the set of bonds 
incident on each vertex.  Given a pair of bonds $b_1=\{u,v_1\}$ and 
$b_2=\{u,v_2\}$ that are incident on a common vertex $u$, we write 
$b_1\preceq b_2$ (and $v_1\preceq v_2$) if $b_1$ is earlier than or equal to 
$b_2$ in that ordering.  Let $\Omega(o,x)$ be the set of nonzero paths from $o$ 
to $x$ each of which may intersect to itself (except for the terminal $x$) but 
does not traverse any bond more than once:  
\begin{align}\lbeq{Omega-def}
\Omega(z,x)=\bigg\{\omega=(\omega_0,\dots,\omega_{|\omega|}):
 \begin{array}{c}
 |\omega|\ge1,~\omega_0=z,~\omega_{|\omega|}=x,~\omega_j\ne x\text{ for }j
  <|\omega|,\\
 \{\omega_{i-1},\omega_i\}\ne\{\omega_{j-1},\omega_j\}\text{ for }1\le i<j
  \le|\omega|
 \end{array}
 \bigg\}.
\end{align}
Given an $\omega\in\Omega(o,x)$, we let 
$B_\omega=\{\{\omega_{j-1},\omega_j\}:j=1,\dots,|\omega|\}$ and 
inductively define the bond set $\tilde B_\omega$ ($\supset B_\omega$) as 
(cf., Figure~\ref{fig:Vdef1})
\begin{align}\lbeq{tildeBdef}
\tilde B_\omega(j)=
 \begin{cases}
 \vno&(j=0),\\
 \Big\{\{\omega_{j-1},v\}\notin\bigcup_{i=0}^{j-1}\tilde B_\omega(i):v
  \preceq\omega_j\Big\}&(j\ge1),
 \end{cases}&&&&
\tilde B_\omega=\bigcup_{j=0}^{|\omega|}\tilde B_\omega(j).
\end{align}
Then we can 
rewrite the random-current representation of $\pi_{\B_\Lambda}^{\sss(0)}(x)$ 
in \refeq{wrongbd0} as 
\begin{align}\lbeq{pi0earliest}
\pi_{\B_\Lambda}^{\sss(0)}(x)&=\delta_{o,x}+\sum_{\substack{\bn\in
 \Z_+^{\B_\Lambda}:\\ \partial\bn=o\vtri x}}\frac{w_{\B_\Lambda}(\bn)}
 {Z_{\B_\Lambda}}\ind{o\db{\bn}{}x}\sum_{\omega\in\Omega(o,x)}
 \ind{\omega\text{ the earliest odd path}}(\bn)\nn\\
&=\delta_{o,x}+\sum_{\substack{\bn\in\Z_+^{\B_\Lambda}:\\
 \partial\bn=o\vtri x}}\frac{w_{\B_\Lambda}(\bn)}{Z_{\B_\Lambda}}\ind{o\db{\bn}
 {}x}\sum_{\omega\in\Omega(o,x)}\prod_{b\in B_\omega}\ind{n_b\text{ odd}}
 \prod_{b'\in\tilde B_\omega\setminus B_\omega}\ind{n_{b'}\text{ even}}.
\end{align}
By splitting the weight as $w_{\B_\Lambda}(\bn)=w_{\tilde B_\omega}(\bm)\,
w_{\B_\Lambda\setminus\tilde B_\omega}(\bk)$, where $\bm$ and $\bk$ are the 
restrictions of $\bn$ on $\tilde B_\omega$ and on 
$\B_\Lambda\setminus\tilde B_\omega$, repectively, we have the rewrite for 
$x\ne o$:
\begin{align}\lbeq{pi0decomp}
\pi_{\B_\Lambda}^{\sss(0)}(x)=\sum_{\omega\in\Omega(o,x)}\sum_{\substack{\bm\in
 \Z_+^{\tilde B_\omega}:\\ \text{odd on }B_\omega,\\ \text{even on }\tilde
 B_\omega\setminus B_\omega}}\frac{w_{\tilde B_\omega}(\bm)}{Z_{\B_\Lambda}}
 \sum_{\substack{\bk\in\Z_+^{\B_\Lambda\setminus\tilde B_\omega}:\\ \partial\bk
 =\vno}}w_{\B_\Lambda\setminus\tilde B_\omega}(\bk)\,\ind{o\db{\bm+\bk}{}x}.
\end{align}

\begin{figure}[t]
\begin{center}
\includegraphics[scale=0.45]{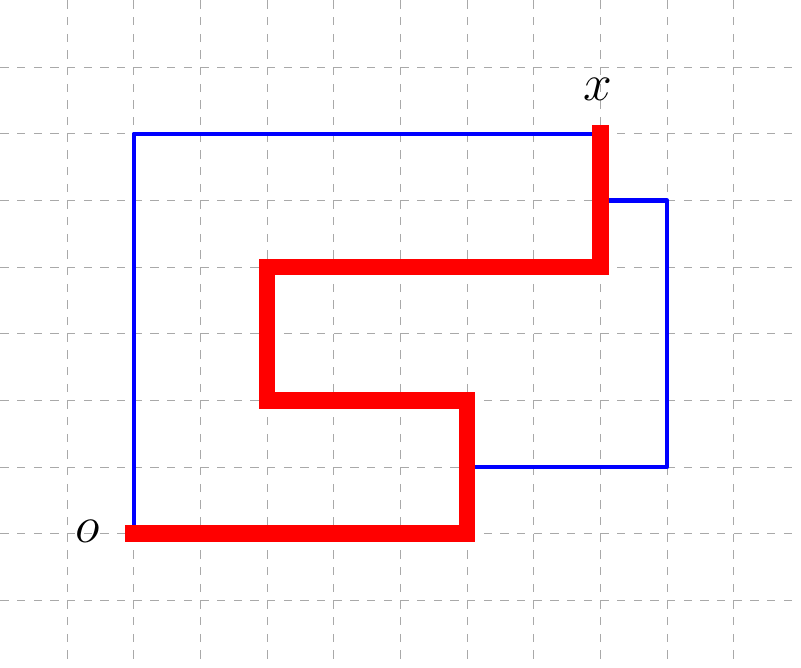}\hskip3pc
\includegraphics[scale=0.45]{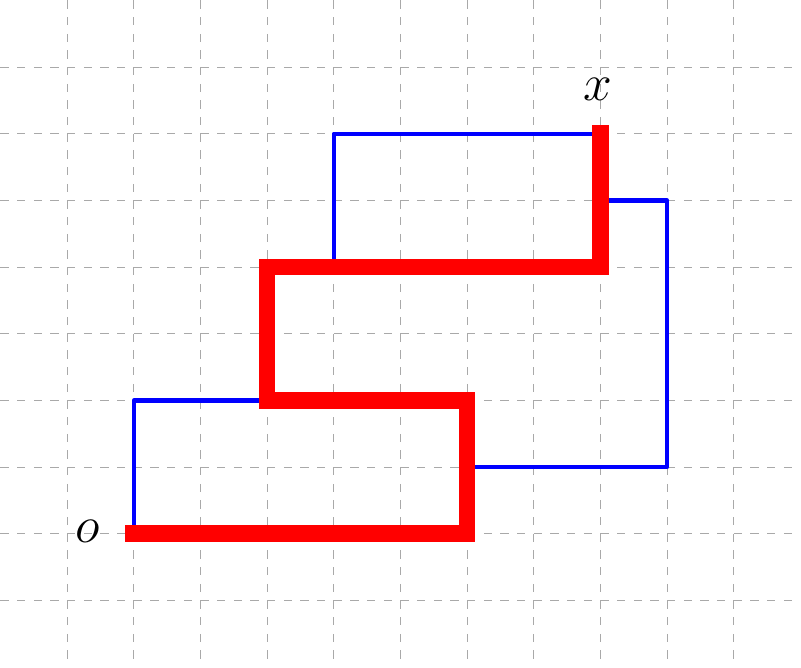}
\end{center}
\caption{Current configurations for $\pi_{\B_\Lambda}^{\sss(0)}(x)$ in 
\refeq{pi0earliest}.  As in Figure~\ref{fig:RC}, bonds with odd current are bold 
(in red), while those with positive-even current are thin-solid (in blue).  
On the left, $o$ and $x$ are still connected by a path of bonds with positive 
current, even after removal of the path of bonds with odd current, which is 
not the case on the right.}
\label{fig:pi0}
\end{figure}

\paragraph{Remark.}
Suppose $\ind{o\db{\bm+\bk}{}x}\le\ind{o\cn{\bk}{}x}$ in \refeq{pi0decomp}, 
i.e., the double connection implies existence of a path from $o$ to $x$ of 
positive current in the restricted region $\B_\Lambda\setminus\tilde B_\omega$ 
(as on the left of Figure~\ref{fig:pi0}, where we regard the set of bold red 
bonds as $\tilde B_\omega$).  Then, by \refeq{lmm0}, the sum over $\bk$ is 
bounded as 
\begin{align}\lbeq{SST}
\sum_{\substack{\bk\in\Z_+^{\B_\Lambda\setminus\tilde B_\omega}:\\ \partial\bk
 =\vno}}w_{\B_\Lambda\setminus\tilde B_\omega}(\bk)\,\ind{o\cn{\bk}{}x}\le G
 (x)^2Z_{\B_\Lambda\setminus\tilde B_\omega}=G(x)^2\sum_{\substack{\bk
 \in\Z_+^{\B_\Lambda\setminus\tilde B_\omega}:\\ \partial\bk=\vno}}w_{\B_\Lambda
 \setminus\tilde B_\omega}(\bk).
\end{align}
As a result, we obtain that, for $x\ne o$,
\begin{align}\lbeq{Remark}
&\sum_{\omega\in\Omega(o,x)}\sum_{\substack{\bm\in\Z_+^{\tilde B_\omega}:\\
 \text{odd on }B_\omega,\\ \text{even on }\tilde B_\omega\setminus
 B_\omega}}\frac{w_{\tilde B_\omega}(\bm)}{Z_{\B_\Lambda}}
 \sum_{\substack{\bk\in\Z_+^{\B_\Lambda\setminus\tilde B_\omega}:\\ \partial\bk
 =\vno}}w_{\B_\Lambda\setminus\tilde B_\omega}(\bk)\,\ind{o\cn{\bk}{}x}\nn\\
&\stackrel{\text{\refeq{SST}}}\le G(x)^2\sum_{\omega\in\Omega(o,x)}
 \sum_{\substack{\bm\in\Z_+^{\tilde B_\omega}:\\ \text{odd on }B_\omega,\\
 \text{even on }\tilde B_\omega\setminus B_\omega}}\sum_{\substack{\bk\in
 \Z_+^{\B_\Lambda\setminus\tilde B_\omega}:\\ \partial\bk=\vno}}\frac{w_{\tilde
 B_\omega}(\bm)\,w_{\B_\Lambda\setminus\tilde B_\omega}(\bk)}
 {Z_{\B_\Lambda}}\nn\\
&~=G(x)^2\underbrace{\sum_{\substack{\bn\in\Z_+^{\B_\Lambda}:\\ \partial\bn
 =o\vtri x}}\frac{w_{\B_\Lambda}(\bn)}{Z_{\B_\Lambda}}}_{\Exp{\vphi_o
 \vphi_x}_{\B_\Lambda}}~\stackrel{\text{\refeq{tildeGdef}}}\le\tilde G(x)^3.
\end{align}
However, the above argument is incomplete, due to the possibility of 
$\ind{o\db{\bm+\bk}{}x}>\ind{o\cn{\bk}{}x}$, as depicted on the right of 
Figure~\ref{fig:pi0}, i.e., $o$ and $x$ may no longer connected after removal 
of $\tilde B_\omega$.  To overcome this problem, we will introduce a notion of 
lace in Step~2.

\Proof{Proof of Lemma~\ref{lmm:lmm2}.}
First, by multiplying $1=\sum_{\partial\bm=\vno}w_B(\bm)/Z_B$, using the 
monotonicity $\{o\cn{\bn}{}x\}\subset\{o\cn{\bm+\bn}{}x\}$ and then the SST, 
we obtain
\begin{eqnarray}
\sum_{\substack{\bn\in\Z_+^B:\\ \partial\bn=\vno}}\frac{w_B(\bn)}{Z_B}\ind{o
 \cn{\bn}{}x\}\cap\{o\cn{\bn}{}y}
&\le&\sum_{\substack{\bm,\bn\in\Z_+^B:\\ \partial\bm=\partial\bn=\vno}}
 \frac{w_B(\bm)}{Z_B}\frac{w_B(\bn)}{Z_B}\ind{o\cn{\bm+\bn}{}x\}\cap\{o
 \cn{\bm+\bn}{}y}\nn\\
&\stackrel{\text{SST}}=&\sum_{\substack{\bm,\bn\in\Z_+^B:\\ \partial\bm
 =\partial\bn=o\vtri x}}\frac{w_B(\bm)}{Z_B}\frac{w_B(\bn)}{Z_B}\ind{o\cn{\bm
 +\bn}{}y}.\lbeq{lmm2-1}
\end{eqnarray}
Now, as explained above in Step~1, we can identify the earliest path 
$\omega\in\Omega(o,x)$ of bonds $b\in B$ with odd $m_b$.  
Splitting the weight $w_B(\bm)$ as 
$w_{\tilde B_\omega}(\bk)\,w_{B\setminus\tilde B_\omega}(\bl)$, we obtain
\begin{align}
\refeq{lmm2-1}=\sum_{\omega\in\Omega(o,x)}\sum_{\substack{\bk\in\Z_+^{\tilde
 B_\omega}:\\ \text{odd on }B_\omega,\\ \text{even on }\tilde B_\omega\setminus
 B_\omega}}\frac{w_{\tilde B_\omega}(\bk)}{Z_B}\sum_{\substack{\bl\in\Z_+^{B
 \setminus\tilde B_\omega}:\\ \partial\bl=\vno}}w_{B\setminus\tilde B_\omega}
 (\bl)\sum_{\substack{\bn\in\Z_+^B:\\ \partial\bn=o\vtri x}}\frac{w_B(\bn)}{Z_B}
 \,\ind{o\cn{\bk+\bl+\bn}{}y}.
\end{align}
Notice that, if $o\cn{\bk+\bl+\bn}{}y$, then there must be a 
$z\in V(\tilde B_\omega)$ such that $o\cn{\bk}{}z$ and $z\cn{\bl+\bn}{}y$ in 
$B\setminus\tilde B_\omega$ (zero length is allowed).  Therefore, the above 
expression is bounded by
\begin{align}\lbeq{lmm2-2}
&\sum_{\omega\in\Omega(o,x)}\sum_{z\in V(\tilde B_\omega)}\sum_{\substack{\bk
 \in\Z_+^{\tilde B_\omega}:\\ \text{odd on }B_\omega,\\ \text{even on }\tilde
 B_\omega\setminus B_\omega}}\frac{w_{\tilde B_\omega}(\bk)}{Z_B}\sum_{
 \substack{\bl\in\Z_+^{B\setminus\tilde B_\omega}:\\ \partial\bl=\vno}}w_{B
 \setminus\tilde B_\omega}(\bl)\sum_{\substack{\bn\in\Z_+^B:\\ \partial\bn=o
 \vtri x}}\frac{w_B(\bn)}{Z_B}\ind{z\cn{\bl+\bn}{}y\text{ in }B\setminus\tilde
 B_\omega}\nn\\
&\stackrel{\text{SST}}=\sum_{\omega\in\Omega(o,x)}\sum_{z\in V(\tilde B_\omega)}
 \sum_{\substack{\bk\in\Z_+^{\tilde B_\omega}:\\ \text{odd on }B_\omega,\\
 \text{even on }\tilde B_\omega\setminus B_\omega}}\frac{w_{\tilde B_\omega}
 (\bk)}{Z_B}\,Z_{B\setminus\tilde B_\omega}\underbrace{\sum_{\substack{\bl\in
 \Z_+^{B\setminus\tilde B_\omega}:\\ \partial\bl=z\vtri y}}\frac{w_{B\setminus
 \tilde B_\omega}(\bl)}{Z_{B\setminus\tilde B_\omega}}}_{\Exp{\vphi_z\vphi_y}_{B
 \setminus\tilde B_\omega}}\,\underbrace{\sum_{\substack{\bn\in\Z_+^B:\\
 \partial\bn=o\vtri x\vtri z\vtri y}}\frac{w_B(\bn)}{Z_B}}_{\Exp{\vphi_o\vphi_x
 \vphi_z\vphi_y}_B~(\because\text{ \refeq{4pt}})}.
\end{align}
Furthermore, by Lebowitz' inequality \cite{l74} and Lemma~\ref{lmm:G}, 
\begin{eqnarray}
\Exp{\vphi_o\vphi_x\vphi_z\vphi_y}_B&\le&\Exp{\vphi_o\vphi_x}_B\,\Exp{\vphi_z
 \vphi_y}_B+\Exp{\vphi_o\vphi_y}_B\,\Exp{\vphi_x\vphi_z}_B+\Exp{\vphi_o
 \vphi_z}_B\,\Exp{\vphi_x\vphi_y}_B\nn\\[5pt]
&\le&G(x)\,G(z-y)+G(y)\,G(z-x)+G(z)\,G(y-x).
\end{eqnarray}
As a result, we obtain
\begin{align}
\refeq{lmm2-2}&\le\sum_z\sum_{\omega\in\Omega(o,x)}\ind{z\in V(\tilde B_\omega)}
 \sum_{\substack{\bk\in\Z_+^{\tilde B_\omega}:\\ \text{odd on }B_\omega,\\
 \text{even on }\tilde B_\omega\setminus B_\omega}}\frac{w_{\tilde B_\omega}
 (\bk)}{Z_B}\,Z_{B\setminus\tilde B_\omega}\,G(z-y)\nn\\
&\qquad\times\Big(G(x)\,G(z-y)+G(y)\,G(z-x)+G(z)\,
 G(y-x)\Big)\nn\\
&\le\sum_z\underbrace{\sum_{\substack{\bn\in\Z_+^B:\\ \partial\bn=o\vtri x}}
 \frac{w_B(\bn)}{Z_B}\,\ind{o\cn{\bn}{}z}}_{\le\,G(z)\,G
 (z-x)~(\because\text{ \refeq{lmm1}})}\,G(z-y)\nn\\
&\qquad\times\Big(G(x)\,G(z-y)+G(y)\,G(z-x)
 +G(z)\,G(y-x)\Big),
\end{align}
as required.
\QED

\begin{figure}[t]
\begin{center}
\includegraphics[scale=0.9]{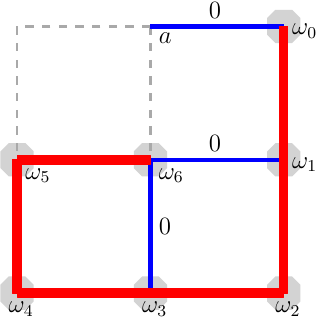}\hskip5pc
\includegraphics[scale=0.9]{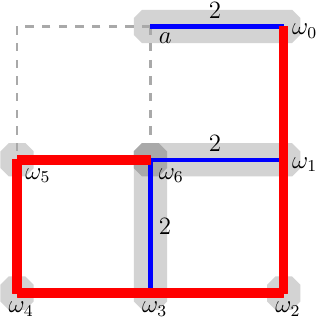}
\end{center}
\caption{Consider the same setting as Figure~\ref{fig:Vdef1} and let $\bm$ 
be a current configuration that assigns odd numbers to the red bonds in 
$B_\omega$.  If $\bm$ assigns zeros to the blue bonds in 
$\tilde B_\omega\setminus B_\omega$ (as depicted in the left figure), then 
$\tilde V_{\bm}(j)=\{\omega_j\}$ (in grey) for all $j=0,1,\dots,6$.  On the 
other hand, if $\bm$ assigns to the blue bonds positive even numbers (say, 2, 
as in the right figure), then $\tilde V_{\bm}(0)=\{\omega_0,a\}$, 
$\tilde V_{\bm}(1)=\{\omega_1,\omega_6\}$ 
and $\tilde V_{\bm}(3)=\{\omega_3,\omega_6\}$.}
\label{fig:Vdef2}
\end{figure}

\paragraph{Step 2.}
To overcome the problem explained below \refeq{Remark}, we use 
$\tilde B_\omega$ as a time line for an expansion, similar to the lace 
expansion, of the sum over $\bk$ in \refeq{pi0decomp}; we call this a double 
expansion.  First, we define the vertex set $\tilde V_{\bm}(j)$ for a current 
configuration $\bm\in\Z_+^{\tilde B_\omega}$ satisfying the constraint in the 
second sum on the right-hand side of \refeq{pi0decomp} as 
(cf., Figure~\ref{fig:Vdef2})
\begin{align}\lbeq{tildeVdef}
\tilde V_{\bm}(j)=
 \begin{cases}
 \{\omega_j\}\cup\Big\{v:\{\omega_j,v\}\in\tilde B_\omega(j+1),~m_{\omega_j,v}
  \text{ is positive-even}\Big\}\quad&(0\le j<|\omega|),\\
 \{\omega_{|\omega|}\}&(j=|\omega|).
 \end{cases}
\end{align}
Given those vertex sets and a $\bk\in\Z_+^{\B_\Lambda\setminus\tilde B_\omega}$ 
in the third sum on the right-hand side of \refeq{pi0decomp}, we uniquely 
define a lace $\sfL_{\bm,\bk}=\{s_jt_j\}_{j=1}^N$ as follows (see 
Figure~\ref{fig:lace}):
\begin{figure}[t]
\begin{center}
\includegraphics[scale=0.5]{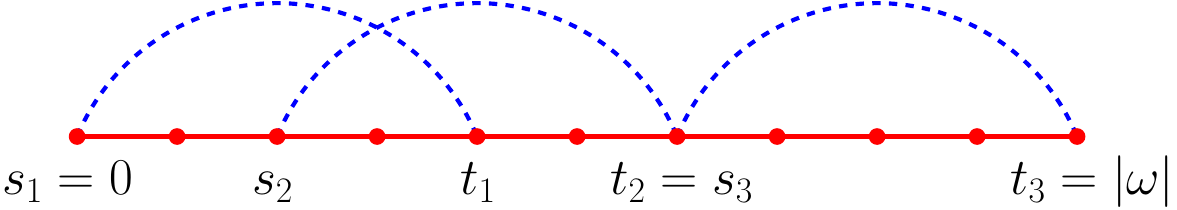}
\end{center}
\caption{An example of a lace consisting of three edges $s_1t_1,s_2t_2,s_3t_3$. 
A dashed arc from $s$ to 
$t$ represents $\tilde V_{\bm}(s)\cn{\bk}{}\tilde V_{\bm}(t)$.}
\label{fig:lace}
\end{figure}
\begin{itemize}
\item
First we define
\begin{align}\lbeq{lacedef1}
s_1=0,&&
t_1=\max\big\{j:\tilde V_{\bm}(0)\cn{\bk}{}\tilde V_{\bm}(j)\big\}.
\end{align}
where $\tilde V_{\bm}(0)\cn{\bk}{}\tilde V_{\bm}(j)$ means either 
$\tilde V_{\bm}(0)\cap\tilde V_{\bm}(j)\ne\vno$ or there is a nonzero 
path of bonds $b$ with $k_b>0$ from a vertex in $\tilde V_{\bm}(0)$ to 
another in $\tilde V_{\bm}(j)$.  If $t_1=|\omega|$ (as on the left of 
Figure~\ref{fig:pi0}), then it is done with $\sfL_{\bm,\bk}=\{0|\omega|\}$ and 
$N=1$.
\item
If $t_1<|\omega|$, then, since $o\db{\bm+\bk}{}x$, there must be an $s_2t_2$ 
uniquely defined as 
\begin{align}
t_2&=\max\big\{j:\exists i\le t_1\text{ s.t. }\tilde V_{\bm}(i)\cn{\bk}{}\tilde
 V_{\bm}(j)\big\},\lbeq{lacedef2}\\
s_2&=\min\big\{i:\tilde V_{\bm}(i)\cn{\bk}{}\tilde V_{\bm}(t_2)\big\}.
 \lbeq{lacedef3}
\end{align}
If $t_2=|\omega|$, then it is done with $\sfL_{\bm,\bk}=\{0t_1,s_2|\omega|\}$ 
and $N=2$.
\item
Repeat this procedure until it reaches $t_N=|\omega|$ with 
$\sfL_{\bm,\bk}=\{s_jt_j\}_{j=1}^N$.
\end{itemize}
Let $\cL_{[0,|\omega|]}^{\sss(N)}$ be the set of $N$-edge graphs 
$\Gamma=\{s_jt_j\}_{j=1}^N$ satisfying 
\begin{align}\lbeq{cLdef}
\begin{cases}
0=s_1<t_1=|\omega|&(N=1),\\
0=s_1<s_2\le t_1<t_2=|\omega|&(N=2),\\
0=s_1<s_2\le\cdots<s_n\le t_{n-1}<s_{n+1}\le t_n<\cdots\le t_{N-1}<t_N=|\omega|
 &(N\ge3).
\end{cases}
\end{align}
Then, we can rewrite the sum over $\bk$ in \refeq{pi0decomp} to obtain that, 
for $x\ne o$,
\begin{align}\lbeq{pi0decompR}
\pi_{\B_\Lambda}^{\sss(0)}(x)&=\sum_{N=1}^\infty\sum_{\omega\in\Omega(o,x)}
 \sum_{\Gamma\in\cL_{[0,|\omega|]}^{\sss(N)}}\sum_{\substack{\bm\in\Z_+^{\tilde
 B_\omega}:\\ \text{odd on }B_\omega,\\ \text{even on }\tilde B_\omega\setminus
 B_\omega}}\frac{w_{\tilde B_\omega}(\bm)}{Z_{\B_\Lambda}}\nn\\
&\qquad\times\sum_{\substack{\bk\in\Z_+^{\B_\Lambda\setminus\tilde B_\omega}:\\
 \partial\bk=\vno}}w_{\B_\Lambda\setminus\tilde B_\omega}(\bk)\,\ind{\sfL_{\bm,
 \bk}=\Gamma}\prod_{st\in\Gamma}\ind{\tilde V_{\bm}(s)\cn{\bk}{}\tilde V_{\bm}
 (t)}.
\end{align}

\begin{figure}[t]
\[ \raisebox{-1pc}{\includegraphics[scale=0.44]{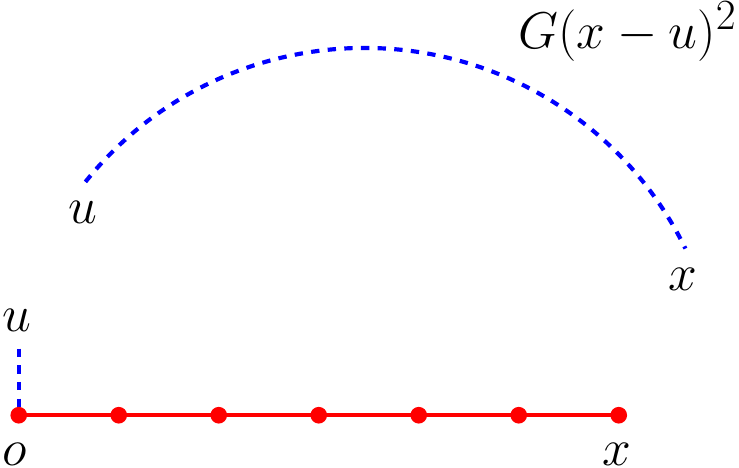}}\qquad\to\hskip3pc
 \raisebox{-1pc}{\includegraphics[scale=0.44]{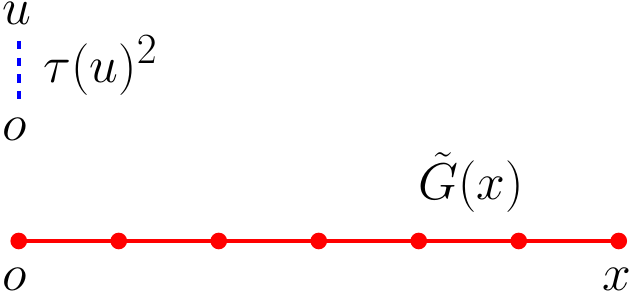}} \]
\caption{Explanation of how to decompose \refeq{pi0decompR:N=1} into two-point 
functions.  The left figure explains the extraction of $G(x-u)^2$ as in 
\refeq{pi0decompR:N=12}, and then the right figure explains the extraction of 
$\tau(u)^2$ and $\tilde G(x)$ after removal of $G(x-u)^2$ as in 
\refeq{tau^20}--\refeq{tau^2}.}
\label{fig:extract}
\end{figure}

\paragraph{Step 3.}
As a practice before considering more complicated cases, we first prove that 
the $N=1$ term in \refeq{pi0decompR} is bounded by 
$2V^1(o,o;x)=2\tilde G(x)^3$, which is 
$2\times\raisebox{-6pt}{\includegraphics[scale=0.27]{pi0diag1}}$ in 
\refeq{pi0diagbd}.  Since $\cL_{[0,|\omega|]}^{\sss(1)}=\{\{0|\omega|\}\}$, 
$\tilde V_{\bm}(|\omega|)=\{x\}$ (cf., \refeq{tildeVdef}) and 
$\ind{\sfL_{\bm,\bk}=\Gamma}\le1$, the $N=1$ term in \refeq{pi0decompR} is 
bounded by
\begin{align}\lbeq{pi0decompR:N=1}
\sum_{\omega\in\Omega(o,x)}\sum_{\substack{\bm\in\Z_+^{\tilde B_\omega}:\\
 \text{odd on }B_\omega,\\ \text{even on }\tilde B_\omega\setminus B_\omega}}
 \frac{w_{\tilde B_\omega}(\bm)}{Z_{\B_\Lambda}}\sum_{\substack{\bk\in
 \Z_+^{\B_\Lambda\setminus\tilde B_\omega}:\\ \partial\bk=\vno}}w_{\B_\Lambda
 \setminus\tilde B_\omega}(\bk)\sum_{u\in\tilde V_{\bm}(0)}\ind{u\cn{\bk}{}x}.
\end{align}
We note that 
\begin{align}\lbeq{pi0decompR:N=1:ind}
\ind{u\in\tilde V_{\bm}(0)}=\delta_{o,u}+\ind{u\in\tilde V_{\bm}(0)\setminus\{o
 \}}.
\end{align}
By \refeq{Remark}, the contribution from $\delta_{o,u}$ is bounded by 
$\tilde G(x)^3$, while the contribution from 
$\ind{u\in\tilde V_{\bm}(0)\setminus\{o\}}$ is bounded as (see Figure~\ref{fig:extract})
\begin{align}\lbeq{pi0decompR:N=12}
&\sum_u\sum_{\omega\in\Omega(o,x)}\sum_{\substack{\bm\in\Z_+^{\tilde
 B_\omega}:\\ \text{odd on }B_\omega,\\ \text{even on }\tilde B_\omega\setminus
 B_\omega}}\frac{w_{\tilde B_\omega}(\bm)}{Z_{\B_\Lambda}}\,\ind{u\in\tilde
 V_{\bm}(0)\setminus\{o\}}\sum_{\substack{\bk\in\Z_+^{\B_\Lambda\setminus\tilde
 B_\omega}:\\ \partial\bk=\vno}}w_{\B_\Lambda\setminus\tilde B_\omega}(\bk)\,
 \ind{u\cn{\bk}{}x}\nn\\
&\stackrel{\text{\refeq{lmm0}}}\le\sum_uG(x-u)^2\sum_{\omega\in\Omega(o,
 x)}\sum_{\substack{\bm\in\Z_+^{\tilde B_\omega}:\\ \text{odd on }B_\omega,\\
 \text{even on }\tilde B_\omega\setminus B_\omega}}\frac{w_{\tilde B_\omega}
 (\bm)}{Z_{\B_\Lambda}}\,Z_{\B_\Lambda\setminus\tilde B_\omega}\,\ind{u\in\tilde
 V_{\bm}(0)\setminus\{o\}}\nn\\
&~\le\sum_uG(x-u)^2\sum_{\substack{\bn\in\Z_+^{\B_\Lambda}:\\ \partial
 \bn=o\vtri x}}\frac{w_{\B_\Lambda}(\bn)}{Z_{\B_\Lambda}}\,\ind{n_{o,u}>0,
 \text{ even}},
\end{align}
where we have used 
$\ind{u\in\tilde V_{\bm}(0)\setminus\{o\}}\le\ind{m_{o,u}>0,\text{ even}}$.  
Notice that
\begin{align}\lbeq{tau^20}
\sum_{\substack{\bn\in\Z_+^{\B_\Lambda}:\\ \partial\bn=o\vtri x}}\frac{w_{
 \B_\Lambda}(\bn)}{Z_{\B_\Lambda}}\,\ind{n_{o,u}>0,\text{ even}}
&=\sum_{n_{o,u}>0,\text{ even}}\frac{(\beta J_{o,u})^{n_{o,u}}}{n_{o,u}!}
 \sum_{\substack{\bn\in\Z_+^{\B_\Lambda\setminus\{o,u\}}:\\ \partial\bn=o\vtri
 x}}\frac{w_{\B_\Lambda\setminus\{o,u\}}(\bn)}{Z_{\B_\Lambda}}\nn\\
&=\big(\cosh(\beta J_{o,u})-1\big)\sum_{\substack{\bn\in\Z_+^{\B_\Lambda
 \setminus\{o,u\}}:\\ \partial\bn=o\vtri x}}\frac{w_{\B_\Lambda\setminus\{o,
 u\}}(\bn)}{Z_{\B_\Lambda}}\nn\\
&=\big(\cosh(\beta J_{o,u})-1\big)\,\frac{Z_{\B_\Lambda\setminus\{o,u\}}\,
 \Exp{\vphi_o\vphi_x}_{\B_\Lambda\setminus\{o,u\}}}{Z_{\B_\Lambda}}.
\end{align}
Since
\begin{align}
Z_{\B_\Lambda}\ge Z_{\B_\Lambda\setminus\{o,u\}}\sum_{n_{o,u}\text{ even}}
 \frac{(\beta J_{o,u})^{n_{o,u}}}{n_{o,u}!}=Z_{\B_\Lambda\setminus\{o,u\}}\cosh
 (\beta J_{o,u}),
\end{align}
and since
\begin{align}
\frac{\cosh(\beta J_{o,u})-1}{\cosh(\beta J_{o,u})}\le\frac{\cosh(\beta J_{o,u})
 -1}{\cosh(\beta J_{o,u})}\underbrace{\frac{\cosh(\beta J_{o,u})+1}{\cosh(\beta
 J_{o,u})}}_{\ge\,1}=\frac{\sinh^2(\beta J_{o,u})}{\cosh^2(\beta J_{o,u})}
 =\tau(u)^2,
\end{align}
we obtain that, for $x\ne o$,
\begin{align}\lbeq{tau^2}
\refeq{pi0decompR:N=12}\le\tilde G(x)\underbrace{\sum_uG(x-u)^2\,\tau(u)^2}_{\le
 \,\tilde G(x)^2}\le\tilde G(x)^3.
\end{align}
As a result, we arrive at
\begin{align}\lbeq{pi0decompR:N=1fin}
\refeq{pi0decompR:N=1}&\le2\tilde G(x)^3=2V^1(o,o;x).
\end{align}

\paragraph{Step 4.}
Next we investigate the contribution to \refeq{pi0decompR} from more general 
$N\ge2$, which is bounded by
\begin{gather}
\sum_{\substack{\{u_j,v_j\}_{j=1}^N\\ (\text{no intersection})}}
 \sum_{\omega\in\Omega(o,x)}\sum_{\{s_jt_j\}_{j=1}^N\in\cL_{[0,|\omega|]}^{\sss
 (N)}}\sum_{\substack{\bm\in\Z_+^{\tilde B_\omega}:\\ \text{odd on }B_\omega,\\
 \text{even on }\tilde B_\omega\setminus B_\omega}}\frac{w_{\tilde B_\omega}
 (\bm)}{Z_{\B_\Lambda}}\prod_{j=1}^N\ind{u_j\in\tilde V_{\bm}(s_j)}\,\ind{v_j\in\tilde
 V_{\bm}(t_j)}\nn\\
\times\sum_{\substack{\bk\in\Z_+^{\B_\Lambda\setminus\tilde B_\omega}:\\
 \partial\bk=\vno}}w_{\B_\Lambda\setminus\tilde B_\omega}(\bk)\,\ind{\sfL_{\bm,
 \bk}=\{s_jt_j\}_{j=1}^N}\prod_{j=1}^N\ind{u_j\cn{\bk}{}v_j},
 \lbeq{pi0decompR:Nge2}
\end{gather}
where the first sum is over $N$ pairs of vertices that do not intersect: 
$\{u_i,v_i\}\cap\{u_j,v_j\}=\vno$ for $i\ne j$. This constraint is due to the 
compatibility with the condition that $\sfL_{\bm,\bk}=\{s_jt_j\}_{j=1}^N$.  
For example, if the lace is given as in Figure~\ref{fig:lace}, then it looks as 
if the coincidence $v_2=u_3$ is possible, but it is not; if $v_2=u_3$, 
then $u_2$ is directly connected to $x$ by using bonds $b$ with $k_b>0$, 
which implies that the number of lace edges is two, not three as suggested in 
Figure~\ref{fig:lace}.  Therefore, the constraint in the first sum in 
\refeq{pi0decompR:Nge2} is unnecessary as long as we keep the indicator 
$\ind{\sfL_{\bm,\bk}=\{s_jt_j\}_{j=1}^N}$.  However, since we drop this 
indicator at some point in the following analysis, we keep this constraint 
for now.

Let $N=2$, for example, and let $\Gamma=\{st,s't'\}=\{0t,s'|\omega|\}$.  
Then the sum over $\bk$ is (n.b., $\tilde V_{\bm}(|\omega|)=\{x\}$)
\begin{align}\lbeq{egGamma=2}
\sum_{\substack{\bk\in\Z_+^{\B_\Lambda\setminus\tilde
 B_\omega}:\\ \partial\bk=\vno}}w_{\B_\Lambda\setminus\tilde B_\omega}(\bk)\,
 \ind{\sfL_{\bm,\bk}=\{0t,s'|\omega|\}}\ind{u_1\cn{\bk}{}v_1}\,\ind{u_2
 \cn{\bk}{}x}.
\end{align}
Notice that, under the constraint $\sfL_{\bm,\bk}=\{0t,s'|\omega|\}$ (recall 
\refeq{lacedef1}--\refeq{lacedef3}, where lace edges are defined by min/max of 
connectivity), the clusters $\cC_{\bk}(u_j)=\{y:u_j\cn{\bk}{}y\}$ are 
disjoint, i.e., $\cC_{\bk}(u_1)\cap\cC_{\bk}(u_2)=\vno$; otherwise the 
number $N$ of lace edges is reduced to 1, which is a contradiction to the 
present situation: $N=2$.  Therefore, we can 
condition on $\cC_{\bk}(u_1)$, say, $\cC_{\bk}(u_1)=A~(\ni v_1)$, split the 
weight $w_{\B_\Lambda\setminus\tilde B_\omega}(\bk)$ as  
$w_{\B_{\bar A}}(\bk_1)\,w_{\B_{A\compl}}(\bk_2)$, where $\B_{\bar A}$ is 
an abbreviation for $\B_\Lambda\setminus\tilde B_\omega\setminus\B_{A\compl}$ 
(recall that $\B_{A\compl}$ is the set of bonds whose end vertices are both in 
$A\compl$, as defined above \refeq{Theta'def}), and then sum over $\bk_2$ to 
obtain 
\begin{eqnarray}
\refeq{egGamma=2}&\le&
 \sum_{A\ni v_1}\sum_{\substack{\bk_1\in
 \Z_+^{\B_{\bar A}}:\\ \partial\bk_1=\vno}}w_{\B_{\bar A}}(\bk_1)\,
 \ind{\cC_{\bk_1}(u_1)=A}\sum_{\substack{\bk_2\in\Z_+^{\B_{A\compl}}:\\ \partial
 \bk_2=\vno}}w_{\B_{A\compl}}(\bk_2)\,\ind{u_2\cn{\bk_2}{}x}\nn\\
&\stackrel{\text{\refeq{lmm0}}}\le&
 G(x-u_2)^2\sum_{A\ni
 v_1}\sum_{\substack{\bk_1\in\Z_+^{\B_{\bar A}}:\\ \partial\bk_1=\vno}}
 w_{\B_{\bar A}}(\bk_1)\,Z_{\B_{A\compl}}\ind{\cC_{\bk_1}(u_1)=A}\nn\\
&=&
 G(x-u_2)^2\sum_{\substack{\bk\in\Z_+^{\B_\Lambda
 \setminus\tilde B_\omega}:\\ \partial\bk=\vno}}w_{\B_\Lambda\setminus\tilde
 B_\omega}(\bk)\,\ind{u_1\cn{\bk}{}v_1}\nn\\
&\stackrel{\text{\refeq{lmm0}}}\le&
 G(x-u_2)^2\,G(v_1-u_1)^2\,Z_{\B_\Lambda\setminus\tilde B_\omega}.
 \lbeq{condoncluster}
\end{eqnarray}

\begin{figure}[t]
\begin{center}
\includegraphics[scale=0.5]{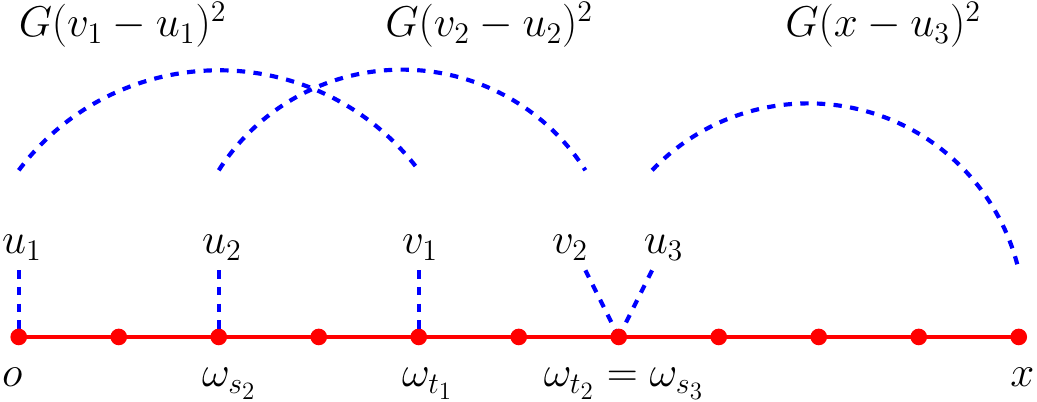}
\end{center}
\caption{Explanation of Step~4: extracting the two-point functions 
$\prod_{i=1}^3G(v_i-u_i)^2$ in \refeq{pi0decompR:Nge2-step4} from 
\refeq{pi0decompR:Nge2} with three lace edges (cf., Figure~\ref{fig:lace}).}
\label{fig:lace-before}
\end{figure}

It is easy to extend the above analysis to more general $N\ge3$.  
As a result, we obtain (see Figure~\ref{fig:lace-before})
\begin{gather}
\refeq{pi0decompR:Nge2}\le
 \sum_{\substack{\{u_j,v_j\}_{j=1}^N\\ (\text{no intersection})}}
 \prod_{i=1}^NG(v_i-u_i)^2\sum_{\omega\in\Omega(o,
 x)}\sum_{\{s_jt_j\}_{j=1}^N\in\cL_{[0,|\omega|]}^{\sss(N)}}\sum_{\substack{\bm
 \in\Z_+^{\tilde B_\omega}:\\ \text{odd on }B_\omega,\\ \text{even on }
 \tilde B_\omega\setminus B_\omega}}\frac{w_{\tilde B_\omega}(\bm)}
 {Z_{\B_\Lambda}}\,Z_{\B_\Lambda\setminus\tilde B_\omega}\nn\\
\times\prod_{j=1}^N\ind{u_j\in\tilde V_{\bm}(s_j)}\,\ind{v_j\in\tilde V_{\bm}
 (t_j)}.\lbeq{pi0decompR:Nge2-step4}
\end{gather}

\paragraph{Step 5.}
We complete the proof for $N\ge2$ by first extracting $2N-1$ factors of 
$\delta+\tau^2$ and then extracting $2N-1$ two-point functions from the sum 
over $\omega\in\Omega(o,x)$ in \refeq{pi0decompR:Nge2-step4}, just as done for 
$N=1$ in \refeq{tau^20}--\refeq{tau^2}.  To do so, we first use 
$\ind{u\in\tilde V_{\bm}(s)}=\delta_{u,\omega_s}+\ind{u\in\tilde V_{\bm}(s)
\setminus\{\omega_s\}}$ (cf., \refeq{pi0decompR:N=1:ind}) to rewrite the sum 
over $\omega\in\Omega(o,x)$ in \refeq{pi0decompR:Nge2-step4} as
\begin{align}\lbeq{pi0decompR:Nge2-step5}
&\sum_{\omega\in\Omega(o,x)}\sum_{\{s_jt_j\}_{j=1}^N\in\cL_{[0,|\omega|]}^{\sss
 (N)}}\sum_{\substack{\bm\in\Z_+^{\tilde B_\omega}:\\ \text{odd on }B_\omega,\\
 \text{even on }\tilde B_\omega\setminus B_\omega}}\frac{w_{\tilde B_\omega}
 (\bm)}{Z_{\B_\Lambda}}\,Z_{\B_\Lambda\setminus\tilde B_\omega}\nn\\
&\qquad\times\prod_{j=1}^N\Big(\delta_{u_j,\omega_{s_j}}+\ind{u_j\in\tilde
 V_{\bm}(s_j)\setminus\{\omega_{s_j}\}}\Big)\Big(\delta_{v_j,\omega_{t_j}}
 +\ind{v_j\in\tilde V_{\bm}(t_j)\setminus\{\omega_{t_j}\}}\Big)\nn\\
&=\sum_{\omega\in\Omega(o,x)}\sum_{\{s_jt_j\}_{j=1}^N\in\cL_{[0,|\omega|]}^{\sss
 (N)}}\sum_{\substack{I,J\subset[N]:\\ N\in J\compl}}\bigg(\prod_{i\in I\compl}
 \delta_{u_i,\omega_{s_i}}\prod_{j\in J\compl}\delta_{v_j,\omega_{t_j}}\bigg)\nn\\
&\qquad\times\sum_{\substack{\bm\in\Z_+^{\tilde B_\omega}:\\ \text{odd on }
 B_\omega,\\ \text{even on }\tilde B_\omega\setminus B_\omega}}\frac{w_{\tilde
 B_\omega}(\bm)}{Z_{\B_\Lambda}}\,Z_{\B_\Lambda\setminus\tilde B_\omega}
 \prod_{\substack{i\in I,\\ j\in J}}\ind{u_i\in\tilde V_{\bm}(s_i)\setminus\{
 \omega_{s_i}\}}\,\ind{v_j\in\tilde V_{\bm}(t_j)\setminus\{\omega_{t_j}\}},
\end{align}
where $[N]=\{1,\dots,N\}$, and $J\compl$ is an abbreviation for 
$[N]\setminus J$.  Since $\{u_j,v_j\}_{j=1}^N$ are $N$ pairs of 
vertices that do not intersect, the bonds in $\tilde B_{I,J}=\{\{\omega_{s_i},
u_i\},\{\omega_{t_j},v_j\}\}_{i\in I,j\in J}\subset\tilde B_\omega\setminus
B_\omega$ (depicted as dashed short line segments in 
Figure~\ref{fig:lace-before}) are distinct.  
Then, by the same analysis as in \refeq{tau^20}--\refeq{tau^2}, the last line 
of \refeq{pi0decompR:Nge2-step5} is bounded above by
\begin{align}
\prod_{\substack{i\in I,\\ j\in J}}\tau(\omega_{s_i}-u_i)^2\,\tau(\omega_{t_j}
 -v_j)^2\sum_{\substack{\bm\in\Z_+^{\tilde B_\omega\setminus\tilde B_{I,J}}:\\
 \text{odd on }B_\omega,\\ \text{even on }\tilde B_\omega\setminus\tilde B_{I,
 J}\setminus B_\omega}}\frac{w_{\tilde B_\omega\setminus\tilde B_{I,J}}(\bm)}
 {Z_{\B_\Lambda\setminus\tilde B_{I,J}}}\,Z_{\B_\Lambda\setminus\tilde
 B_\omega}.
\end{align}
Therefore, by changing the order of summations, we obtain
\begin{align}\lbeq{pi0decompR:Nge2-step51}
\refeq{pi0decompR:Nge2-step5}&\le\sum_{\substack{y_1,\dots,y_N,\\ z_1,\dots,
 z_N:\\ y_1=o,\\ z_N=x}}\sum_{I\subset[N]}\bigg(\prod_{i\in I\compl}\delta_{u_i,
 y_i}\prod_{i'\in I}\tau(y_{i'}-u_{i'})^2\bigg)\sum_{\substack{J\subset[N]:\\
 N\in J\compl}}\bigg(\prod_{j\in J\compl}\delta_{v_j,z_j}\prod_{j'\in J}
 \tau(z_{j'}-v_{j'})^2\bigg)\nn\\
&\quad\times\sum_{\omega\in\Omega(o,x)}\sum_{\{s_jt_j\}_{j=1}^N\in\cL_{[0,
 |\omega|]}^{\sss(N)}}\prod_{j=1}^N\delta_{\omega_{s_j},y_j}\delta_{\omega_{t_j}
 ,z_j}\sum_{\substack{\bm\in\Z_+^{\tilde B_\omega\setminus\tilde B_{I,J}}:\\
 \text{odd on }B_\omega,\\ \text{even on }\tilde B_\omega\setminus\tilde B_{I,
 J}\setminus B_\omega}}\frac{w_{\tilde B_\omega\setminus\tilde B_{I,J}}(\bm)}
 {Z_{\B_\Lambda\setminus\tilde B_{I,J}}}\,Z_{\B_\Lambda\setminus\tilde
 B_\omega}.
\end{align}

The remaining task is to extract $2N-1$ two-point functions one by one from the 
second line of \refeq{pi0decompR:Nge2-step51}.  To do so, we first split the 
sum over $\omega\in\Omega(o,x)$ into $\xi\in\Omega(o,z_{N-1})$ with 
$\xi\not\ni x$ and $\eta\in\tilde\Omega_{\xi,I,J}(z_{N-1},x)$, where
\begin{align}
\tilde\Omega_{\xi,I,J}(z,x)=\big\{\eta\in\Omega(z,x):\forall j=1,\dots,|\eta|,~
 \{\eta_{j-1},\eta_j\}\notin\tilde B_\xi\cup\tilde B_{I,J}\big\},
\end{align}
which is the set of nonzero paths from $z$ to $x~(\ne z$, due to the definition of 
$\Omega(z,x)$ in \refeq{Omega-def}) consisting of bonds that are not yet 
explored by $\xi$ or used to extract $\tau^2$ as in 
\refeq{pi0decompR:Nge2-step51}.  Then, by splitting the weight 
$w_{\tilde B_\omega\setminus\tilde B_{I,J}}(\bm)$ as 
$w_{\tilde B_\xi\setminus\tilde B_{I,J}}(\bk)\,w_{\tilde B_\eta\setminus(\tilde
 B_\xi\cup\tilde B_{I,J})}(\bm)$ and multiplying 
$1= Z_{\B_\Lambda\setminus(\tilde B_\xi\cup\tilde B_{I,J})}
/Z_{\B_\Lambda\setminus(\tilde B_\xi\cup\tilde B_{I,J})}$, the second line 
of \refeq{pi0decompR:Nge2-step51} equals
\begin{align}\lbeq{pi0decompR:Nge2-step52}
&\sum_{\substack{\xi\in\Omega(o,z_{N-1}):\\ \xi\not\ni x}}\sum_{\substack{\{s_jt_j\}_{j=1}^{N-1}\in\cL_{[0,
 |\xi|]}^{\sss(N-1)},\\ s_N\in(t_{N-2},t_{N-1}]}}\prod_{j=1}^{N-1}\delta_{
 \xi_{s_j},y_j}\delta_{\xi_{t_j},z_j}\delta_{\xi_{s_N},y_N}\sum_{\substack{\bk
 \in\Z_+^{\tilde B_\xi\setminus\tilde B_{I,J}}:\\ \text{odd on }B_\xi,\\
 \text{even on }\tilde B_\xi\setminus\tilde B_{I,J}\setminus B_\xi}}
 \frac{w_{\tilde B_\xi\setminus\tilde B_{I,J}}(\bk)}{Z_{\B_\Lambda\setminus
 \tilde B_{I,J}}}\,Z_{\B_\Lambda\setminus(\tilde B_\xi\cup\tilde B_{I,J})}
 \nn\\
&\quad\times\ind{z_{N-1}\ne x}\sum_{\eta\in\tilde\Omega_{\xi,I,J}(z_{N-1},x)}
 \sum_{\substack{\bm\in\Z_+^{\tilde B_\eta\setminus(\tilde B_\xi\cup\tilde
 B_{I,J})}:\\ \text{odd on }B_\eta,\\ \text{even on }\tilde B_\eta\setminus
 (\tilde B_\xi\cup\tilde B_{I,J})\setminus B_\eta}}\frac{w_{\tilde B_\eta
 \setminus(\tilde B_\xi\cup\tilde B_{I,J})}(\bm)}{Z_{\B_\Lambda\setminus
 (\tilde B_\xi\cup\tilde B_{I,J})}}\,Z_{\B_\Lambda\setminus(\tilde B_\xi\cup
 \tilde B_\eta)}.
\end{align}
Since
\begin{align}
&\big(\tilde B_\eta\setminus(\tilde B_\xi\cup\tilde B_{I,J})\big)\cap\big(
 \B_\Lambda\setminus(\tilde B_\xi\cup\tilde B_\eta)\big)=\vno,\\
&\big(\tilde B_\eta\setminus(\tilde B_\xi\cup\tilde B_{I,J})\big)\cup\big(
 \B_\Lambda\setminus(\tilde B_\xi\cup\tilde B_\eta)\big)=\B_\Lambda\setminus
 (\tilde B_\xi\cup\tilde B_{I,J}),
\end{align} 
and since $\eta$ is the earliest path of odd current in the restricted region 
$\B_\Lambda\setminus(\tilde B_\xi\cup\tilde B_{I,J})$,  the last line 
of \refeq{pi0decompR:Nge2-step52} is exactly equal to $\Exp{\vphi_{z_{N-1}}
\vphi_x}_{\B_\Lambda\setminus\tilde B_{I,J}\setminus\tilde B_\xi}$, 
which is further bounded by $\tilde G(x-z_{N-1})$ for $z_{N-1}\ne x$.  
Repeating the above analysis to extract two-point functions one by one 
and using $\ind{\xi\not\ni x}\le\ind{\xi_{s_N}\ne x}$, we obtain
\begin{align}\lbeq{pi0decompR:Nge2-step52bd}
\refeq{pi0decompR:Nge2-step52}\le\tilde G(y_2)\prod_{i=1}^{N-1}G(
 z_i-y_{i+1})\prod_{j=2}^{N-1}\tilde G(y_{j+1}-z_{j-1})\,\tilde G(x-z_{N
 -1})\,\ind{y_N\ne x},
\end{align}
where we have used \refeq{tildeGdef} to gain $\tilde G$ instead of $G$ for 
$y_2\ne o$ and $y_{j+1}\ne z_{j-1}$ for all $j=2,\dots,N-1$, due to 
the construction \refeq{cLdef} of lace edges: each $\tilde G$ (resp., $G$) in 
\refeq{pi0decompR:Nge2-step52bd} corresponds to a strict inequality 
(resp., an inequality) in \refeq{cLdef}.  The empty product $\prod_{j=2}^0$ is 
regarded as 1 by convention, as always.  As a result, we obtain
\begin{align}\lbeq{pi0decompR:Nge2-step53}
\refeq{pi0decompR:Nge2-step51}&\le\sum_{\substack{y_1,\dots,y_N,\\ z_0,\dots,
 z_N:\\ y_1=z_0=o,\\ z_N=x}}\prod_{j=1}^{N-1}G(z_j-y_{j+1})\,\tilde G(y_{j+1}
 -z_{j-1})\,\tilde G(x-z_{N-1})\,\ind{y_N\ne x}\nn\\
&\quad\times\underbrace{\sum_{I\subset[N]}\bigg(\prod_{i\in I\compl}\delta_{u_i,
 y_i}\prod_{i'\in I}\tau(y_{i'}-u_{i'})^2\bigg)}_{\prod_{j=1}^N(\delta
 +\tau^2)(y_j-u_j)}\underbrace{\sum_{\substack{J\subset[N]:\\ N\in J
 \compl}}\bigg(\prod_{j\in J\compl}\delta_{v_j,z_j}\prod_{j'\in J}\tau
 (z_{j'}-v_{j'})^2\bigg)}_{\prod_{j=1}^{N-1}(\delta+\tau^2)(z_j-v_j)\,
 \delta_{v_N,x}},
\end{align}
hence
\begin{align}\lbeq{pi0decompR:Nge2-step54}
\refeq{pi0decompR:Nge2-step4}&\le\sum_{\substack{y_1,\dots,y_N,\\ z_0,\dots,
 z_{N-1}:\\ y_1=z_0=o}}\prod_{j=1}^{N-1}\underbrace{\Big((\delta+\tau^2)*G^2*
 (\delta+\tau^2)\Big)(z_j-y_j)\,G(z_j-y_{j+1})\,\tilde G(y_{j+1}-z_{j-1})}_{\le
 \,U^1(z_{j-1},y_j;z_j,y_{j+1})~(\because\text{ \refeq{tildeGdef}})}\nn\\
&\hskip5pc\times\underbrace{\tilde G(x-z_{N-1})\,\big((\delta+\tau^2)*G^2\big)
 (x-y_N)\,\ind{y_N\ne x}}_{\le\,2V^1(z_{N-1},y_N;x)~(\because
 \text{ \refeq{pi0decompR:N=1fin}})}\nn\\
&\le2\big((U^1)^{\star(N-1)}\star V^1\big)_{o,x}.
\end{align}
Combining this for $N\ge2$ with \refeq{pi0decompR:N=1fin} for $N=1$ and 
recalling the definition \refeq{Xdef} of $X_{o,x}^1$, we complete the 
proof of Theorem~\ref{thm:pi0bd}.
\QED

\subsection{Proof of Theorem~\ref{thm:Theta'bd}}\label{ss:thm2}
Recall the definition \refeq{Theta'def} of $\Theta'_{o,x;A}$:
\begin{align}\lbeq{Theta'redef}
\Theta'_{o,x;A}=\sum_{\substack{\bl\in\Z_+^{\B_{A\compl}}:\\ \partial\bl=\vno}}
 \frac{w_{\B_{A\compl}}(\bl)}{Z_{\B_{A\compl}}}\sum_{\substack{\bn\in
 \Z_+^{\B_\Lambda}:\\ \partial\bn=o\vtri x}}\frac{w_{\B_\Lambda}(\bn)}
 {Z_{\B_\Lambda}}\ind{o\db{\bn+\bl}{A}x}.
\end{align}
Since it is similar to $\pi_{\B_\Lambda}^{\sss(0)}(x)$, we can follow the 
same line of proof as explained in the previous subsection, by taking note of 
the following two differences:
\begin{enumerate}[(i)]
\item
All paths from $o$ to $x$ with positive current in the superposition of two 
current configurations must go through the vertex set $A$, so that the 
earliest path $\omega\in\Omega(o,x)$ of odd current also contains a vertex 
in $A$.
\item
A double connection from $o$ to $x$ is achieved by the superposition of two 
current configurations, not by a single current configuration as in 
$\pi_{\B_\Lambda}^{\sss(0)}(x)$, and one of them is defined in the restricted 
region $\B_{A\compl}$.
\end{enumerate}

Now we begin the proof of Theorem~\ref{thm:Theta'bd}.  First, by identifying 
the earliest path $\omega\in\Omega(o,x)$ of bonds $b$ with odd $n_b$ (as done 
in Step~1 of the previous subsection) and then relaxing the through-$A$ 
condition to $\omega\cap A\ne\vno$, we obtain the following inequality similar 
to \refeq{pi0decomp}:
\begin{align}\lbeq{Theta'decomp}
\Theta'_{o,x;A}\le\sum_{\substack{\omega\in\Omega(o,x):\\ \omega\cap A\ne\vno}}
 \sum_{\substack{\bm\in\Z_+^{\tilde B_\omega}:\\ \text{odd on }B_\omega,\\
 \text{even on }\tilde B_\omega\setminus B_\omega}}\frac{w_{\tilde B_\omega}
 (\bm)}{Z_{\B_\Lambda}}\sum_{\substack{\bk\in\Z_+^{\B_\Lambda\setminus\tilde
 B_\omega}:\\ \partial\bk=\vno}}w_{\B_\Lambda\setminus\tilde B_\omega}(\bk)
 \sum_{\substack{\bl\in\Z_+^{\B_{A\compl}}:\\ \partial\bl=\vno}}\frac{w_{\B_{A
 \compl}}(\bl)}{Z_{\B_{A\compl}}}\,\ind{o\db{\bm+\bk+\bl}{}x}.
\end{align}
Then, by using the double expansion with a lace $\sfL_{\bm,\bk+\bl}$ (as done 
in Step~2 of the previous subsection), we obtain the following inequality 
similar to \refeq{pi0decompR}:
\begin{align}\lbeq{Theta'bd1}
&\Theta'_{o,x;A}\le\sum_{N=1}^\infty\sum_{\substack{\omega\in\Omega(o,x):\\
 \omega\cap A\ne\vno}}\sum_{\Gamma\in\cL_{[0,|\omega|]}^{\sss(N)}}
 \sum_{\substack{\bm\in\Z_+^{\tilde B_\omega}:\\ \text{odd on }B_\omega,\\
 \text{even on }\tilde B_\omega\setminus B_\omega}}\frac{w_{\tilde B_\omega}
 (\bm)}{Z_{\B_\Lambda}}\sum_{\substack{\bk\in\Z_+^{\B_\Lambda\setminus\tilde
 B_\omega}:\\ \partial\bk=\vno}}w_{\B_\Lambda\setminus\tilde B_\omega}(\bk)
 \sum_{\substack{\bl\in\Z_+^{\B_{A\compl}}:\\ \partial\bl=\vno}}\frac{w_{\B_{A
 \compl}}(\bl)}{Z_{\B_{A\compl}}}\nn\\
&\hskip4pc\times\ind{\sfL_{\bm,\bk+\bl}=\Gamma}\prod_{st\in\Gamma}\ind{\tilde
 V_{\bm}(s)\cn{\bk+\bl}{}\tilde V_{\bm}(t)\text{ in }\B_\Lambda\setminus\tilde
 B_\omega}\nn\\
&~\le\sum_{N=1}^\infty\!\!\!
 \sum_{\substack{\{u_j,v_j\}_{j=1}^N\\ (\text{no intersection})}}
 \sum_{\substack{\omega\in\Omega(o,x):\\ \omega\cap A\ne\vno}}\sum_{\{s_j
 t_j\}_{j=1}^N\in\cL_{[0,|\omega|]}^{\sss(N)}}\!\!\!\sum_{\substack{\bm\in
 \Z_+^{\tilde B_\omega}:\\ \text{odd on }B_\omega,\\ \text{even on }
 \tilde B_\omega\setminus B_\omega}}\!\!\!\!\!\frac{w_{\tilde B_\omega}
 (\bm)}{Z_{\B_\Lambda}}\prod_{j=1}^N\ind{u_j\in\tilde V_{\bm}(s_j)}\,\ind{v_j\in
 \tilde V_{\bm}(t_j)}\nn\\
&\qquad\times\sum_{\substack{\bk\in\Z_+^{\B_\Lambda\setminus\tilde
 B_\omega}:\\ \partial\bk=\vno}}w_{\B_\Lambda\setminus\tilde B_\omega}(\bk)
 \sum_{\substack{\bl\in\Z_+^{\B_{A\compl}}:\\ \partial\bl=\vno}}\frac{w_{\B_{A
 \compl}}(\bl)}{Z_{\B_{A\compl}}}\,\ind{\sfL_{\bm,\bk+\bl}=\{s_jt_j\}_{j=1}^N}
 \prod_{j=1}^N\ind{u_j\cn{\bk+\bl}{}v_j\text{ in }\B_\Lambda\setminus\tilde
 B_\omega}.
\end{align}
However, we cannot use \refeq{lmm0} here to extract 
$Z_{\B_\Lambda\setminus\tilde B_\omega}\prod_{j=1}^NG(v_j-u_j)^2$ from 
the double sum over $\bk,\bl$ (as done in Step~3 and Step~4 in the previous 
subsection), due to the difference (ii) mentioned above.  Instead, as described 
in \refeq{bubblechain}, the last line of \refeq{Theta'bd1} is bounded by chains 
of nonzero bubbles $Z_{\B_\Lambda\setminus\tilde B_\omega}\prod_{j=1}^N
\sum_{i=0}^\infty(\tilde G^2)^{*i}(v_j-u_j)$.  Then, we obtain the following 
inequality similar to \refeq{pi0decompR:Nge2-step4}:
\begin{gather}
\Theta'_{o,x;A}\le\sum_{N=1}^\infty\!\!\!
 \sum_{\substack{\{u_j,v_j\}_{j=1}^N\\ (\text{no intersection})}}\!\!\!
 \prod_{j=1}^N\sum_{i_j=0}^\infty(\tilde G^2)^{*i_j}(v_j-u_j)
 \sum_{\substack{\omega\in\Omega(o,x):\\ \omega\cap A\ne\vno}}\sum_{\{s_j
 t_j\}_{j=1}^N\in\cL_{[0,|\omega|]}^{\sss(N)}}\!\!\!\sum_{\substack{\bm\in\Z_+^{\tilde
 B_\omega}:\\ \text{odd on }B_\omega,\\ \text{even on }\tilde B_\omega
 \setminus B_\omega}}\!\!\!\!\!\frac{w_{\tilde B_\omega}(\bm)}{Z_{\B_\Lambda}}\nn\\
\times Z_{\B_\Lambda\setminus\tilde B_\omega}\prod_{j=1}^N\ind{u_j\in\tilde
 V_{\bm}(s_j)}\,\ind{v_j\in\tilde V_{\bm}(t_j)}.\lbeq{Theta'bd2}
\end{gather}

The remaining task is to extract two-point functions and factors of 
$\delta+\tau^2$ from the above sum over $\omega$, as done in Step~5 of the 
previous subsection.  However, since $\omega\cap A\ne\vno$, among $2N-1$ 
segments $\{\omega_{[0,s_2)},\omega_{[s_2,t_1)},\omega_{[t_1,s_3)},\dots,
\omega_{[s_N,t_{N-1})},\omega_{[t_{N-1},|\omega|]}\}$, where 
$\omega_{[s,t)}=(\omega_s,\omega_{s+1},\dots,\omega_{t-1})$ and $\omega_{[t_{N
-1},|\omega|]}=\omega_{[t_{N-1},|\omega|)}\cup\{\omega_{|\omega|}\}$, there is 
always a segment that contains a vertex $a\in A$.  Therefore, to bound 
\refeq{Theta'bd2}, we replace the product of $2N-1$ two-point functions in 
\refeq{pi0decompR:Nge2-step53}, i.e., 
\begin{align}
\prod_{j=1}^{N-1}G(z_j-y_{j+1})\,\tilde G(y_{j+1}-z_{j-1})\,\tilde
 G(x-z_{N-1}),\qquad\text{where }~z_0=o,
\end{align}
by
\begin{align}\lbeq{G*G}
&\sum_{a\in A}\bigg(\prod_{j=1}^{N-1}G(z_j-y_{j+1})\,\tilde G(y_{j+1}-z_{j-1})
 \Big(G(a-z_{N-1})\,\tilde G(x-a)+G(x-z_{N-1})\,\delta_{a,x}\Big)\nn\\
&\quad+\sum_{i=1}^{N-1}\Big(G(a-y_{i+1})\,\tilde G(z_i-a)\,\tilde G(y_{i+1}
 -z_{i-1})+G(z_i-y_{i+1})\,G(a-z_{i-1})\,\tilde G(y_{i+1}-a)\Big)\nn\\
&\qquad\times\prod_{\substack{j=1\\ (j\ne i)}}^{N-1}G(z_j-y_{j+1})\,
 \tilde G(y_{j+1}-z_{j-1})\,\tilde G(x-z_{N-1})\bigg),
\end{align}
which is obtained by applying \refeq{lmm1} to each segment.  Assembling all 
the above estimates yields the wanted bound \refeq{Theta'bd}, just as done 
in \refeq{pi0decompR:Nge2-step54} for $\pi_{\B_\Lambda}^{\sss(0)}(x)$.
\QED

\subsection{Proof of Theorem~\ref{thm:pi0'bd}}\label{ss:thm3}
Recall the definition \refeq{tildepi0def} of 
$\tilde\pi_{\B_\Lambda;y}^{\sss(0)}(x)$:
\begin{align}\lbeq{tildepi0redef}
\tilde\pi_{\B_\Lambda;y}^{\sss(0)}(x)=\sum_{\substack{\bn\in
 \Z_+^{\B_\Lambda}:\\ \partial\bn=o\vtri x}}\frac{w_{\B_\Lambda}(\bn)}
 {Z_{\B_\Lambda}}\ind{o\db{\bn}{}x\}\cap\{o\cn{\bn}{}y}.
\end{align}
This looks simpler than $\Theta'_{o,x;A}$, as we only need to control one 
current configuration, not two.  It turns out to be a little more involved, 
due to the extra $\ind{o\cn{\bn}{}y}$, as explained now.

First, by identifying the earliest path $\omega\in\Omega(o,x)$ of bonds $b$ 
with odd $n_b$ (as done in Step~1 of Section~\ref{ss:thm1}), we can rewrite 
$\tilde\pi_{\B_\Lambda;y}^{\sss(0)}(x)$ as (cf., \refeq{pi0decomp})
\begin{align}\lbeq{pi0'decomp}
\tilde\pi_{\B_\Lambda;y}^{\sss(0)}(x)=\sum_{\omega\in\Omega(o,x)}
 \sum_{\substack{\bm\in\Z_+^{\tilde B_\omega}:\\ \text{odd on }B_\omega,\\
 \text{even on }\tilde B_\omega\setminus B_\omega}}\frac{w_{\tilde B_\omega}
 (\bm)}{Z_{\B_\Lambda}}\sum_{\substack{\bk\in\Z_+^{\B_\Lambda\setminus\tilde
 B_\omega}:\\ \partial\bk=\vno}}w_{\B_\Lambda\setminus\tilde B_\omega}(\bk)\,
 \ind{o\db{\bm+\bk}{}x\}\cap\{o\cn{\bm+\bk}{}y}.
\end{align}
Then, by the double expansion as in Step 2 of Section~\ref{ss:thm1}, 
we obtain (see \refeq{pi0decompR} for the equality below and then 
\refeq{pi0decompR:Nge2} for the inequality)
\begin{align}\lbeq{pi0'lace}
\refeq{pi0'decomp}&=\sum_{N=1}^\infty\sum_{\omega\in\Omega(o,x)}\sum_{\Gamma
 \in\cL_{[0,|\omega|]}^{\sss(N)}}\sum_{\substack{\bm\in\Z_+^{\tilde B_\omega}:\\
 \text{odd on }B_\omega,\\ \text{even on }\tilde B_\omega\setminus
 B_\omega}}\frac{w_{\tilde B_\omega}(\bm)}{Z_{\B_\Lambda}}\nn\\
&\qquad\times\sum_{\substack{\bk\in\Z_+^{\B_\Lambda\setminus\tilde B_\omega}:\\
 \partial\bk=\vno}}w_{\B_\Lambda\setminus\tilde B_\omega}(\bk)\,\ind{o\cn{\bm
 +\bk}{}y}\,\ind{\sfL_{\bm,\bk}=\Gamma}\prod_{st\in\Gamma}\ind{\tilde V_{\bm}(s)
 \cn{\bk}{}\tilde V_{\bm}(t)}\nn\\
&\le\sum_{N=1}^\infty\!\!\!
 \sum_{\substack{\{u_j,v_j\}_{j=1}^N\\ (\text{no intersection})}}\!\!\!
 \sum_{\omega\in\Omega(o,x)}\sum_{\{s_jt_j\}_{j=1}^N\in\cL_{[0,|\omega|]}^{\sss
 (N)}}\!\!\!\sum_{\substack{\bm\in\Z_+^{\tilde B_\omega}:\\ \text{odd on }
 B_\omega,\\ \text{even on }\tilde B_\omega\setminus B_\omega}}
 \!\!\!\!\!\frac{w_{\tilde B_\omega}(\bm)}{Z_{\B_\Lambda}}\prod_{j=1}^N\ind{u_j
 \in\tilde V_{\bm}(s_j)}\,\ind{v_j\in\tilde V_{\bm}(t_j)}\nn\\
&\qquad\times\sum_{\substack{\bk\in\Z_+^{\B_\Lambda\setminus\tilde B_\omega}:\\
 \partial\bk=\vno}}w_{\B_\Lambda\setminus\tilde B_\omega}(\bk)\,\ind{o\cn{\bm
 +\bk}{}y}\,\ind{\sfL_{\bm,\bk}=\{s_jt_j\}_{j=1}^N}\prod_{st\in\Gamma}\ind{u_j
 \cn{\bk}{}v_j}.
\end{align}

Next we investigate the effect of the indicator $\ind{o\cn{\bm+\bk}{}y}$.  
Since $\{\cC_{\bk}(u_j)\}_{j=1}^N$ are disjoint, i.e., 
$\cC_{\bk}(u_i)\cap\cC_{\bk}(u_j)=\vno$ for $i\ne j$, we have the rewrite
\begin{align}\lbeq{inddecomp}
\ind{o\cn{\bm+\bk}{}y}=\sum_{i=1}^N\ind{u_i\cn{\bk}{}y}+\ind{o\cn{\bm+\bk}
 {}y\}\setminus\bigcup_{i=1}^N\{u_i\cn{\bk}{}y}.
\end{align}
By conditioning on clusters (as done in Steps~3 \& 4 in Section~\ref{ss:thm1}) 
and using Lemma~\ref{lmm:lmm2}, the contribution to \refeq{pi0'lace} from 
$\ind{u_N\cn{\bk}{}y}$ is bounded by (cf., \refeq{pi0decompR:Nge2-step54})
\begin{align}\lbeq{tildepibd2}
\big((U^1)^{\star(N-1)}\star\DDotV^1_y\big)_{o,x}=\sum_{\substack{y_1,
 \dots,y_N,\\ z_0,\dots,z_{N-1}:\\ y_1=z_0=o}}\prod_{j=1}^{N-1}U^1(z_{j-1},y_j;
 z_j,y_{j+1})\,\DDotV^1_y(z_{N-1},y_N;x),
\end{align}
while the contribution from each $\ind{u_i\cn{\bk}{}y}$ with $i<N$ is bounded 
by
\begin{align}\lbeq{tildepibd1}
2\Big((U^1)^{\star(i-1)}\star\DDotU^1_y\star(U^1)^{\star(N-1-i)}\star V^1
 \Big)_{o,x},
\end{align}
where $\DDotU^1_y$ and $\DDotV^1_y$ are defined in 
\refeq{ddotUdef}--\refeq{ddotVdef}.

To bound the contribution to \refeq{pi0'lace} from $\ind{o\cn{\bm+\bk}{}y\}
\setminus\bigcup_{j=1}^N\{u_j\cn{\bk}{}y}$ in \refeq{inddecomp} is not much 
difficult, as we can follow the same line up to \refeq{pi0decompR:Nge2-step51}, 
with its second line replaced by 
\begin{align}\lbeq{tildepibd3}
\sum_{\omega\in\Omega(o,x)}\sum_{\{s_jt_j\}_{j=1}^N\in\cL_{[0,|\omega|]}^{\sss
 (N)}}\prod_{j=1}^N\delta_{\omega_{s_j},y_j}\delta_{\omega_{t_j},z_j}\sum_{
 \substack{\bm\in\Z_+^{\tilde B_\omega\setminus\tilde B_{I,J}}:\\ \text{odd on }
 B_\omega,\\ \text{even on }\tilde B_\omega\setminus\tilde B_{I,J}\setminus
 B_\omega}}\frac{w_{\tilde B_\omega\setminus\tilde B_{I,J}}(\bm)}{Z_{\B_\Lambda
 \setminus\tilde B_{I,J}}}\nn\\
\times\sum_{\substack{\bk\in\Z_+^{\B_\Lambda\setminus\tilde B_\omega}:\\
 \partial\bk=\vno}}w_{\B_\Lambda\setminus\tilde B_\omega}(\bk)\sum_{i=1}^{2N-1}
 \ind{\omega_{I_i}\cn{\bm+\bk}{}y\}\setminus\bigcup_{j>i}\{\omega_{I_j}\cn{\bm
 +\bk}{}y},
\end{align}
where $\omega_I=(\omega_s,\dots,\omega_{t-1})$ for $I=[s,t)$ (cf., below 
\refeq{Theta'bd2}) and
\begin{align}
(I_1,I_2,I_3,\dots,I_{2N-2},I_{2N-1})=([0,s_2),[s_2,t_1),[t_1,s_3),\dots,[s_N,
 t_{N-1}),[t_{N-1},|\omega|]). 
\end{align}
Then, by repeated applications of conditioning on clusters to extract 
two-point functions one by one (as done in showing 
\refeq{pi0decompR:Nge2-step52bd}) and using \refeq{lmm1} to deal with the 
indicator $\ind{\omega_{I_i}\cn{\bm+\bk}{}y}$, we can bound 
\refeq{tildepibd3} by \refeq{G*G} with $A$ replaced by a singleton $\{y\}$.  
Therefore, the contribution to \refeq{pi0'lace} from 
$\ind{o\cn{\bm+\bk}{}y\}\setminus\bigcup_{j=1}^N\{u_j\cn{\bk}{}y}$ in 
\refeq{inddecomp} obeys the same bound as $\Theta'_{o,x;\{y\}}$, with 
$U^\infty,\DotU^\infty_a,V^\infty,\DotV^\infty_a$ replaced by 
$U^1,\DotU^1_y,V^1,\DotV^1_y$, respectively.  Combining this with 
\refeq{tildepibd2}--\refeq{tildepibd1}, 
we complete the proof of Theorem~\ref{thm:pi0'bd}.
\QED

\subsection{Proof of Theorem~\ref{thm:Theta''bd}}\label{ss:thm5}
Recall the definition \refeq{Theta''def} of $\Theta''_{o,x,y;A}$:
\begin{align}\lbeq{Theta''redef}
\Theta''_{o,x,y;A}=\sum_{\substack{\bl\in\Z_+^{\B_{A\compl}}:\\ \partial\bl
 =\vno}}\frac{w_{\B_{A\compl}}(\bl)}{Z_{\B_{A\compl}}}\sum_{\substack{\bn\in
 \Z_+^{\B_\Lambda}:\\ \partial\bn=o\vtri x}}\frac{w_{\B_\Lambda}(\bn)}
 {Z_{\B_\Lambda}}\ind{o\db{\bn+\bl}{A}x\}\cap\{o\cn{\bn+\bl}{}y}.
\end{align}
First, by identifying the earliest path $\omega\in\Omega(o,x)$ of bonds $b$ 
with odd $n_b$ (as done in Step~1 of Section~\ref{ss:thm1}; cf., 
\refeq{Theta'decomp} and \refeq{pi0'decomp}), we can rewrite 
$\Theta''_{o,x,y;A}$ as
\begin{align}\lbeq{Theta''decomp}
\Theta''_{o,x,y;A}=\sum_{\omega\in\Omega(o,x)}\sum_{\substack{\bm\in\Z_+^{\tilde
 B_\omega}:\\ \text{odd on }B_\omega,\\ \text{even on }\tilde B_\omega
 \setminus B_\omega}}&\frac{w_{\tilde B_\omega}(\bm)}{Z_{\B_\Lambda}}
 \sum_{\substack{\bk\in\Z_+^{\B_\Lambda\setminus\tilde B_\omega}:\\ \partial
 \bk=\vno}}w_{\B_\Lambda\setminus\tilde B_\omega}(\bk)\nn\\
&\times\sum_{\substack{\bl\in\Z_+^{\B_{A\compl}}:\\ \partial\bl=\vno}}
 \frac{w_{\B_{A\compl}}(\bl)}{Z_{\B_{A\compl}}}\,\ind{o\db{\bm+\bk+\bl}{A}x\}
 \cap\{o\cn{\bm+\bk+\bl}{}y}.
\end{align}
Then, by the double expansion (as done in Step~2 of Section~\ref{ss:thm1}), we 
obtain the following inequality that is a mixture of \refeq{Theta'bd1} and 
\refeq{pi0'lace}:
\begin{gather}
\Theta''_{o,x,y;A}\le\sum_{N=1}^\infty\!\!\!
 \sum_{\substack{\{u_j,v_j\}_{j=1}^N\\ (\text{no intersection})}}\!\!\!
 \sum_{\omega\in\Omega(o,x)}\sum_{\{s_jt_j\}_{j=1}^N\in\cL_{[0,|\omega|]}^{\sss
 (N)}}\!\!\!\sum_{\substack{\bm\in\Z_+^{\tilde B_\omega}:\\ \text{odd on }
 B_\omega,\\ \text{even on }\tilde B_\omega\setminus B_\omega}}
 \!\!\!\!\!\frac{w_{\tilde B_\omega}(\bm)}{Z_{\B_\Lambda}}\prod_{j=1}^N
 \ind{u_j\in\tilde V_{\bm}(s_j)}\ind{v_j\in\tilde V_{\bm}(t_j)}\nn\\
\times\sum_{\substack{\bk\in\Z_+^{\B_\Lambda\setminus\tilde B_\omega}:\\
 \partial\bk=\vno}}w_{\B_\Lambda\setminus\tilde B_\omega}(\bk)\sum_{\substack{
 \bl\in\Z_+^{\B_{A\compl}}:\\ \partial\bl=\vno}}\frac{w_{\B_{A\compl}}(\bl)}
 {Z_{\B_{A\compl}}}\,\ind{o\cn{\bm+\bk+\bl}{A}x}\,\ind{o\cn{\bm+\bk+\bl}{}y}
 \nn\\
\times\ind{\sfL_{\bm,\bk+\bl}=\{s_jt_j\}_{j=1}^N}\prod_{j=1}^N\ind{u_j\cn{\bk
 +\bl}{}v_j\text{ in }\B_\Lambda\setminus\tilde B_\omega}.\lbeq{Theta''lace}
\end{gather}
Then we rewrite $\ind{o\cn{\bm+\bk+\bl}{}y}$, by using \refeq{inddecomp} with 
$\bk$ replaced by $\bk+\bl$, as
\begin{align}\lbeq{inddecomp2}
\ind{o\cn{\bm+\bk+\bl}{}y}=\sum_{i=1}^N\ind{u_i\cn{\bk+\bl}{}y}+\ind{o\cn{\bm
 +\bk+\bl}{}y\}\setminus\bigcup_{i=1}^N\{u_i\cn{\bk+\bl}{}y}.
\end{align}

For the contribution from $\sum_{i=1}^N\ind{u_i\cn{\bk+\bl}{}y}$, we ignore 
$\ind{o\cn{\bm+\bk+\bl}{A}x}$ and impose the through-$A$ condition only 
on $\bm$ by replacing the sum over $\omega$ by 
$\sum_{\omega:\omega\cap A\ne\vno}$, as done in \refeq{Theta'decomp}. 
Then the contribution from $\sum_{i=1}^N\ind{u_i\cn{\bk+\bl}{}y}$ is bounded 
by the right-hand side of \refeq{Theta'bd2} with 
$\prod_{j=1}^N\sum_{i_j=0}^\infty(\tilde G^2)^{*i_j}(v_j-u_j)$ replaced by 
\begin{align}\lbeq{haircomingout}
\sum_{i=1}^N\prod_{\substack{j=1\\ (j\ne i)}}^N\sum_{t_j=0}^\infty(\tilde
 G^2)^{*t_j}(v_j-u_j)&\sum_{\substack{z_1,z_2,z_3,\\ z'_1,z'_2,z'_3}}(\delta
 +\tau^2)(u_i-z_1)\,(\delta+\tau^2)(v_i-z_2)\,\delta_{z_3,y}\nn\\
&\times\prod_{k=1}^3\sum_{t_k=0}^\infty(\tilde G^2)^{*t_k}(z'_i-z_i)\,T(z'_1,z'_2,
 z'_3).
\end{align}
The rest is the same as described in the last paragraph of 
Section~\ref{ss:thm2}.  Consequently, the contribution from 
$\sum_{i=1}^N\ind{u_i\cn{\bk+\bl}{}y}$ is bounded by
\begin{align}\lbeq{Theta''bd1}
2\sum_{a\in A}&\bigg(\DDotX^\infty_{o,x;y}\delta_{a,x}+\sum_{i=0}^\infty
 \Big((U^\infty)^{\star i}\star\tfrac12\DDDotV^\infty_{a,y}\Big)_{o,x}
 +\sum_{i,j=0}^\infty\Big((U^\infty)^{\star i}\star\DDDotU^\infty_{a,y}\star
 (U^\infty)^{\star j}\star V^\infty\Big)_{o,x}\nn\\
&+\sum_{i,j=0}^\infty\Big((U^\infty)^{\star i}\star\DotU^\infty_a\star
 (U^\infty)^{\star j}\star\tfrac12\DDotV^\infty_y\Big)_{o,x}+\sum_{i,j
 =0}^\infty\Big((U^\infty)^{\star i}\star\DDotU^\infty_y\star(U^\infty)^{\star
 j}\star\DotV^\infty_a\Big)_{o,x}\nn\\
&+\sum_{i,j,k=0}^\infty\Big((U^\infty)^{\star i}\star\DotU^\infty_a\star
 (U^\infty)^{\star j}\star\DDotU^\infty_y\star(U^\infty)^{\star k}\star V^\infty
 \Big)_{o,x}\nn\\
&+\sum_{i,j,k=0}^\infty\Big((U^\infty)^{\star i}\star\DDotU^\infty_y\star
 (U^\infty)^{\star j}\star\DotU^\infty_a\star(U^\infty)^{\star k}\star V^\infty
 \Big)_{o,x}\bigg).
\end{align}

For the contribution from 
$\ind{o\cn{\bm+\bk+\bl}{}y\}\setminus\bigcup_{i=1}^N\{u_i\cn{\bk+\bl}{}y}$ in 
\refeq{inddecomp2}, on the other hand, we again ignore the indicator 
$\ind{o\cn{\bm+\bk+\bl}{A}x}$, but impose the through-$A$ condition on 
$\bk+\bl$ by replacing $\prod_{j=1}^N\ind{u_j\cn{\bk+\bl}{}v_j\text{ in }
\B_\Lambda\setminus\tilde B_\omega}$ in \refeq{Theta''lace} by
\begin{align}
\sum_{a\in A}\sum_{i=1}^N\prod_{\substack{j=1\\ (j\ne i)}}^N\ind{u_j\cn{\bk
 +\bl}{}v_j\text{ in }\B_\Lambda\setminus\tilde B_\omega}\,\ind{u_i\cn{\bk+\bl}
 {}v_i\text{ in }\B_\Lambda\setminus\tilde B_\omega\}\cap\{u_i\cn{\bk+\bl}{}a
 \text{ in }\B_\Lambda\setminus\tilde B_\omega},
\end{align}
which yields \refeq{haircomingout} with $\delta_{z_3,y}$ replaced by 
$\delta_{z_3,a}$ and summed over $a\in A$.  Then the rest is almost the same as 
described in the last paragraph of Section~\ref{ss:thm3}, except for two 
things: there are three current configurations involved, instead of two as in 
\refeq{tildepibd3}, and one of them is restricted on $\B_{A\compl}$.  Taking 
them into account, we can conclude that the contribution from 
$\ind{o\cn{\bm+\bk+\bl}{}y\}\setminus\bigcup_{i=1}^N\{u_i\cn{\bk+\bl}{}y}$ in 
\refeq{inddecomp2} is bounded by \refeq{Theta''bd1} with $a$ and $y$ swapped, 
$y$ replaced by $y'$ and then multiplied by 
$\sum_{i=0}^\infty(\tilde G^2)^{*i}(y-y')$.  This completes the proof of 
Theorem~\ref{thm:Theta''bd}.
\QED

\section{Application to the spread-out model}\label{s:appl2spread-out}
In this section, we demonstrate how to use the diagrammatic bounds proven in 
the previous section to derive the wanted $x$-space decay (see \refeq{pi0Qbd}, 
\refeq{pi1Qbd} and \refeq{pijQbd} below) for the spread-out model with $L\gg1$ in dimensions $d>4$.  
To do so, we repeatedly use the following convolution bound 
\cite[Lemma~3.2(i)]{cs15}, which is an improved version of 
\cite[Proposition~1.7]{hhs03}.

\begin{shaded}
\begin{lmm}[Convolution bound for the spread-out model \cite{cs15}]
Let
\begin{align}
\veee{x}_L=|x|\vee L.
\end{align}
For any $a\ge b>0$ with $a+b>d$, there is an $L$-independent constant 
$C=C(a,b,d)<\infty$ such that
\begin{align}\lbeq{convbd}
\sum_{y\in\Zd}\veee{x-y}_L^{-a}\,\veee{y}_L^{-b}\le
 \begin{cases}
 CL^{d-a}\veee{x}_L^{-b}\quad&(a>d),\\
 C\veee{x}_L^{d-a-b}&(a<d).
 \end{cases}
\end{align}
\end{lmm}
\end{shaded}

Throughout this section, we assume the following bound on $\|\tau\|_1$ and $G$.

\begin{shaded}
\begin{ass}\label{ass:bootstrap}
Let $J_{o,x}$ be the spread-out interaction \refeq{spread-out} and 
define $\theta=O(L^{-2})$ as in \refeq{theta-def}.  We assume
\begin{align}\lbeq{hyp1}
\|\tau\|_1\vee\sup_{x\ne o}\frac{G(x)}{\theta\veee{x}_L^{2-d}}\le2.
\end{align}
\end{ass}
\end{shaded}

As explained earlier, if $d>4$, $\theta\ll1$ and \refeq{aboveassumption} 
holds uniformly in $x\in\Lambda\subset\Zd$ and $\beta\le\betac$, then 
$G_{\betac}(x)\sim A'S_1(x)$ as $|x|\uparrow\infty$, where 
$A'\stackrel{\text{\refeq{A-def}}}=A\sigma^2=(1+O(L^{-2}))/\|\tau_{\betac}\|_1$ 
(the latest reference is \cite[(1.41)]{cs19} with $\alpha=\infty$) and 
$\|\tau_{\betac}\|_1=\|\Pi_{\betac}\|_1^{-1}=1+O(L^{-d})$ (due to 
\refeq{aboveassumption}).  As a result, the assumption \refeq{hyp1} indeed 
holds at $\betac$.  In fact, we can show that \refeq{hyp1} holds uniformly in 
$\betamf\le\beta<\betac$, where $\betamf$ is the mean-field critical point 
(i.e., $\|\tau_{\betamf}\|_1=1$), and the result at $\betac$ is obtained by the 
continuity in $\beta$ of the left-hand side of \refeq{hyp1}.  For more details, 
see, e.g., \cite[Theorem~3.3]{cs15} with $\alpha=\infty$.

Now we are left to show the inequality \refeq{aboveassumption} for each 
$\beta<\betac$, under Assumption~\ref{ass:bootstrap}.  To do so, we will 
frequently use the following bound and notation:

\begin{shaded}
\begin{lmm}\label{lmm:tildeGbd}
Under Assumption~\ref{ass:bootstrap}, we have
\begin{align}\lbeq{hyp2}
\sup_x\frac{\tilde G(x)}{\veee{x}_L^{2-d}}\le O(\theta),
\end{align}
where the implicit constant in $O(\theta)$ is independent of $L$.  This means 
that $\tilde G$ obeys the same $x$-space bound on $G$, modulo $L$-independent 
constant multiplication.  We denote this by
\begin{align}\lbeq{hyp3pre}
\tilde G(x)\lesssim G(x).
\end{align}
\end{lmm}
\end{shaded}

Notice that, by repeated use of \refeq{hyp3pre}, we have
\begin{align}\lbeq{hyp3}
(\tau^{*2}*\tilde G)(x)\lesssim(\tau^{*2}*G)(x)=(\tau*\tilde G)(x)\lesssim
 (\tau*G)(x)=\tilde G(x).
\end{align}
We will use this relation to bound the diagrammatic bounds on the expansion 
coefficients.

\Proof{Proof of Lemma~\ref{lmm:tildeGbd}.}
Since $h$ in \refeq{spread-out} is bounded and supported on $[-1,1]^d$, 
\begin{align}
\tau(x)=O(L^{-d})\,\ind{\|x\|_\infty\le L}\le\frac{O(L^2)}{\veee{x}_L^{d+2}}.
\end{align}
By \refeq{convbd} and \refeq{hyp1}, we obtain
\begin{align}
\tilde G(x)=\tau(x)+\sum_{y\ne o}\tau(x-y)\,G(y)\le\frac{O(\theta)}
 {\veee{x}_L^{d-2}}.
\end{align}
In particular, $\tilde G(o)=O(L^{-d})$, while $G(o)=1$.  This implies the 
relation \refeq{hyp3pre}.
\QED

\subsection{Bound on the $0^\text{th}$-order expansion coefficient}\label{ss:pi0}
First we recall Theorem~\ref{thm:pi0bd}.  The following proposition provides a 
bound on $\pi_{\B_\Lambda}^{\sss(0)}(x)$ for $x\ne o$.
\begin{shaded}
\begin{prp}\label{prp:xspacebd1}
Under Assumption~\ref{ass:bootstrap}, if $d>4$ and $\theta\ll1$, then, for any 
$m\ge1$, 
\begin{align}\lbeq{UmVmbds}
U^m(y,z;y',z')\lesssim U^0(y,z;y',z'),&&
V^m(y,z;x)\lesssim V^1(y,z;x),
\end{align}
i.e., $U^m$ and $V^m$ obey the same $x$-space bounds on $U^0$ and $V^1$, 
respectively, modulo $L$-independent constant multiplication. 
As a result, for $x\ne o$ and $m\ge1$,
\begin{align}\lbeq{Xmbd}
X^m_{o,x}\lesssim V^1(o,o;x)=\tilde G(x)^3.
\end{align}
\end{prp}
\end{shaded}

The following is an immediate consequence of \refeq{Xmbd} and 
Theorem~\ref{thm:pi0bd}.

\begin{shaded}
\begin{cor}[cf., (3.3) of \cite{s07}]\label{cor:pi0bd}
Under Assumption~\ref{ass:bootstrap}, if $d>4$ and $\theta\ll1$, then 
\begin{align}\lbeq{pi0Qbd}
\delta_{o,x}\le\pi_{\B_\Lambda}^{\sss(0)}(x)\le\delta_{o,x}+\frac{O(\theta^3)}
 {\veee{x}_L^{3(d-2)}},
\end{align}
where the implicit constant in $O(\theta^3)$ may depend on $d$, but not on $L$.
\end{cor}
\end{shaded}

\Proof{Proof of Proposition~\ref{prp:xspacebd1}.}
First we prove \refeq{UmVmbds}.  By the convolution bound 
\refeq{convbd}, degree-4 vertices can be eliminated one by one when $d>4$, 
as follows.  Since $|u-x|\vee|x-u'|\ge|u-u'|/2$ and 
$|v-x|\vee|x-v'|\ge|v-v'|/2$, we have
\begin{align}
&\sum_x\tilde G(u-x)\,\tilde G(x-u')\,\tilde G(v-x)\,\tilde G(x-v')\nn\\
&\stackrel{\text{\refeq{hyp2}}}\le\sum_x\frac{O(\theta)}{\veee{u-x}_L^{d-2}}
 \frac{O(\theta)}{\veee{x-u'}_L^{d-2}}\frac{O(\theta)}{\veee{v-x}_L^{d-2}}
 \frac{O(\theta)}{\veee{x-v'}_L^{d-2}}\nn\\
&~\:\le\frac{O(\theta)^4}{\veee{u-u'}_L^{d-2}\veee{v-v'}_L^{d-2}}\Bigg(
 \underbrace{\sum_x\frac{\ind{|u-x|\le|x-u'|}\,\ind{|v-x|\le|x-v'|}}{\veee{u
 -x}_L^{d-2}\veee{v-x}_L^{d-2}}}_{\le\,C\veee{u-v}_L^{4-d}\,\le\,CL^{4-d}~
 (\because\text{ \refeq{convbd} \& }d>4)}+~\text{[3 others]}\Bigg)\nn\\
&~\:\le\frac{O(\theta)}{\veee{u-u'}_L^{d-2}}\frac{O(\theta)}
 {\veee{v-v'}_L^{d-2}}\;O(\theta^2L^{4-d}),
\end{align}
where [3 others] is the sum of the contributions from 
$\ind{|u-x|\le|x-u'|}\ind{|v-x|>|x-v'|}$,
from $\ind{|u-x|>|x-u'|}\ind{|v-x|\le|x-v'|}$ and 
from $\ind{|u-x|>|x-u'|}\ind{|v-x|>|x-v'|}$.
Since $\theta=O(L^{-2})$, this may be depicted as
\begin{align}\lbeq{depicted0}
\raisebox{-1.2pc}{\includegraphics[scale=0.3]{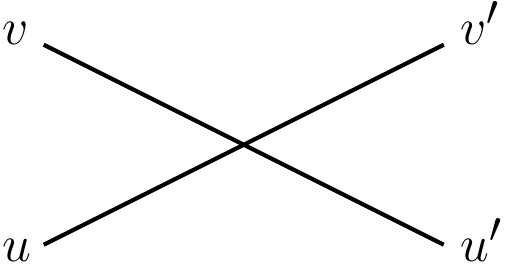}}~~\lesssim~~
 \raisebox{-1.2pc}{\includegraphics[scale=0.3]{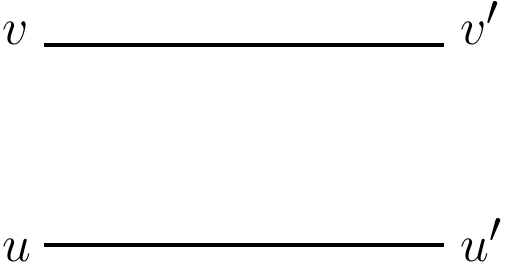}}~\times L^{-d}.
\end{align}
By taking $u=v$ and $u'=v'$, we obtain 
$\tilde G^2*\tilde G^2\lesssim L^{-d}\,\tilde G^2$.  By repeated use of 
this relation, we obtain 
\begin{align}\lbeq{bubblechainbound}
\sum_{j=1}^m(\tilde G^2)^{*j}\lesssim\tilde G^2\sum_{j=1}^\infty L^{-d(j-1)}
 \lesssim\tilde G^2,
\end{align}
which proves the second relation in \refeq{UmVmbds}.  Similarly, 
$(\delta+\tau^2)*\sum_{j=0}^m(\tilde G^2)^{*j}*(\delta+\tau^2)$ in 
\refeq{Um-def} can be evaluated, by using $\tau\le\tilde G$, 
\refeq{bubblechainbound} and then \refeq{hyp3pre}, as
\begin{eqnarray}
(\delta+\tau^2)^{*2}*\sum_{j=0}^m(\tilde G^2)^{*j}
&\stackrel{\tau\le\tilde G}\le&(\delta+\tilde G^2)^{*2}*\bigg(\delta+\sum_{j
 =1}^m(\tilde G^2)^{*j}\bigg)\nn\\
&\stackrel{\text{\refeq{bubblechainbound}}}\lesssim&(\delta+\tilde G^2)^{*
 3}\nn\\
&\stackrel{\text{\refeq{depicted0}}}\lesssim&\delta+\tilde G^2\nn\\[5pt]
&\le&(\delta+\tilde G)^2\nn\\
&\stackrel{\text{\refeq{hyp3pre}}}\lesssim&G^2.\lbeq{psi1}
\end{eqnarray}
This implies that the sum of the bubble chain in Figure~\ref{fig:UVdef} 
(including the black disks at both ends of the chain) can be replaced by 
$G^2$, at the cost of $L$-independent constant multiplication.  
This proves the first relation in \refeq{UmVmbds}.

As a result, we have
\begin{align}\lbeq{Xmlesssimbd}
X_{o,x}^m\lesssim\sum_{i=0}^\infty\big((U^0)^{\star i}\star V^1\big)_{o,x}.
\end{align}
To prove \refeq{Xmbd}, we repeatedly use the convolution bound \refeq{convbd} 
to eliminate all diagram vertices of degree 4 one by one.  For example, if one 
of the four line segments in \refeq{depicted0}, say, between $u$ and $x$, is 
slashed, then we use \refeq{tildeGdef} to replace $G(u-x)$ by 
$\delta_{u,x}+\tilde G(u-x)$.  The contribution from $\tilde G(u-x)$ is 
identical to \refeq{depicted0}.  The contribution from 
$\delta_{u,x}$ is equal to $\tilde G(u-u')\,\tilde G(v-u)\,\tilde G(u-v')$.  
However, since $|v-u|\vee|u-v'|\ge|v-v'|/2$, we have 
$\tilde G(v-u)\,\tilde G(u-v')\lesssim L^{-d}\,\tilde G(v-v')$.  Therefore,
\begin{align}\lbeq{depicted1}
\raisebox{-1.2pc}{\includegraphics[scale=0.3]{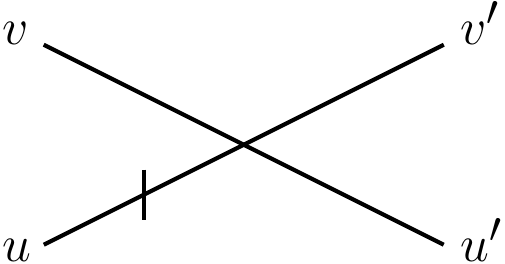}}~~\lesssim~~
 \raisebox{-1.2pc}{\includegraphics[scale=0.3]{0slash0}}~\times L^{-d}.
\end{align}
Similarly, we obtain
\begin{align}\lbeq{depicted2}
\raisebox{-1.2pc}{\includegraphics[scale=0.3]{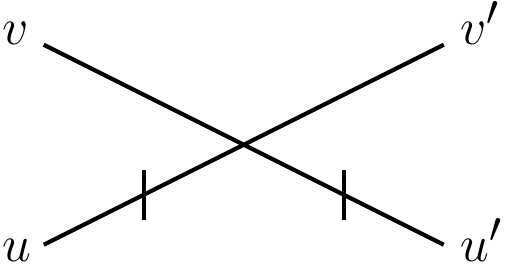}}~~&\lesssim~~
 \raisebox{-1.2pc}{\includegraphics[scale=0.3]{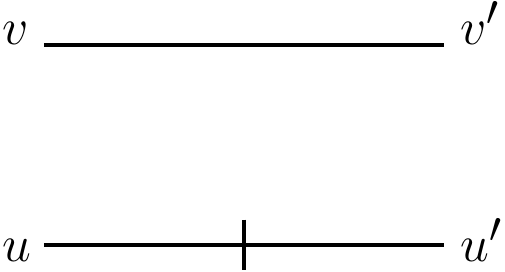}}~\times L^{-d},
\end{align}
\begin{align}\lbeq{depicted3}
\left.\begin{array}{c}
 \raisebox{-1.2pc}{\includegraphics[scale=0.3]{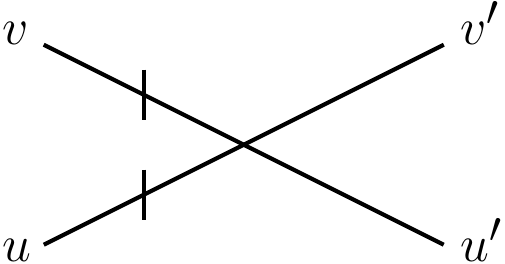}}\\[2pc]
 \raisebox{-1.2pc}{\includegraphics[scale=0.3]{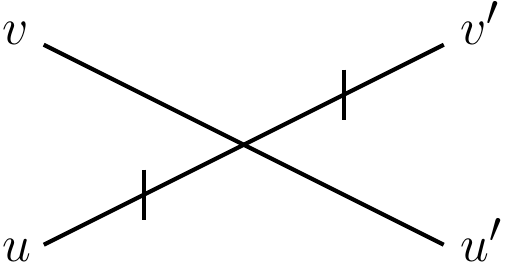}}
 \end{array}\right\}&\lesssim~~
 \raisebox{-1.2pc}{\includegraphics[scale=0.3]{0slash0}}~\times1,
\end{align}
\begin{align}\lbeq{depicted4}
\raisebox{-1.2pc}{\includegraphics[scale=0.3]{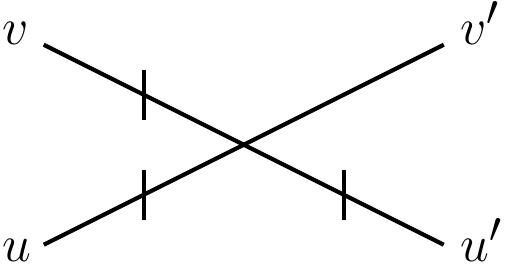}}~~&\lesssim~~
 \raisebox{-1.2pc}{\includegraphics[scale=0.3]{1slash0}}~\times1,
\end{align}
\begin{align}\lbeq{depicted5}
\raisebox{-1.2pc}{\includegraphics[scale=0.3]{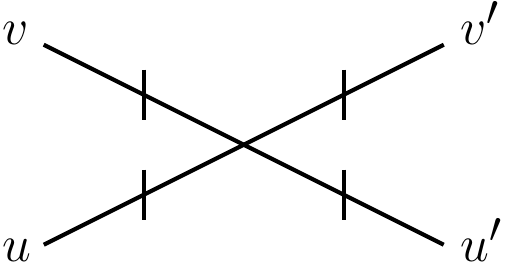}}~~&\lesssim~~
 \raisebox{-1.2pc}{\includegraphics[scale=0.3]{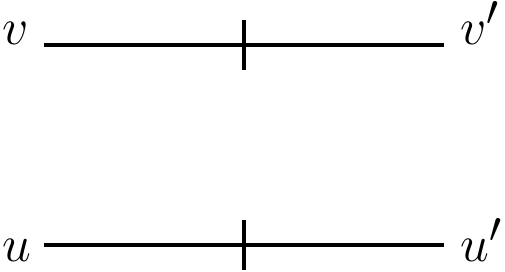}}~\times1.
\end{align}
As a rule of thumb, the factor of $L^{-d}$ arises when at least one of 
those two removed line segments is unslashed.  

\begin{figure}[t]
\begin{align*}
\raisebox{-1.2pc}{\includegraphics[scale=0.6]{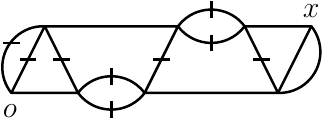}}~\lesssim~
 \raisebox{-7pt}{\includegraphics[scale=0.6]{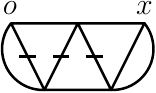}}\times1^3\lesssim~
 \raisebox{-1.2pc}{\includegraphics[scale=0.6]{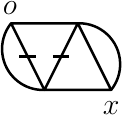}}\times1^3L^{-d}
  \lesssim~
 \raisebox{-7pt}{\includegraphics[scale=0.6]{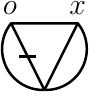}}\times1^3(L^{-d})^2
\end{align*}
\caption{Reduction of the simplified version of the $n=3$ term in 
\refeq{Xdef} to even simpler diagrams, by using \refeq{depicted4} three times 
and then using \refeq{depicted1} twice.  Using \refeq{depicted1} once more 
yields $V^1(o,o;x)=\tilde G(x)^3$ multiplied by a factor of $(L^{-d})^3$
.}
\label{fig:reduction}
\end{figure}

To evaluate the $i^\text{th}$ term in \refeq{Xmlesssimbd}, we first use 
\refeq{depicted4} with $u=v$ to eliminate all bubbles ($=G^2$) and 
then use \refeq{depicted1} to eliminate all degree-4 vertices (see 
Figure~\ref{fig:reduction}).  As a result, the $i^\text{th}$ term in 
\refeq{Xmlesssimbd} is reduced to the simplest diagram 
$V^1(o,o;x)=\tilde G(x)^3$ multiplied by a factor of $L^{-di}$, 
which is summable in $i$ if $L\gg1$.  This completes the proof of 
\refeq{Xmbd}.
\QED

\subsection{Bound on the $1^\text{st}$-order expansion coefficient}\label{ss:pi1}
Recall \refeq{pi1prebd} and \refeq{tildepibd}, where $X,\DotX,\DDotX$ are 
involved.  We have already shown that 
$X_{o,x}^m\lesssim V^1(o,o;x)=\tilde G(x)^3$ if $d>4$ and $\theta\ll1$ under 
Assumption~\ref{ass:bootstrap}.  It remains to investigate $\DotX$ and $\DDotX$.

First we investigate $\DotX$.  By the same reason as in 
Proposition~\ref{prp:xspacebd1}, $\DotU^m_a$ and $\DotV^m_a$ obey the same 
$x$-space bounds on $\DotU^0_a$ and $\DotV^1_a$, respectively, modulo 
$L$-independent constant multiplication, if $d>4$ and $L\gg1$ under 
Assumption~\ref{ass:bootstrap}.  Then, by repeated use of the 
convolution bound \refeq{convbd} (as in Figure~\ref{fig:reduction}), we can 
show that $(U^m)^{\star i}\star\DotV^m_a$ and 
$(U^m)^{\star i}\star\DotU^m_a\star(U^m)^{\star j}\star V^m$ in 
\refeq{dotXdef} obey the same $x$-space bound on $\DotV^1_a(o,o;x)$ 
multiplied by factors of $L^{-di}$ and $L^{-d(i+j)}$, respectively, 
which are summable if $L\gg1$.  As a result, we obtain the following:

\begin{shaded}
\begin{prp}\label{prp:xspacebd2}
Under Assumption~\ref{ass:bootstrap}, if $d>4$ and $\theta\ll1$, then, for any 
$m\ge1$, 
\begin{align}\lbeq{dotUmdotVmbds}
\DotU^m_a(y,z;y',z')\lesssim\DotU^0_a(y,z;y',z'),&&
\DotV^m_a(y,z;x)\lesssim\DotV^1_a(y,z;x).
\end{align}
As a result, for $x\ne o$,
\begin{align}\lbeq{dotXmbd}
\DotX^m_{o,x;a}\lesssim\DotV^1_a(o,o;x)=\tilde G(x)^2\,G(a)\,\tilde G(x-a).
\end{align}
\end{prp}
\end{shaded}

Recall \refeq{Theta'bd}, \refeq{Xmbd} and \refeq{dotXmbd}.  
Since, for $x\ne o$, 
\begin{align}\lbeq{xneobd}
 \left.\begin{array}{l}
 V^1(o,o;x)\delta_{x,a}=\tilde G(x)^3\delta_{x,a}\\
 \DotV^1_a(o,o;x)=\tilde G(x)^2\,G(a)\,\tilde
  G(x-a)
 \end{array}\right\}
 \stackrel{\text{\refeq{hyp3pre}}}\lesssim\tilde G(x)^2\,G(a)\,G(x-a),
\end{align}
and since $\Theta'_{x,x;A}=\ind{x\in A}$, we obtain the following:

\begin{shaded}
\begin{cor}\label{cor:Theta'xbd}
Under Assumption~\ref{ass:bootstrap}, if $d>4$ and $\theta\ll1$, then
\begin{align}\lbeq{Theta'xbd}
\Theta'_{o,x;A}
 \lesssim\sum_{a\in A}\raisebox{-1.5pc}{\includegraphics
 [scale=0.25]{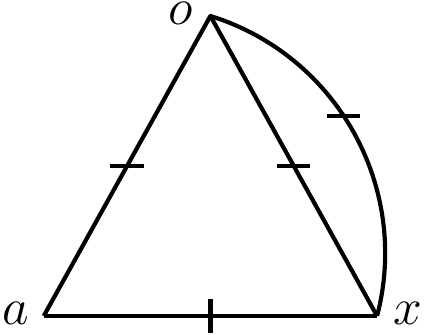}}~=\sum_{a\in A}G(x)^2\,G(a)\,G(x-a).
\end{align}
\end{cor}
\end{shaded}

\begin{figure}[t]
\begin{align*}
\DDotU^0_a(y,z;y',z')=~\raisebox{-1.7pc}{\includegraphics[scale=0.3]
 {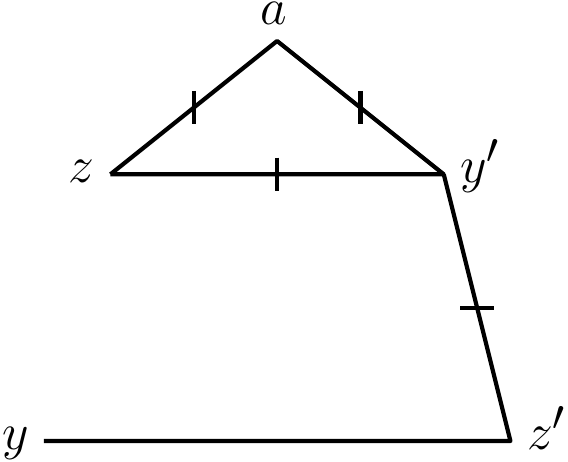}}&&&&
\DDotV^0_a(y,z;x)=~\raisebox{-1.8pc}{\includegraphics[scale=0.3]
 {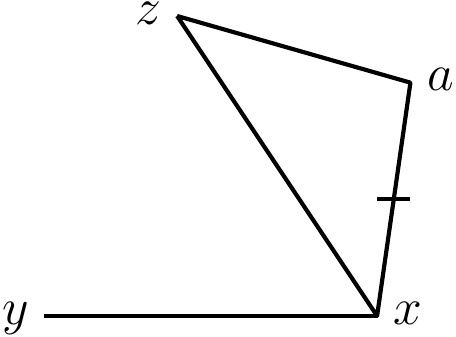}}
\end{align*}
\caption{Schematic representations of $\DDotU^0_a(y,z;y',z')$ and 
$\DDotV^0_a(y,z;x)$ in \refeq{ddotU0}--\refeq{ddotV0}.}
\label{fig:ddotU0V0}
\end{figure}

Next we investigate $\DDotX$.  Again, by repeated use of the convolution 
bound \refeq{convbd}, we can show that, if $d>4$ and $L\gg1$ under 
Assumption~\ref{ass:bootstrap}, $\DDotU^m_a$ and $\DDotV^m_a$ obey the same 
$x$-space bounds on $\DDotU^1_a$ and $\DDotV^1_a$, respectively, where
\begin{align}
\DDotU^1_a(y,z;y',z')=\tilde G(z'-y)\,G(z'-y')&\sum_{v,v'}(\delta+\tau^2)(z-v)
 \,(\delta+\tau^2)(y'-v')\,T(v,v',a),\\
\DDotV^1_a(y,z;x)=\ind{z\ne x}\,\tilde G(x-y)&\sum_v(\delta+\tau^2)(z-v)\,
 T(v,x,a).\lbeq{ddotV1}
\end{align}
Moreover, by using \refeq{depicted5} once, we have
\begin{align}
T(v,x,a)\lesssim G(x-v)\,G(a-v)\,G(x-a).
\end{align}
Plugging this back to \refeq{ddotV1} yields
\begin{align}
\DDotV^1_a(y,z;x)\lesssim\tilde G(x-y)\bigg(&\ind{z\ne x}\,\underbrace{G(x
 -z)}_{\le\,\tilde G(x-z)}\,\underbrace{G(a-z)\,G(x-a)}_{\le\,2\tilde G(a-z)
 \,G(x-a)}\nn\\
&+\underbrace{\sum_v\tau(z-v)^2\,G(x-v)\,G(a-v)}_{\le\,\tilde
 G(x-z)\,\tilde G(a-z)}\,G(x-a)\bigg),
\end{align}
where we have used 
\begin{eqnarray}
G(a-z)\,G(x-a)&\le&\big(\delta_{z,a}+\tilde G(a-z)\big)\,G(x-a)\nn\\
&=&G(x-z)+\tilde
 G(a-z)\,G(x-a)\nn\\
&\stackrel{z\ne x}\le&\tilde G(x-z)+\tilde G(a-z)\,G(x-a)\nn\\
&\stackrel{\delta\le G}\le&2\tilde G(a-z)\,G(x-a).
\end{eqnarray}
Similarly, we can show (cf., \refeq{psi1})
\begin{align}
\DDotU^1_a(y,z;y',z')\lesssim\tilde G(z'-y)\,G(z'-y')\,G(y'-z)\,G(a-z)\,G(y'-a).
\end{align}
Let (see Figure~\ref{fig:ddotU0V0})
\begin{align}
\DDotU^0_a(y,z;y',z')&=\tilde G(z'-y)\,G(z'-y')\,G(y'-z)\,G(a-z)\,G(y'-a),
 \lbeq{ddotU0}\\
\DDotV^0_a(y,z;x)&=\tilde G(x-y)\,\tilde G(x-z)\,\tilde G(a-z)\,G(x-a).
 \lbeq{ddotV0}
\end{align}
Then, $\DDotX^m_{o,x;y}$ in \refeq{ddotXdef} obeys the same $x$-space bound 
on $\DDotX^0_{o,x;y}$.  Repeated applications of the convolution bound 
\refeq{convbd} to $\DDotX^0_{o,x;y}$ (as in Figure~\ref{fig:reduction}), 
we can show that $\DDotX^0_{o,x;y}$ obeys the same $x$-space bound on 
$\DDotV^0_y(o,o;x)$, if $d>4$ and $L\gg1$ under Assumption~\ref{ass:bootstrap}. 
As a result, we obtain the following:

\begin{shaded}
\begin{prp}\label{prp:xspacebd3}
Under Assumption~\ref{ass:bootstrap}, if $d>4$ and $\theta\ll1$, then, for any 
$m\ge1$, 
\begin{align}\lbeq{ddotUmdotVmbds}
\DDotU^m_a(y,z;y',z')\lesssim\DDotU^0_a(y,z;y',z'),&&
\DDotV^m_a(y,z;x)\lesssim\DDotV^0_a(y,z;x).
\end{align}
As a result, for $x\ne o$,
\begin{align}\lbeq{ddotXmbd}
\DDotX^m_{o,x;y}\lesssim\DDotV^0_y(o,o;x)=\tilde G(x)^2\,\tilde G(y)\,G(x-y).
\end{align}
\end{prp}
\end{shaded}

Recall \refeq{tildepibd}, \refeq{Xmbd} and \refeq{dotXmbd} (also 
\refeq{xneobd}).  Since $\tilde\pi_{\B_\Lambda;y}^{\sss(0)}(o)\le G(y)^2$ 
(cf., \refeq{lmm0}), we readily obtain the following:

\begin{shaded}
\begin{cor}\label{cor:tildepi0xbd}
Under Assumption~\ref{ass:bootstrap}, if $d>4$ and $\theta\ll1$, then
\begin{align}\lbeq{tildepi0xbd}
\tilde\pi_{\B_\Lambda;y}^{\sss(0)}(x)\lesssim~\raisebox{-1.5pc}{\includegraphics
 [scale=0.25]{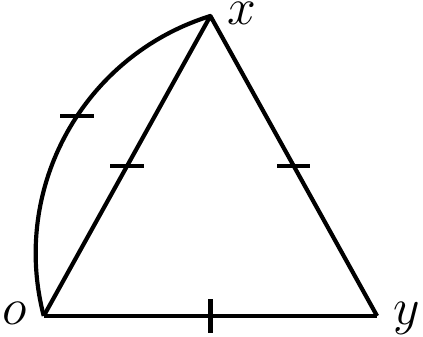}}~= G(x)^2\,G(y)\,G(x-y).
\end{align}
\end{cor}
\end{shaded}

Substituting this back into \refeq{pi1prebd} and using \refeq{depicted4} and 
\refeq{depicted5}, we can conclude the following wanted $x$-space decay of 
$\pi_{\B_\Lambda}^{\sss(1)}$:

\begin{shaded}
\begin{cor}[cf., (3.3) of \cite{s07}]
Under Assumption~\ref{ass:bootstrap}, if $d>4$ and $\theta\ll1$, then
\begin{align}\lbeq{pi1Qbd}
\pi_{\B_\Lambda}^{\sss(1)}(x)\lesssim~
 \raisebox{-1.5pc}{\includegraphics[scale=0.25]{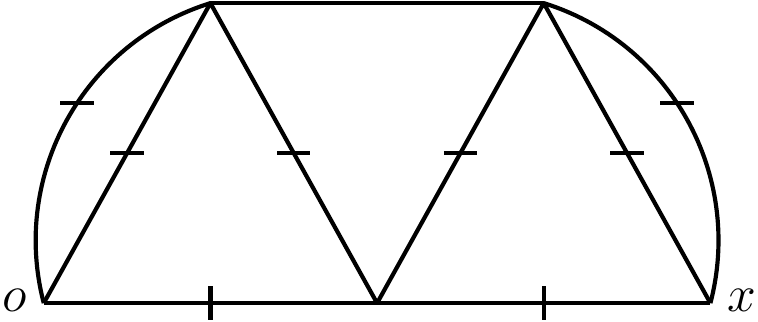}}
 ~\le O(L^{-d})\,\delta_{o,x}+\frac{O(\theta^3)}{\veee{x}_L^{3(d-2)}},
\end{align}
where the implicit constants in $O(L^{-d})$ and $O(\theta^3)$ may depend 
on $d$, but not on $L$.
\end{cor}
\end{shaded}

\subsection{Bound on the higher-order expansion coefficient}\label{ss:higher}
Recall \refeq{pi2prebd} and \refeq{Theta''bd}, where $X,\DotX,\DDotX,\DDDotX$ 
are involved.  The first three obey the bounds in \refeq{Xmbd}, \refeq{dotXmbd} 
and \refeq{ddotXmbd}.  It remains to investigate $\DDDotX$ in 
\refeq{DDDotXdef}.  By Propositions~\ref{prp:xspacebd1}, \ref{prp:xspacebd2} 
and \ref{prp:xspacebd3}, we can reduce $U^m,V^m,\DotU^m,\DotV^m,\DDotU^m$ and 
$\DDotV^m$ in \refeq{DDDotXdef} to simpler $U^0,V^1,\DotU^0,\DotV^1,\DDotU^0$ 
and $\DDotV^0$, respectively, if $d>4$ and $L\gg1$ under Assumption~\ref{ass:bootstrap}.  
Similarly, we can reduce $\DDDotU^m$ and $\DDDotV^m$ to $\DDDotU^0$ and 
$\DDDotV^0$, respectively, where
\begin{align}
\DDDotU^0_{a,v}(y,z;y',z')&=\Big(G(a-y)\,\tilde G(z'-a)\,G(z'-y')+\tilde G(z'-y)
 \,\tilde G(a-y')\,G(z'-a)\Big)\nn\\
&\quad\times G(y'-z)\,G(v-z)\,G(y'-v),\lbeq{dddotU0}\\
\DDDotV^0_{a,v}(y,z;x)&=G(a-y)\,\tilde G(x-a)\,\tilde G(x-z)\,\tilde G(v-z)\,G(x
 -v).\lbeq{dddotV0}
\end{align}
Moreover, due to the observation below \refeq{depicted5}, 
the sums over $i,j,k$ are convergent if $d>4$ and $L\gg1$, and the dominant 
terms come from the $i=j=k=0$ case.  Among those six terms, the largest (modulo 
$L$-independent constant multilpication) is $\DDDotV^0_{a,y}(o,o;x)$.  This is 
summarised as follows:

\begin{shaded}
\begin{prp}\label{prp:xspacebd4}
Under Assumption~\ref{ass:bootstrap}, if $d>4$ and $\theta\ll1$, then, for any 
$m\ge1$, 
\begin{align}\lbeq{dddotUmdotVmbds}
\DDDotU^m_{a,v}(y,z;y',z')\lesssim\DDDotU^0_{a,v}(y,z;y',z'),&&
\DDDotV^m_{a,v}(y,z;x)\lesssim\DDDotV^0_{a,v}(y,z;x).
\end{align}
As a result, for $x\ne o$,
\begin{align}\lbeq{dddotXmbd}
\DDDotX^m_{o,x;a,y}\lesssim\DDDotV^0_{a,y}(o,o;x)=G(a)\,\tilde G(x-a)\,\tilde
 G(x)\,\tilde G(y)\,G(x-y).
\end{align}
\end{prp}
\end{shaded}

Since $\Theta''_{x,x,y;A}\le\ind{x\in A}\sum_{j=0}^\infty(\tilde G^2)^{*j}(y)
\lesssim\ind{x\in A}\,G(y)^2$, we readily obtain the following:

\begin{shaded}
\begin{cor}
Under Assumption~\ref{ass:bootstrap}, if $d>4$ and $\theta\ll1$, then
\begin{align}
\Theta''_{o,x,y;A}\lesssim\sum_{a\in A}~\raisebox{-1.5pc}{\includegraphics
 [scale=0.25]{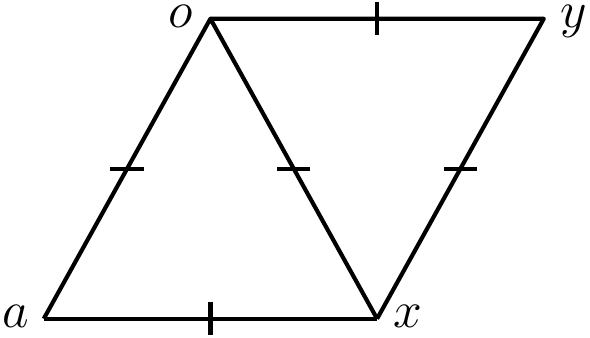}}~=\sum_{a\in A}G(a)\,G(x-a)\,G(x)\,G(y)\,G(x-y).
\end{align}
\end{cor}
\end{shaded}

Substituting this back into \refeq{pi2prebd}, generalizing it to 
$\pi_{\B_\Lambda}^{\sss(j)}(x)$ for $j\ge2$, and repeatedly using the 
convolution bounds \refeq{depicted2} and \refeq{depicted4}, 
we can conclude the following:

\begin{shaded}
\begin{cor}\label{cor:pige2bd}
Under Assumption~\ref{ass:bootstrap}, if $d>4$ and $\theta\ll1$, then, for 
$j\ge2$,
\begin{align}\lbeq{pijQbd}
\pi_{\B_\Lambda}^{\sss(j)}(x)&\lesssim\sum_{\substack{y_1,\dots,y_j,\\ z_1,
 \dots,z_j}}\raisebox{-1.5pc}{\includegraphics[scale=0.25]{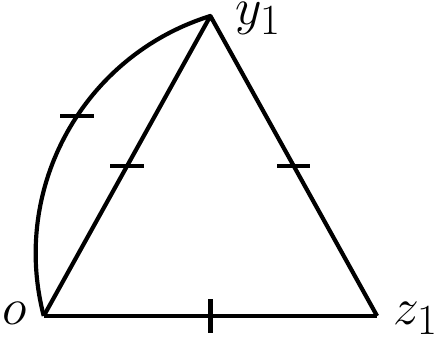}}\prod_{i
 =1}^{j-1}\left(\raisebox{-1.5pc}{\includegraphics[scale=0.25]{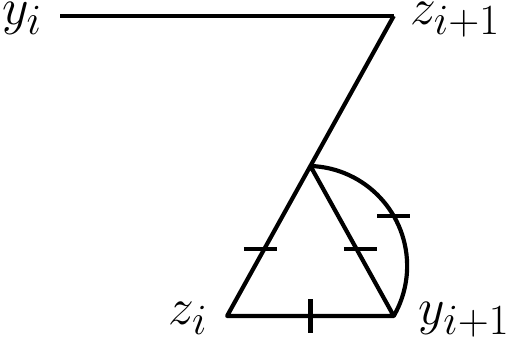}}
 +\raisebox{-1.5pc}{\includegraphics[scale=0.25]{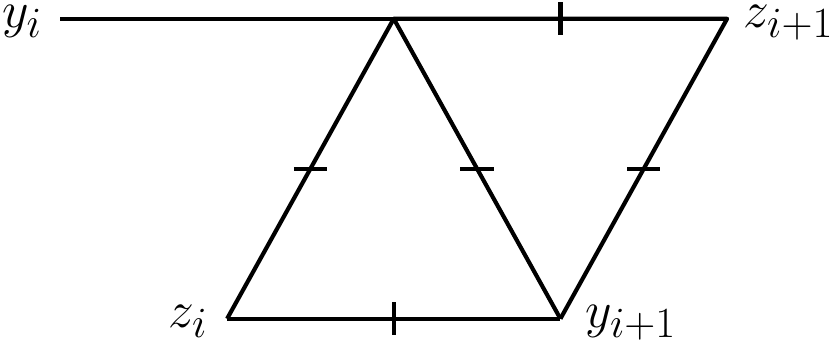}}\right)
 \raisebox{-1.5pc}{\includegraphics[scale=0.25]{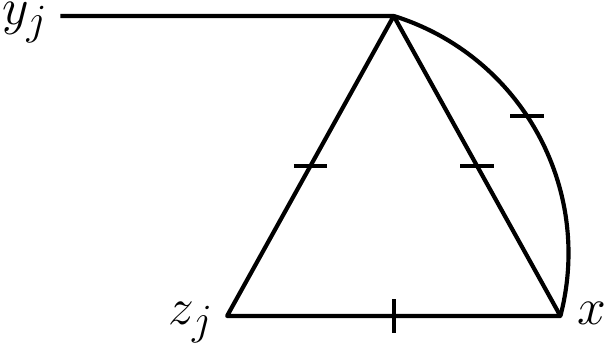}}\nn\\
&\le O(L^{-jd})\,\delta_{o,x}+\frac{O(L^{-d(j-2)}\theta^3)}{\veee{x}_L^{3(d-2)}},
\end{align}
where the implicit constants in $O(L^{-jd})$ and $O(L^{-d(j-2)}\theta^3)$ may 
depend on $d$, but not on $L$.  This is an improved version of \cite[(3.3)]{s07} 
(see also \cite[(3.4)]{cs15} and \cite[(3.22)]{cs19}.
\end{cor}
\end{shaded}

\Proof{Proof of the last line of \refeq{pijQbd}.}
First we note that, by \refeq{depicted4},
\begin{align}
\raisebox{-1.5pc}{\includegraphics[scale=0.25]{pijblock21}}~
 +~\raisebox{-1.5pc}{\includegraphics[scale=0.25]{pijblock22}}~\stackrel{d>4}
 \lesssim~\raisebox{-1.8pc}{\includegraphics[scale=0.25]{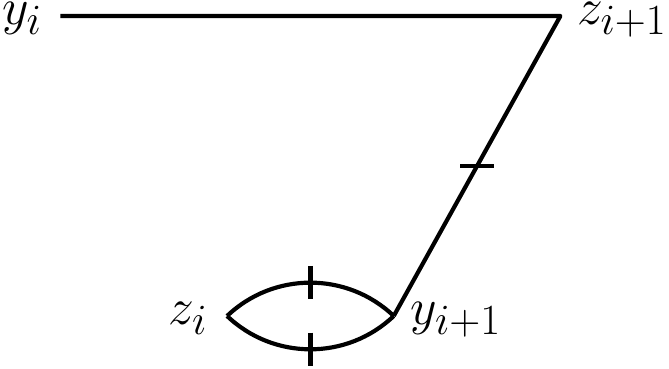}}~
 \stackrel{d>4}\lesssim~\raisebox{-1.5pc}{\includegraphics[scale=0.25]
 {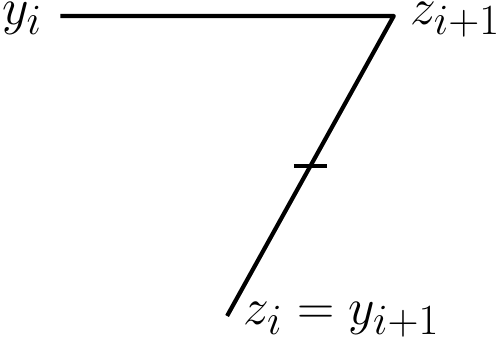}}~.
\end{align}
Moreover, by using \refeq{depicted2} twice, we have
\begin{align}
\raisebox{-1.2pc}{\includegraphics[scale=0.25]{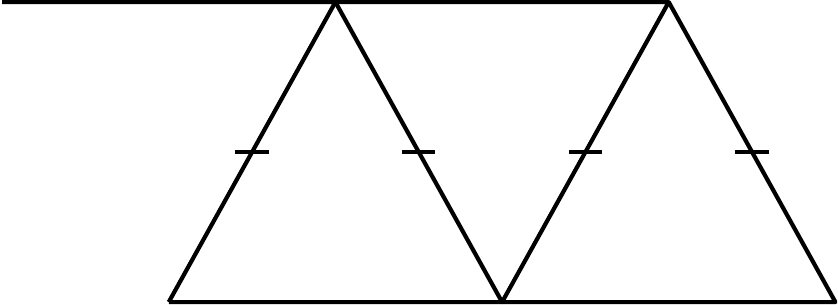}}~\stackrel{d>4}
 \lesssim~\raisebox{-1.6pc}{\includegraphics[scale=0.25]{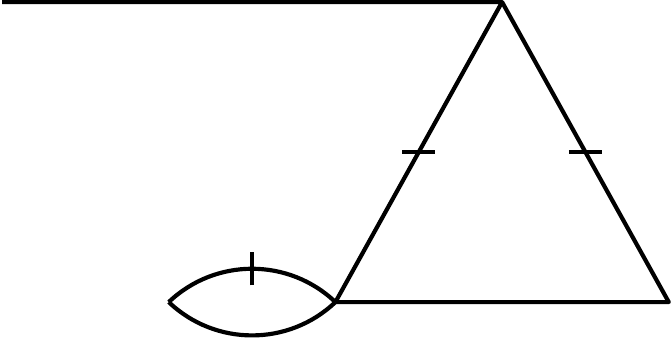}}~\times L^{-d}
 \stackrel{d>4}\lesssim~\raisebox{-1.2pc}{\includegraphics[scale=0.25]{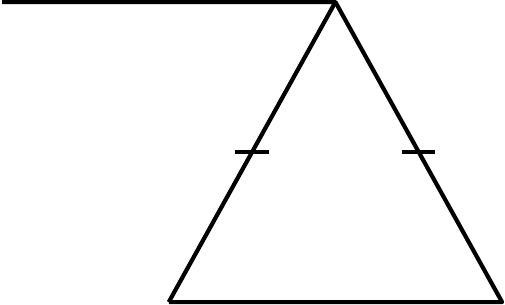}}
 ~\times(L^{-d})^2,
\end{align}
hence the recurrence formula $\pi_{\B_\Lambda}^{\sss(j)}(x)\lesssim(L^{-d})^2
\,\pi_{\B_\Lambda}^{\sss(j-2)}(x)$ for $j\ge4$.  However, by repeated use of 
\refeq{depicted4}, we obtain 
$\pi_{\B_\Lambda}^{\sss(2)}(x)\lesssim\tilde G(x)^2\,G(x)$ and 
$\pi_{\B_\Lambda}^{\sss(3)}(x)\lesssim L^{-d}\,\tilde G(x)^2\,G(x)$.  Therefore,
\begin{align}
\pi_{\B_\Lambda}^{\sss(j)}(x)\lesssim(L^{-d})^{j-2}\,\tilde G(x)^2\,G(x)
 \stackrel{\text{\refeq{tildeGdef}}}\le(L^{-d})^{j-2}\big(\underbrace{\tilde G
 (o)^2}_{\lesssim\,L^{-2d}}\delta_{o,x}+\tilde G(x)^3\big),
\end{align}
as required.
\QED

\section*{Acknowledgements}
This work was supported by JSPS KAKENHI Grant Number 18K03406.  
I would like to thank Hugo Duminil-Copin for drawing my atention to 
the issue discussed in this paper. 
I would also like to thank Satoshi Handa for working together in the 
early stage of this project.

\end{document}